%% file: TOPQ-2017-14-PAPER.tex
\pdfoutput=1
\newcommand*{\ATLASLATEXPATH}{}
\documentclass[cernpreprint, UKenglish, texlive=2016]{\ATLASLATEXPATH atlasdoc}
 
\usepackage{\ATLASLATEXPATH atlaspackage}
\usepackage{\ATLASLATEXPATH atlasbiblatex}
 
\usepackage{\ATLASLATEXPATH atlasphysics}
\usepackage{multirow}
\usepackage{amssymb}
\usepackage{url}
 
\graphicspath{{logos/}{figures/}}
 
\usepackage{TOPQ-2017-14-PAPER-defs}

\AtlasTitle{Measurements of inclusive and differential fiducial cross-sections of $t\bar{t}\gamma$ production in leptonic final states
at $\sqrt{s}=13~\text{TeV}$ in ATLAS}

\AtlasAbstract{
Inclusive and differential cross-sections for the production of a top-quark pair in association with a photon are measured with proton-proton collision data corresponding to an integrated luminosity of 36.1~\mbox{fb$^{-1}$}, collected by the ATLAS detector at the LHC in 2015 and 2016 at a centre-of-mass energy of 13~\ensuremath{\text{Te\kern -0.1em V}}. The measurements are performed in single-lepton and dilepton final states in a fiducial volume. Events with exactly one photon, one or two leptons, a channel-dependent minimum number of jets, and at least one $b$-jet are selected. Neural network algorithms are used to separate the signal from the backgrounds. The fiducial cross-sections are measured to be $521 \pm 9\text{(stat.)} \pm 41\text{(sys.)}~\text{fb}$ and $69 \pm 3\text{(stat.)} \pm 4\text{(sys.)}~\text{fb}$ for the single-lepton and dilepton channels, respectively. The differential cross-sections are measured as a function of photon transverse momentum, photon absolute pseudorapidity, and angular distance between the photon and its closest lepton in both channels, as well as azimuthal opening angle and absolute pseudorapidity difference between the two leptons in the dilepton channel. All measurements are in agreement with the theoretical predictions.
}
 
\author{The ATLAS Collaboration}

\PreprintIdNumber{CERN-EP-2018-302}

\AtlasJournalRef{\EPJC 79 (2019) 382}
\AtlasDOI{10.1140/epjc/s10052-019-6849-6}

\hypersetup{pdftitle={ATLAS document},pdfauthor={The ATLAS Collaboration}}
 
\begin{document}
 
\maketitle

\section{Introduction}
 
Measurements of top-quark properties play an important role in testing the Standard Model (SM) and its possible extensions. Studies of the production and kinematic properties of a top-quark pair in association with a photon (\ttg) probe the $t\gamma$ electroweak coupling. For instance, deviations in the transverse momentum (\pT) spectrum of the photon from the SM prediction could point to new physics through anomalous dipole moments of the top quark~\cite{Baur:2004uw, Bouzas:2012av, Schulze:2016qas}. A precision measurement of the \ttg production cross-section could effectively constrain some of the Wilson coefficients in top-quark effective field theories~\cite{Bylund:2016phk}. Furthermore, differential distributions of photon production in $\ttbar$ events could provide insight on the $\ttbar$ production mechanism, in particular about the $\ttbar$ spin correlation and the charge asymmetry~\cite{Aguilar-Saavedra:2014vta}.
 
Evidence for the production of a top-quark pair in association with an energetic, isolated photon was found in proton-antiproton (\ppbar) collisions at the Tevatron collider at a centre-of-mass energy of $\sqrt{s} = 1.96~\TeV$ by the CDF Collaboration~\cite{Aaltonen:2011sp}. Observation of the \ttg process was reported by the ATLAS Collaboration in proton-proton ($pp$) collisions at $\sqrt{s}=7~\TeV$~\cite{TOPQ-2012-07}. Recently, both the ATLAS and CMS Collaborations measured the \ttg cross-section at $\sqrt{s}=8~\TeV$~\cite{TOPQ-2015-21,CMS-TOP-14-008}. In the ATLAS measurement, the differential cross-sections with respect to the transverse momentum $p_T$ and absolute pseudorapidity $|\eta|$\footnote{ATLAS uses a right-handed coordinate system with its origin at the nominal interaction point (IP) in the centre of the detector and the $z$-axis along the beam pipe. The $x$-axis points from the IP to the centre of the LHC ring, and the $y$-axis points upward. Cylindrical coordinates $(r,\phi)$ are used in the transverse plane, $\phi$ being the azimuthal angle around the $z$-axis. The pseudorapidity is defined in terms of the polar angle $\theta$ as $\eta=-\ln\tan(\theta/2)$.} of the photon were reported. In the CMS measurement, the ratio of the \ttg fiducial cross-section to the \ttbar total cross-section was measured.
 
This \paperornote describes a measurement of the \ttg production cross-section in final states with one or two leptons, electron or muons, referred to as the \chljets or \chll channel, based on a data set recorded at the LHC in 2015 and 2016 at a centre-of-mass energy of $\sqrt{s}=13~\TeV$ and corresponding to an integrated luminosity of 36.1~\ifb. The photon can originate not only from a top quark, but also from its charged decay products, including a charged fermion (quark or lepton) from the decay of the $W$-boson. In addition, it can be radiated from an incoming charged parton. In this analysis, no attempt is made to separate these different sources of photons, but criteria are applied to suppress those radiated from top-quark decay products: e.g. by requiring the photon to have a large angular distance from the lepton(s). In each channel, the fiducial inclusive cross-section, referred to as fiducial cross-section in the following for simplicity, is measured with a likelihood fit to the output of a neural network trained to differentiate between signal and background
events. In both channels, differential cross-sections, normalized to unity, are measured in the same fiducial region without performing the likelihood fit, as a function of the photon \pt, the photon $|\eta|$, and the distance $\Delta R$ between the photon and its closest lepton. The distance $\Delta R$ between two objects is defined as the quadratic sum of their pseudorapidity difference $\Delta\eta$ and azimuthal opening angle $\Delta\phi$. In the \chll channel, the normalized differential cross-sections are also measured as a function of the absolute pseudorapidity difference $|\Delta\eta|$ and $\Delta\phi$ between the two leptons, the latter being sensitive to the spin correlation of the \ttbar pair .The measured cross-sections are compared to predictions from leading order (LO) generators. The predictions for the inclusive cross-sections are corrected by next-to-leading order (NLO) $k$-factors~\cite{melnikov} in the strong interaction, calculated at parton level.
 
This \paperornote is organized as follows. The ATLAS detector is briefly introduced in Section~\ref{sec:detector}. The data and simulation samples used are listed in Section~\ref{sec:samples}. The derivation of the NLO correction to the LO cross section is described in Section~\ref{sec:ttgnlokfactor}. The object and event selection, and the neural-network algorithms are presented in Section~\ref{sec:selection}. The estimation of the backgrounds are introduced in Section~\ref{sec:bkg}. The definition of the fiducial region and the strategies to extract the fiducial and differential cross-sections are described in Section~\ref{sec:strategy}. The evaluation of the systematics uncertainties are discussed in Section~\ref{sec:syst}. Section~\ref{sec:result} gives the final results, and Section~\ref{sec:conclusion} presents the conclusion.
 
\section{ATLAS detector}
\label{sec:detector}
 
The ATLAS detector~\cite{PERF-2007-01} consists of three main components. The innermost component is the Inner Detector (ID), which is used for tracking charged particles. It surrounds the beam pipe and is located inside a superconducting solenoid, operating with a magnetic field of $\SI{2}{~T}$. An additional silicon pixel layer, the insertable B-layer, was added between 3 and $\SI{4}{~cm}$ from the beam line to improve $b$-hadron tagging~\cite{ATLAS-TDR-19,Abbott:2018ikt} for Run 2. The calorimeter outside the ID is divided into two subsystems. The inner subsystem is the electromagnetic calorimeter (ECAL) and the second is the hadronic calorimeter (HCAL). The outermost layer is the third main component of the ATLAS detector: the muon spectrometer (MS), which is within a magnetic field provided by air-core toroid magnets with a bending integral of about $\SI{2.5}{~Tm}$ in the barrel and up to $\SI{6}{~Tm}$ in the end-caps. The ID provides tracking information from silicon pixel and silicon microstrip detectors in the pseudorapidity range $| \eta | <$~2.5 and from a transition radiation tracker (TRT) covering $| \eta | <$~2.0. The magnetic field of the superconducting solenoid bends charged particles for the momentum measurement. The ECAL uses lead absorbers and liquid argon (LAr) as active medium and is divided into barrel ($| \eta |<$~1.475) and end-cap (1.375~$<| \eta |<$~3.2) regions. The HCAL is composed of a steel/scintillating-tile calorimeter, segmented into three barrel structures within $| \eta |<$~1.7, and two copper/LAr hadronic endcap calorimeters, that cover the region 1.5~$<| \eta |<$~3.2. The solid angle coverage is completed with forward copper/LAr and tungsten/LAr calorimeter modules, optimised for electromagnetic and hadronic measurements respectively, and covering the region 3.1~$<| \eta |<$~4.9. The MS measures the deflection of muon tracks within $| \eta |<$ 2.7 using multiple layers of high-precision tracking chambers in toroidal fields of approximately $0.5~\text{T}$ and $1~\text{T}$ in the central and end-cap regions, respectively. The MS is instrumented with separate trigger chambers covering $| \eta |<$2.4.
 
Data are selected from inclusive $pp$ interactions using a two-level trigger system~\cite{TRIG-2016-01}. A hardware-based trigger uses custom hardware and coarser-granularity detector data to initially reduce the trigger rate to approximately 100 kHz from the original 40 MHz LHC proton bunch crossing rate. Next, a software-based high-level trigger, which has access to the full detector granularity, is applied to further reduce the event rate to 1 kHz.
 
\section{Data and simulation samples}
\label{sec:samples}
 
The data used for this analysis were recorded by the ATLAS detector in 2015 and 2016 at a centre-of-mass energy of 13~$\TeV$, corresponding to an integrated luminosity of 36.1~\ifb. Only the data-taking periods in which all detector systems were operating normally are considered. Candidate events were collected using \chljets triggers, designed to select events with at least one isolated high-\pT electron or muon.
 
The signal and background processes were modelled using Monte Carlo (MC) generators and passed through a detector simulation using \GEANT~4~\cite{Agostinelli:2002hh,SOFT-2010-01}. The simulated events were reconstructed with the same software algorithms as data. To account for overlapping $pp$ collisions (pile-up), multiple interactions were simulated with the soft QCD processes of \PYTHIA~v8.186 \cite{Sjostrand:2007gs} using the set of tuned parameters called A2~\cite{ATL-PHYS-PUB-2012-003} and the MSTW2008LO parton distribution functions (PDF) set~\cite{Martin:2009iq}.
 
The \ttg signal sample was simulated as a $2\to 7$ process for the semileptonic and \chllic decay channels of the \ttbar system at LO by \MGMCatNLO v2.33~\cite{Alwall:2014hca} (denoted as \textsc{MG5}\_\textsc{aMC} in the following) interfaced with \PYTHIA~v8.212~\cite{Sjostrand:2014zea}, using the A14 set of tuned parameters~\cite{ATL-PHYS-PUB-2014-021} and the NNPDF2.3LO PDF set~\cite{Pumplin:2002vw}. The photon could be radiated from an initial charged parton, an intermediate top quark, or any of the charged final state particles. The top-quark mass, top-quark decay width, $W$-boson decay width, and fine structure constant were set to 172.5~\GeV, 1.320~\GeV, 2.085~\GeV, and 1/137, respectively. The five-flavour scheme was used where all the quark masses are set to zero, except for the top quark. The renormalization and the factorization scales were set to 0.5$\times \sum_i \sqrt{m^2_i+p^2_{T,i}}$, where the sum runs over all the particles generated from the matrix
element calculation. The photon was requested to have $\pT > 15~\GeV$ and $|\eta| < 5.0$. At least one lepton with $\pT > 15~\GeV$ was required, with all the leptons satisfying $|\eta| < 5.0$. The $\Delta R$ between the photon and any of the charged particles among the seven final-state particles were required to be greater than 0.2. The resulting total cross-section of the sample was calculated to be 4.62~pb. The NLO $k$-factors, introduced in Section~\ref{sec:ttgnlokfactor}, were applied to correct the fiducial cross-sections and acceptances to NLO.
 
The inclusive \ttbar sample~\cite{ATL-PHYS-PUB-2016-004} was generated with \POWHEGBOX~v2~\cite{powheg} using the NNPDF3.0NLO PDF set~\cite{Ball:2014uwa}, and interfaced with \PYTHIA~v8.210 using the A14 tune set and the NNPDF2.3LO PDF set. The $h_{\textit{damp}}$ parameter, which controls the \pT of the first additional parton emission beyond Born level in \POWHEG, was set to 1.5 times the top-quark mass. The production of a vector boson ($V = W, Z$) in association with a photon (\Vgamma) was simulated with \SHERPA~v2.2.2~\cite{Gleisberg:2008ta}, and the inclusive production of \Vjets~\cite{ATL-PHYS-PUB-2016-003} was simulated with \SHERPA~v2.2.1, both using the NNPDF3.0NLO PDF set. The $s$-channel single top quark and $tW$ samples~\cite{ATL-PHYS-PUB-2016-004} were produced with \POWHEGBOX~v1 using the CT10 (NLO) PDF set~\cite{PhysRevD.82.074024}, interfaced with \PYTHIA~v6.428 using the \Perugia~2012 tune set~\cite{Skands:2010ak} and the CTEQ6L1 PDF
set~\cite{Pumplin:2002vw}. The $t$-channel single top quark was produced with the same generator and parton shower, with the four-flavour scheme and the corresponding CT104fs PDF set~\cite{Pumplin:2002vw}. The diboson samples of $WW$, $WZ$ and $ZZ$~\cite{ATL-PHYS-PUB-2016-002} were generated by \SHERPA~v2.1, using the CT10 (NLO) PDF set. The $t\bar{t}V$ samples~\cite{ATL-PHYS-PUB-2016-005} were generated with \MGaMC~v2.2 using the NNPDF3.0NLO PDF set, interfaced with \PYTHIA~v8.210 using the NNPDF2.3LO PDF set. For all samples without photon radiation in the matrix element calculation, the radiation was simulated by the corresponding parton shower. The EvtGen program~\cite{LANGE2001152} was used to simulate the decay of bottom and charm hadrons, except for the \SHERPA samples. All these samples were generated with NLO precision in QCD and, in the case of \SHERPA samples, the NLO calculations were performed for up to one or two additional partons.
 
To assess the effects of initial- and final-state radiation (ISR and FSR), alternative signal samples were produced with the relevant \PYTHIA8 A14 Var3c tune parameters~\cite{ATL-PHYS-PUB-2014-021} varied to increase or decrease the parton radiation. The effect of the choice of parton shower algorithm for the signal is evaluated with a sample generated using \Herwig~v7.0.1~\cite{Bellm:2015jjp} instead of \PYTHIA~v8.212. An alternative \ttbar sample was generated to enhance the parton shower radiation, with the renormalization and factorization scales varied down by a factor of two, the high radiation variation of the A14 Var3c tune parameter and the $h_{\textit{damp}}$ value increased by a factor of two. A corresponding \ttbar sample with reduced parton shower radiation was generated with the renormalization and factorization scales multiplied by a factor of two and the low radiation variation of the A14 Var3c tune parameter. The uncertainty arising from the choice of \ttbar generator is evaluated using a \SHERPA~v2.2 sample. An alternative \Zgamma sample, generated by \MGaMC~v2.33 interfaced with \PYTHIA~v8.212, is used to evaluate the modelling uncertainty of the \Zgamma background estimate.
 
The \ttbar and \Vjets samples contain events already accounted for by the \ttg and \Vgamma samples. Based on truth information, the overlap is removed by vetoing the events where the selected photon originates from the hard interaction in the \ttbar and \Vjets samples.
 
\section[Next-to-leading order $k$-factor for \texorpdfstring{\ttg}{ttgamma}]{Next-to-leading order $k$-factor for \ttg}
\label{sec:ttgnlokfactor}
 
Calculations at NLO precision in QCD are available for the \ttg process at a centre-of-mass energy of $\sqrt{s}=14$~$\TeV$~\cite{melnikov}, extending results performed using the approximation of stable top quarks~\cite{PengFei:2009ph}. A dedicated calculation at $\sqrt{s}=13$~$\TeV$ has been performed for both \chljets and \chll channels, by the authors of Ref.~\cite{melnikov}. The renormalization and factorization scales are both set to the top-quark mass, while the rest of the parameters are set to the same values used by the \MGaMC \ttg MC sample, as described in Section~\ref{sec:samples}. These calculations are used to derive corrections at parton level to the normalization of the LO \ttg MC sample.
 
The NLO calculation is performed at parton level in a phase space very close to the fiducial region defined in Section~\ref{sec:fiducial}. The lepton (at least one lepton) is required to have $\pt > 25~\GeV$ for the \chljets (\chll) channel, and all leptons must have $|\eta| < 2.5$. The photon is required to have $\pt > 20~\GeV$ and $|\eta| < 2.37$. Jets are reconstructed from quarks and gluons using the anti-$k_{t}$ algorithm~\cite{Cacciari:2008gp} with a radius parameter of $R=0.4$, and they are required to have $\pt > 25~\GeV$ and $|\eta| < 2.5$. At least four (two) jets are required for the \chljets (\chll) channel. All jets are required to be separated from the photon by $\Delta R(\gamma,\text{jet})\textgreater $ 0.4. Leptons are required to be separated from the photon by $\Delta R(\gamma,\ell)\textgreater $ 1.0. The leptons in the \chll channel are required to be separated from the jets by $\Delta R(\text{jet},\ell)\textgreater $ 0.4.
 
The LO cross-sections are calculated using the \MGaMC LO sample at particle level in the same phase space as above to derive the NLO $k$-factors for the \chljets and \chll channels. Since the kinematic properties of all the objects used in the NLO theoretical calculation are taken from parton level, the photons, leptons, and jets of the LO MC sample must be defined carefully to correspond to those at the parton level. This is achieved by requiring the photon and the leptons to be produced from the matrix element rather than from the parton shower and adding the QED radiation simulated by the parton shower back to the leptons. The anti-$k_{t}$ algorithm with $R=0.4$ is used for jet clustering, using all the final state particles, excluding the above photon, leptons, and their corresponding neutrinos.
 
The calculated NLO cross-sections are 120~fb and 31~fb for the \chejets and \chemu channels, while the calculated LO cross-sections are 92~fb and 21~fb for the same channels, respectively, resulting in NLO $k$-factors of 1.30 and 1.44. These $k$-factors are applied to other \chljets or \chll channels. Statistical uncertainties are negligible. Systematic uncertainties of the $k$-factors receive contributions from two sources. For the NLO theoretical cross-section, the relative uncertainty due to the QCD scale and PDF choices are 14\% (13\%) for the \chljets (\chll) channel, with the QCD scale uncertainty dominating. For the LO MC cross-section, the non-perturbative effects in the parton shower model are studied by turning off the multiple parton interaction and hadronization of \PYTHIA8 separately, resulting in an uncertainty of 8\% (4\%) for the \chljets (\chll) channel. In addition, the jet cone size is varied from 0.4 to 0.3 or 0.5 separately to evaluate the impact of additional QCD radiation on the reconstruction of the particle-level jet. The resulting uncertainties are 11\% (6\%) for the \chljets (\chll) channel. Summing up the components in quadrature, the total relative uncertainty on the $k$-factor is 20\% (15\%) for the \chljets (\chll) channel.
 
\section{Object and event selection}
\label{sec:selection}
 
The object and event selection at the detector level are introduced in Section~\ref{sec:objselection} and Section~\ref{sec:evtselection} respectively. The neural-network algorithms used in the analysis are described in Section~\ref{sec:nn}. In Section~\ref{sec:fiducial}, the fiducial region at particle level is defined.
 
\subsection{Object selection}
\label{sec:objselection}
 
Electron candidates are reconstructed from energy deposits in the central region of the ECAL associated with reconstructed tracks from the ID \cite{Aaboud:2019ynx} and are required to have a $\pt > 25~\GeV$ and an absolute calorimeter cluster pseudorapidity $|\eta_{\textrm{cluster}}| < 2.47$, excluding the transition region between the barrel and endcap calorimeters ($|\eta_{\textrm{cluster}}|$ $\not\in$ $[1.37,1.52]$). ``Tight'' likelihood-based identification criteria are applied, which correspond to an efficiency between 80\% and 90\% for electrons in different \pT and $\eta$ ranges measured in $Z\to ee$ events~\cite{Aaboud:2019ynx}. Muon candidates are reconstructed by an algorithm that combines the track segments in the various layers of the MS with the tracks in the ID \cite{PERF-2015-10} and are required to have a $\pt > 25~\GeV$ and $\abseta < 2.5$. ``Medium'' cut-based identification criteria are required, which correspond to an average efficiency
around 96\% in \ttbar events for muons in different \pT and $\eta$ ranges~\cite{PERF-2015-10}. Isolation criteria are applied to both the electron and muon candidates using calorimeter- and track-based information to obtain 90\% efficiency for leptons with $\pt = \SI{25}{~GeV}$, rising to 99\% efficiency at $\pt = \SI{60}{~GeV}$ in $Z\to\ell\ell$ events. The transverse impact parameter divided by its estimated uncertainty $|d_0|/\sigma(d_0)$ is required to be lower than five for electron candidates and three for muon candidates. The longitudinal impact parameter must satisfy $|z_0 \sin(\theta)| < \SI{0.5}{~mm}$ for both. The lepton reconstruction and identification efficiencies in simulation are corrected to match the corresponding values in data~\cite{Aaboud:2019ynx,PERF-2015-10}.
 
A photon could convert into an electron positron pair when it traverses the material before entering the active volume of the ECAL.
Photon candidates are reconstructed from energy deposits in the central region of the ECAL~\cite{PERF-2017-02} and classified as unconverted if there is no matching track or reconstructed conversion vertex or as converted if there is a matching reconstructed conversion vertex or a matching track consistent with originating from a photon conversion. They must have a $\pt > 20~\GeV$ and $|\eta_{\textrm{cluster}}| < 2.37$, excluding the transition region between the barrel and endcap. ``Tight'' cut-based identification criteria, based on discriminating variables and corresponding to an efficiency around 85\% at 40~\GeV, are applied~\cite{PERF-2017-02}. Cut-based \pT-dependent isolation criteria are applied using calorimeter- and track-based information and correspond to an efficiency between 75\% and 90\% for prompt photons (photons not from hadron decays) in $Z\to\ell\ell\gamma$ events. The photon reconstruction and identification efficiencies in simulation are
corrected to match the corresponding values in data~\cite{PERF-2017-02}.
 
Jets are reconstructed using the anti-$k_{t}$ algorithm with a radius parameter of $R=0.4$ from topological clusters of energy deposits in the calorimeter~\cite{ATL-PHYS-PUB-2015-036}. The jet energy scale and jet energy resolution are calibrated using energy- and $\eta$-dependent calibration schemes resulting from simulation and \textit{in situ} corrections based on data \cite{PERF-2016-04}. The jets are required to have a $\pt > 25~\GeV$ and $|\eta| < 2.5$. Jets likely to originate from pile-up are suppressed by using the output of a multivariate jet-vertex-tagger (JVT)~\cite{ATLAS-CONF-2014-018}. Scale factors are used to correct the selection efficiency in simulation to match data. Jets containing $b$-hadrons ($b$-jets) are identified with a $b$-tagging algorithm using a multivariate discriminant that combines information about secondary vertices and track impact parameters (MV2c10)~\cite{PERF-2016-05,ATL-PHYS-PUB-2016-012}. The operating point used corresponds to an overall 77\% $b$-tagging efficiency in \ttbar events, with a corresponding rejection of $c$-jets (light-jets) by a factor of 6 (134). Efficiencies to tag $b$-, $c$-, and light-jets in the simulation are scaled by \pT- and $\eta$-dependent factors~\cite{PERF-2016-05} to match the efficiencies in data.
 
The transverse energy carried by the neutrinos is accounted for in the reconstructed missing transverse momentum $\met$ \cite{Aaboud:2018tkc}, which is computed as the transverse component of the negative vector sum of all the selected electrons, muons, photons, and jets, as well as ID tracks associated with the primary vertex but not with any of the above objects, which is called track-based soft term.
 
An overlap removal procedure is applied to avoid the same calorimeter energy deposit or the same track being reconstructed as two different objects. Electrons sharing their track with a muon candidate are removed. Jets within a $\Delta R = 0.2$ cone of an electron are removed. After that, electrons within a $\Delta R = 0.4$ cone of a remaining jet are removed. When a muon and a jet are close, the jet is removed if it has no more than two associated tracks and is within $\Delta R < 0.2$ of the muon, otherwise the muon is removed if it is within $\Delta R < 0.4$ of the jet and the jet has more than two associated tracks. Photons within a $\Delta R = 0.4$ cone of a remaining electron or muon are removed. Finally, the jets within a $\Delta R = 0.4$ cone of a remaining photon are removed.
 
\subsection{Event selection}
\label{sec:evtselection}
 
The events must have at least one primary vertex with at least two associated tracks, each with $\pt > 400~\MeV$. Primary vertices are formed from reconstructed tracks spatially compatible with the interaction region. The primary vertex with the highest sum of $p_{\text{T}}^{2}$ over all associated tracks is chosen. Events are categorized into the \chljets channel if their final state contains exactly one lepton (electron or muon), and into the \chll channel if they contain two electrons, two muons, or one electron and one muon, with each pair required to be of opposite charge. The lepton (at least one of the leptons) must be matched to a fired \chljets trigger for the \chljets (\chll) channel. The \pT of the electron (muon) that fired the trigger has to be larger than 27 (27.5)~$\GeV$ in order to match the higher lepton \pT trigger threshold in 2016. The selected events must have at least four (two) jets in the \chljets (\chll) channel, at least one of which is $b$-tagged, and exactly one photon. A $Z$-boson veto is applied in the single electron channel by excluding events with invariant mass of the system of the electron and the photon around the $Z$-boson mass ($|m(e,\gamma)-m(Z)|<5~\GeV$), where $m(Z) = 91.188$~\GeV. In the \chll channel when the two leptons have the same flavour, events are excluded if the \chll invariant mass or the invariant mass of the system of the two leptons and the photon is between 85 and 95~\GeV, and $\met$ is required to be larger than 30~\GeV. The \chll invariant mass is required to be higher than $15~\GeV$ to suppress events from $J/\psi$, $\Upsilon$ and $\gamma^*$ decays. Finally, to suppress photons radiated from lepton(s), the $\Delta R$ between the selected photon and lepton(s) must be greater than 1.0. The event selection is summarized in Table~\ref{tab:eventsel}.
 
\begin{table}[htbp]
\caption{Summary of the event selection. ``OS'' means the charges of the two leptons must have opposite signs.}
\centering
\begin{tabular}{c|c|c|c|c}
\toprule
\chejets & \chmujets & \chee & \chmumu & \chemu \\
\hline
\multicolumn{5}{c} {Primary vertex} \\
\hline
1 $e$ & 1 $\mu$ & 2 $e$, OS & 2 $\mu$, OS & 1 $e$ + 1 $\mu$, OS \\
\hline
\multicolumn{5}{c} {Trigger match} \\
\hline
\multicolumn{2}{c|} {$\geq$ 4 jets} & \multicolumn{3}{c} {$\geq$ 2 jets} \\
\hline
\multicolumn{5}{c} {$\geq$ 1 $b$-jet} \\
\hline
\multicolumn{5}{c} {1 $\gamma$} \\
\hline
$|m(e, \gamma) - m(Z)| > 5$~\GeV & \multicolumn{4}{c} {-} \\
\hline
\multicolumn{2}{c|} {-} & \multicolumn{2}{c|} {$m(\ell, \ell)$ $\not\in$ [85,95] \GeV} & - \\
\hline
\multicolumn{2}{c|} {-} & \multicolumn{2}{c|} {$m(\ell, \ell, \gamma)$ $\not\in$ [85,95] \GeV} & -\\
\hline
\multicolumn{2}{c|} {-} & \multicolumn{2}{c|} {\MET $>$ 30 \GeV} & -\\
\hline
\multicolumn{2}{c|} {-} & \multicolumn{3}{c} {$m(\ell, \ell)>$ 15 \GeV} \\
\hline
\multicolumn{5}{c} {$\Delta R(\gamma, \ell)$ $>$ 1.0} \\
\bottomrule
\end{tabular}
\label{tab:eventsel}
\end{table}
 
There are four types of backgrounds to the selected \ttg candidates, three of which are events with a misidentified object. The contribution from events in which the selected photon candidate originates from a jet or a non-prompt photon from hadron decays, referred to as hadronic-fake background, is estimated following the method outlined in Section~\ref{sec:hfake}. The contribution from events in which the selected photon candidate originates from an electron, referred to as electron-fake background, is estimated following the method outlined in Section~\ref{sec:efake}. The contribution from events in which the selected lepton candidate originates from a jet or a non-prompt lepton from heavy-flavour decays, referred to as fake-lepton background, is estimated following the method outlined in Section~\ref{sec:nonprompt}. Finally, the contribution from events with a prompt photon (excluding the \ttg signal and the fake-lepton background with prompt photon radiation), referred to as prompt-photon background, is estimated following the method outlined in Section~\ref{sec:prompt}. In the \chljets channel, the main backgrounds are from events with a hadronic-fake or electron-fake photon and \Wgamma production, while in the \chll channel, \Zgamma production and events with a hadronic-fake photon are the dominant backgrounds.
 
A total number of $11\,662$ and 902 candidate events are selected for the \chljets and \chll channels, respectively, with expected numbers of 6490 $\pm$ 420 and 720 $\pm$ 34 signal events, where the corresponding NLO $k$-factors are applied and the uncertainties include the simulation statistical uncertainty and all systematic uncertainties introduced in Section~\ref{sec:syst}. Table~\ref{tab:prefiteventYieldswSFs} summarizes the observed data and the expected event yields for the signal and background processes. Figures~\ref{fig:prefitPlotsSL} and \ref{fig:prefitPlotsDL} show comparisons of the data with the expected simulated distributions. The simulation is corrected with data-driven corrections. The statistical uncertainty of data and systematic uncertainties are included. The signals are scaled by the NLO $k$-factors.
 
\begin{table}[!h]
\begin{center}
\caption{The observed data and the expected event yields for the signal and backgrounds in the \chljets and \chll channels. All data-driven corrections and systematic uncertainties are included. The signals are scaled by the NLO $k$-factors. The fake-lepton background in the dilepton channel is negligible, represented by a ``-''. The \Zgamma (\Wgamma) background in the single-lepton (dilepton) channel is included in ``\Other''. The uncertainty of the \Wgamma background in the single-lepton channel is not given since the normalization of this background is a free parameter in the likelihood fit.}
\scalebox{1.0}{
\begin{tabular}{l|r@{$\,\pm\,$}r|r@{$\,\pm\,$}r}
\toprule
Channel  & \multicolumn{2}{c|}{Single lepton} & \multicolumn{2}{c}{Dilepton}  \\ \hline
\ttg & $6\,490$ & 420 &  720 & 34 \\ \hline
Hadronic-fake & $1\,440$ & 290 &  49 & 27 \\ \hline
Electron-fake & $1\,650$ & 170 &  2 & 1 \\ \hline
Fake lepton & 360 &  200 &  \multicolumn{2}{c}{-} \\ \hline
\Wgamma & \multicolumn{2}{l|}{$ \enspace 1\,130$} & \multicolumn{2}{c}{} \\ \hline
\Zgamma & \multicolumn{2}{c|}{} & 75 & 52 \\ \hline
\Other & 690 & 260 &  18 & 7 \\ \hline
\hline
Total & $11\,750$ & 710 &  863 &   78 \\ \hline
Data & \multicolumn{2}{l|}{$11\,662$} &  \multicolumn{2}{l}{902} \\
\bottomrule
\end{tabular}
}
\label{tab:prefiteventYieldswSFs}
\end{center}
\end{table}
 
\FloatBarrier
 
\begin{figure}[!htbp]
\centering
\subfloat[]{
\includegraphics[width=0.45\linewidth]{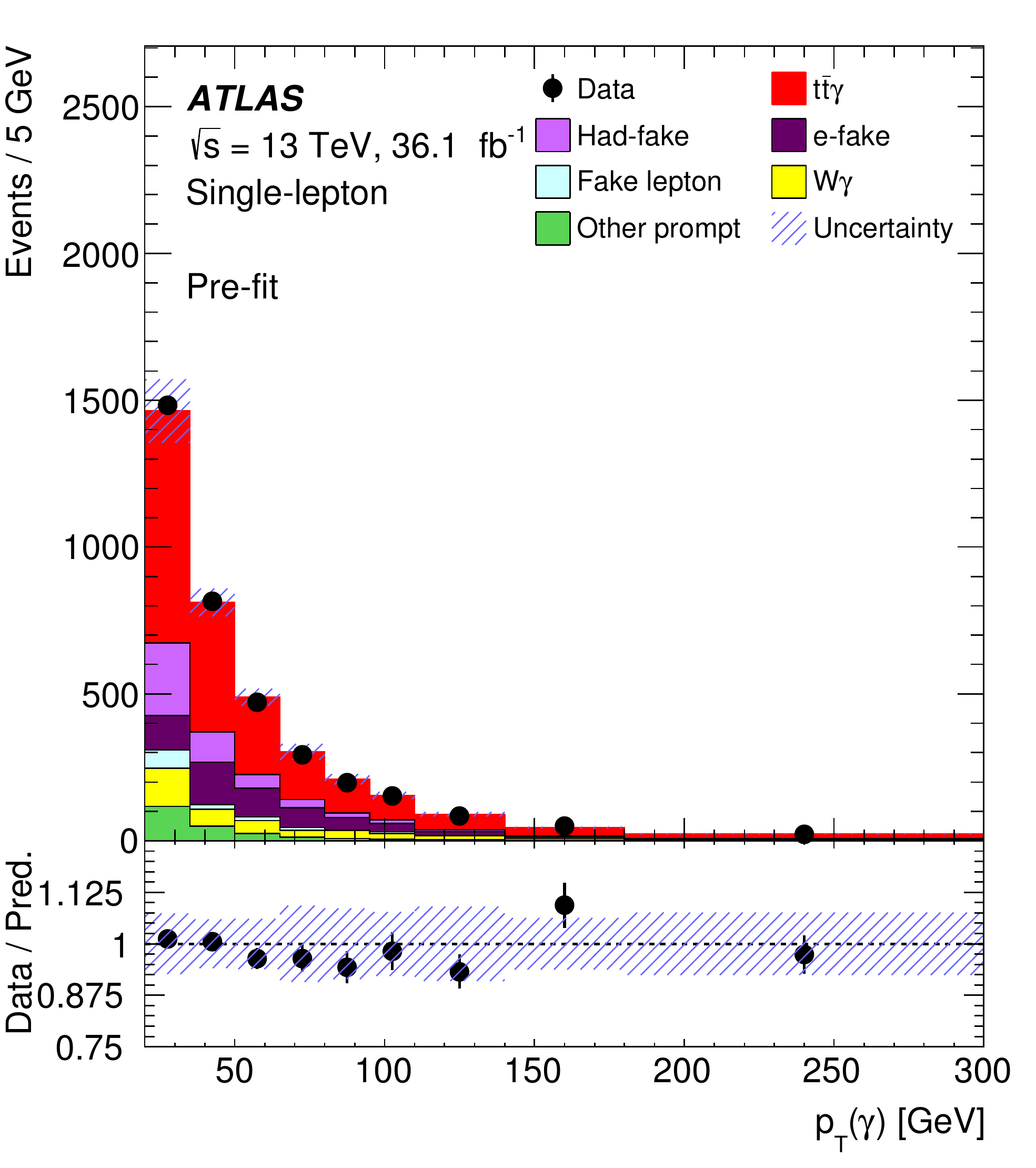}
}
\subfloat[]{
\includegraphics[width=0.45\linewidth]{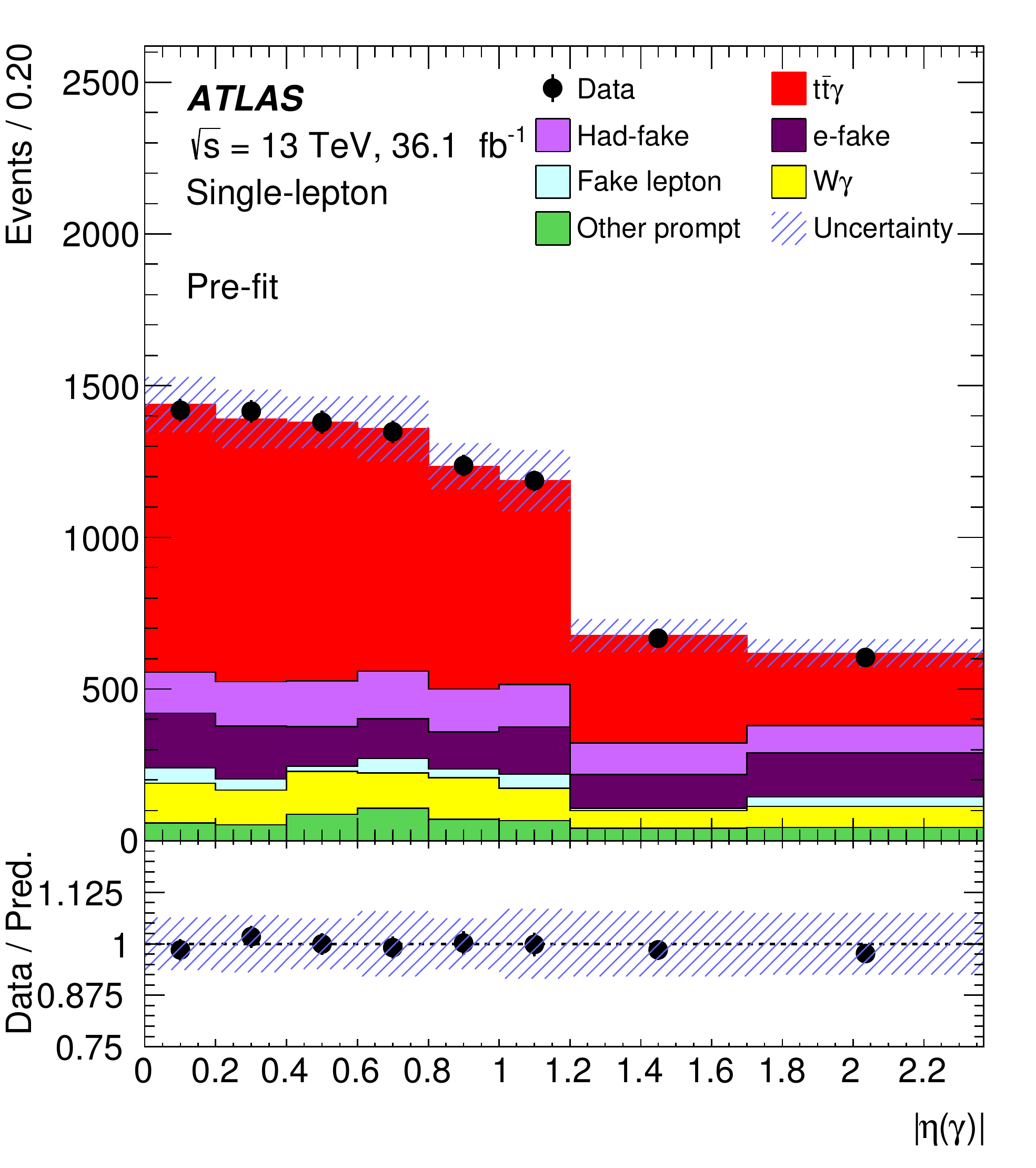}
}
 
\subfloat[]{
\includegraphics[width=0.45\linewidth]{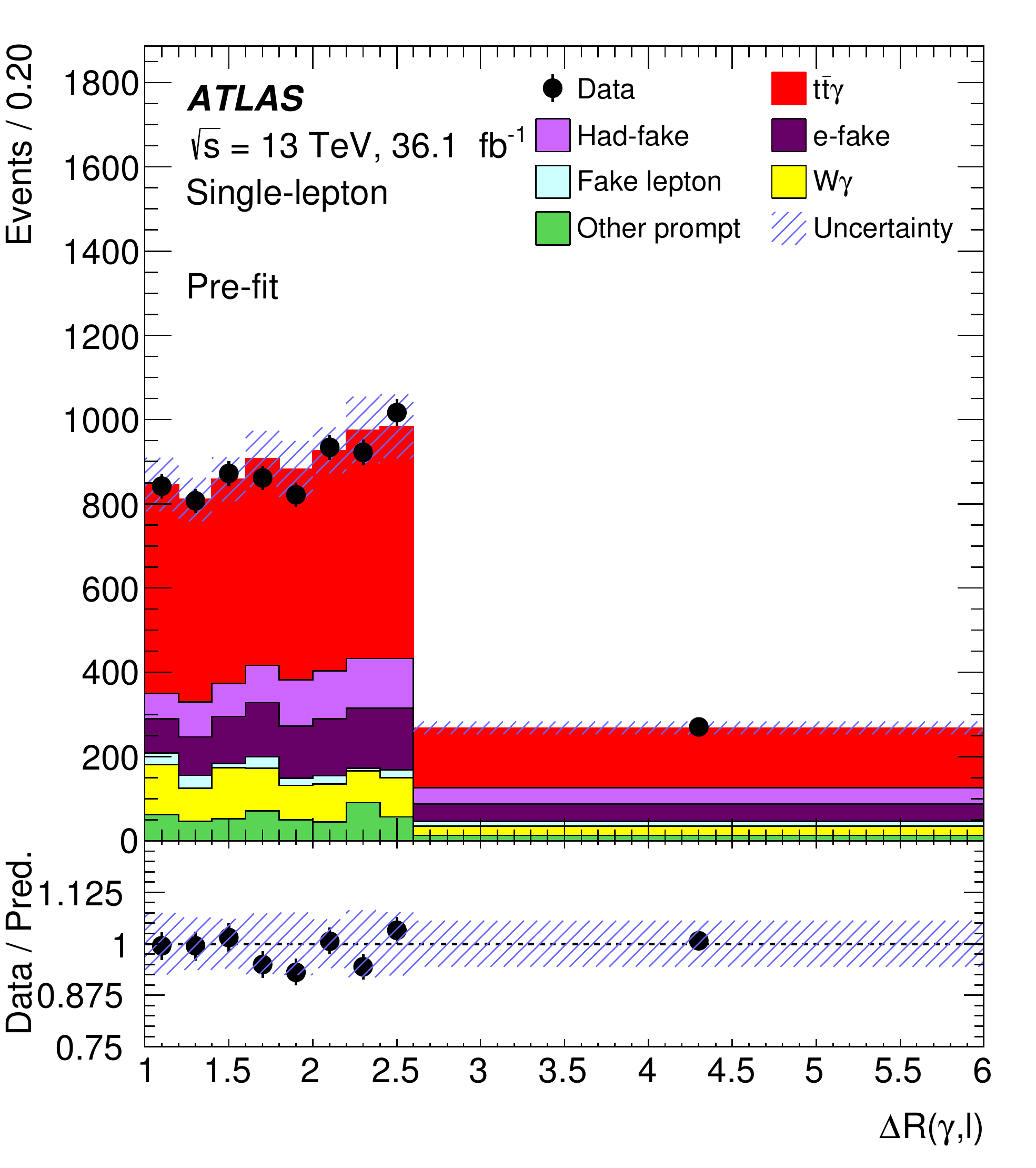}
}
\caption [] {Distributions of the (a) photon \pT, (b) photon $|\eta|$, and (c) $\Delta R (\gamma,\ell)$ in the \chljets channel after event selection and before likelihood fit. All data-driven corrections and systematic uncertainties are included. Overflow events are included in the last bin.}
\label{fig:prefitPlotsSL}
\end{figure}
 
\begin{figure}[!htbp]
\centering
\subfloat[]{
\includegraphics[width=0.34\linewidth]{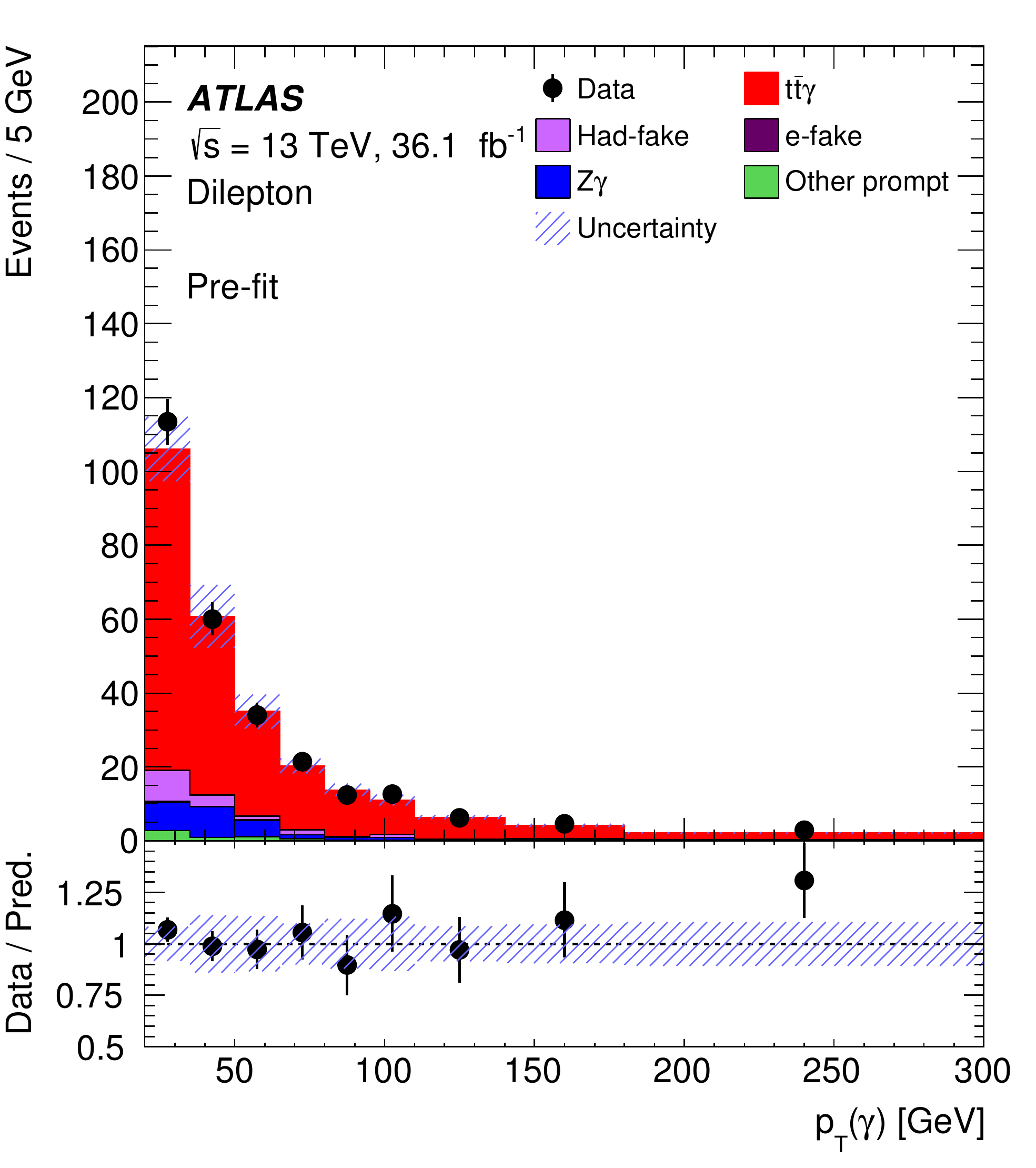}
}
\subfloat[]{
\includegraphics[width=0.34\linewidth]{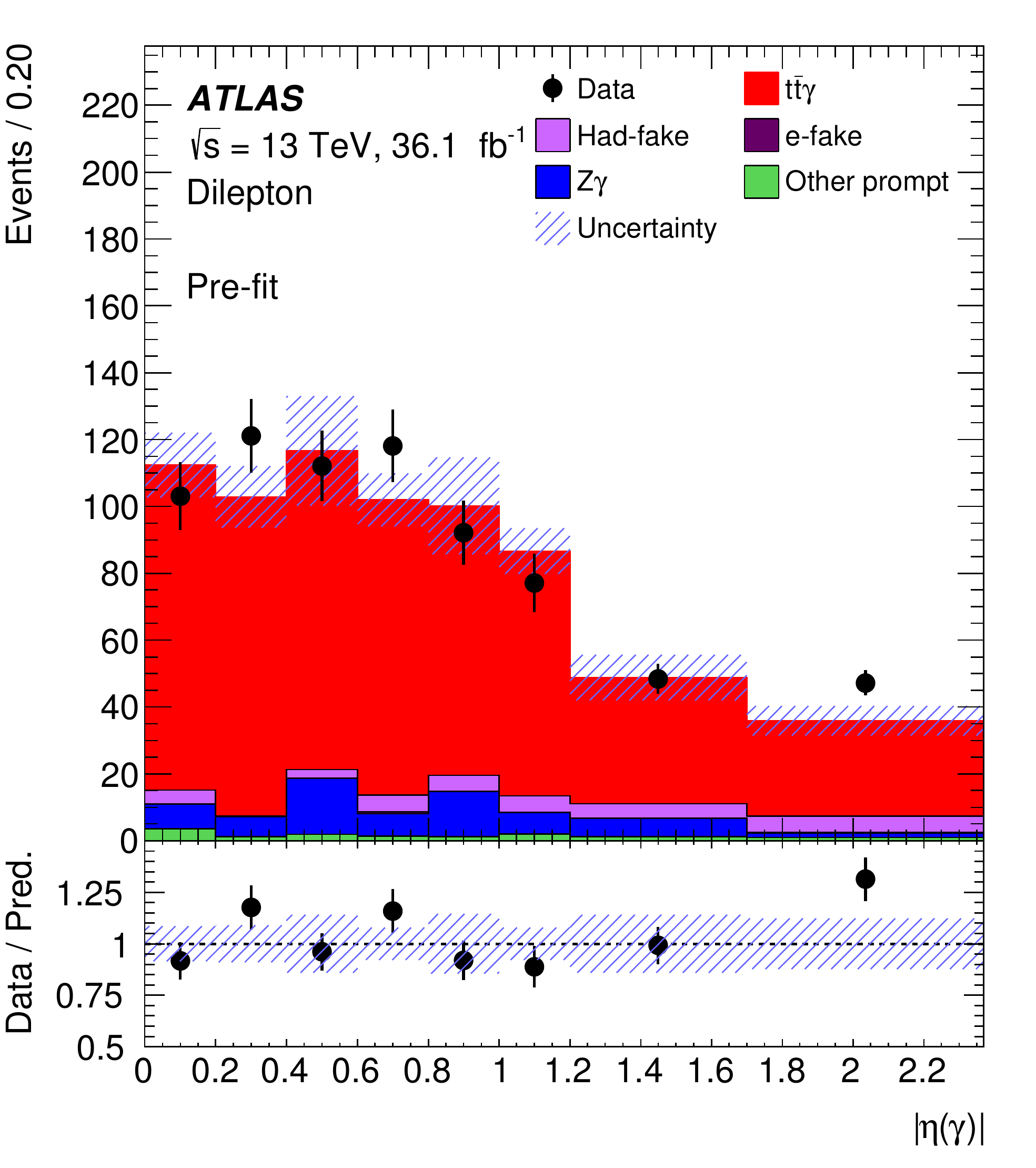}
}
 
\subfloat[]{
\includegraphics[width=0.34\linewidth]{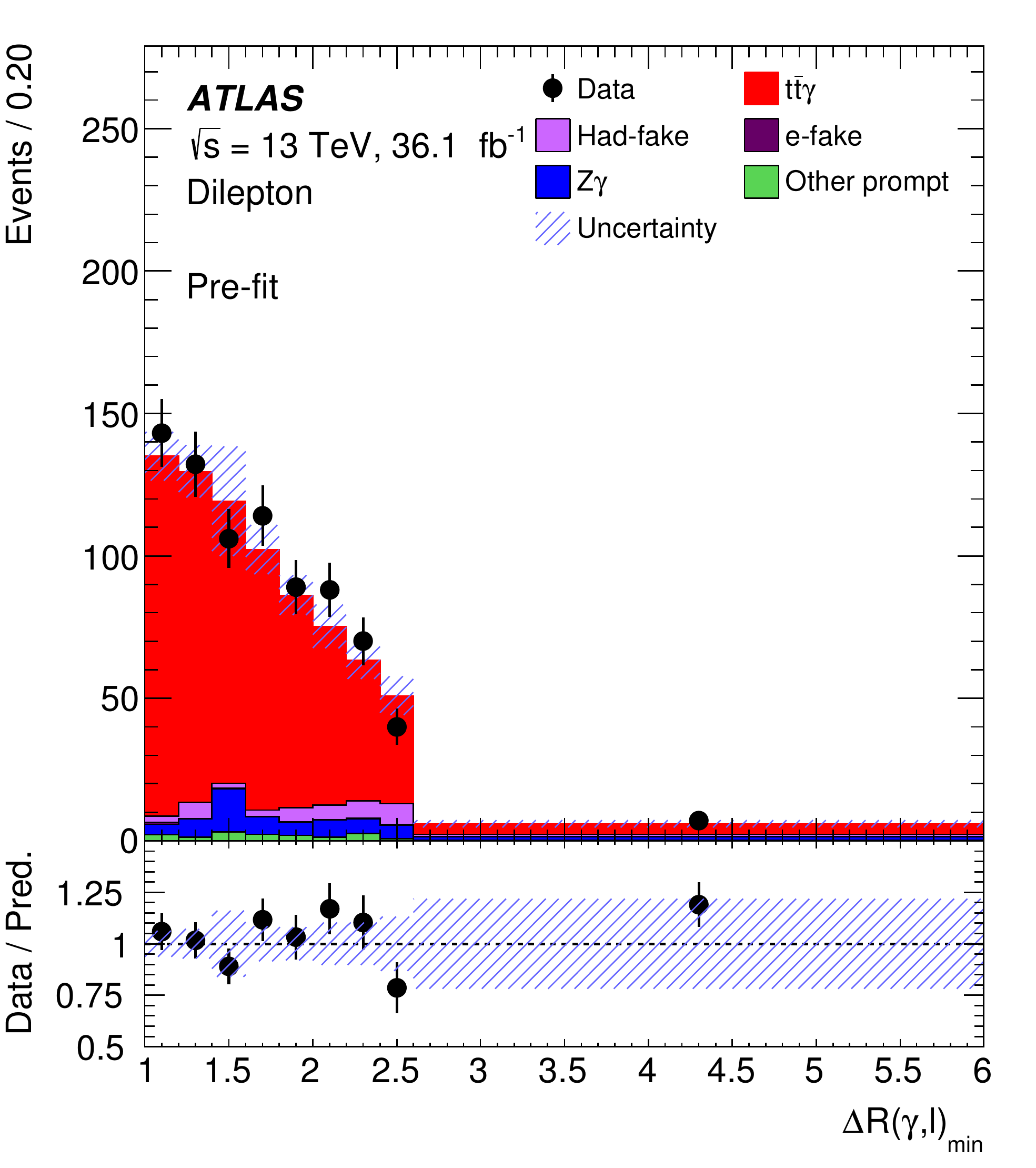}
}
\subfloat[]{
\includegraphics[width=0.34\linewidth]{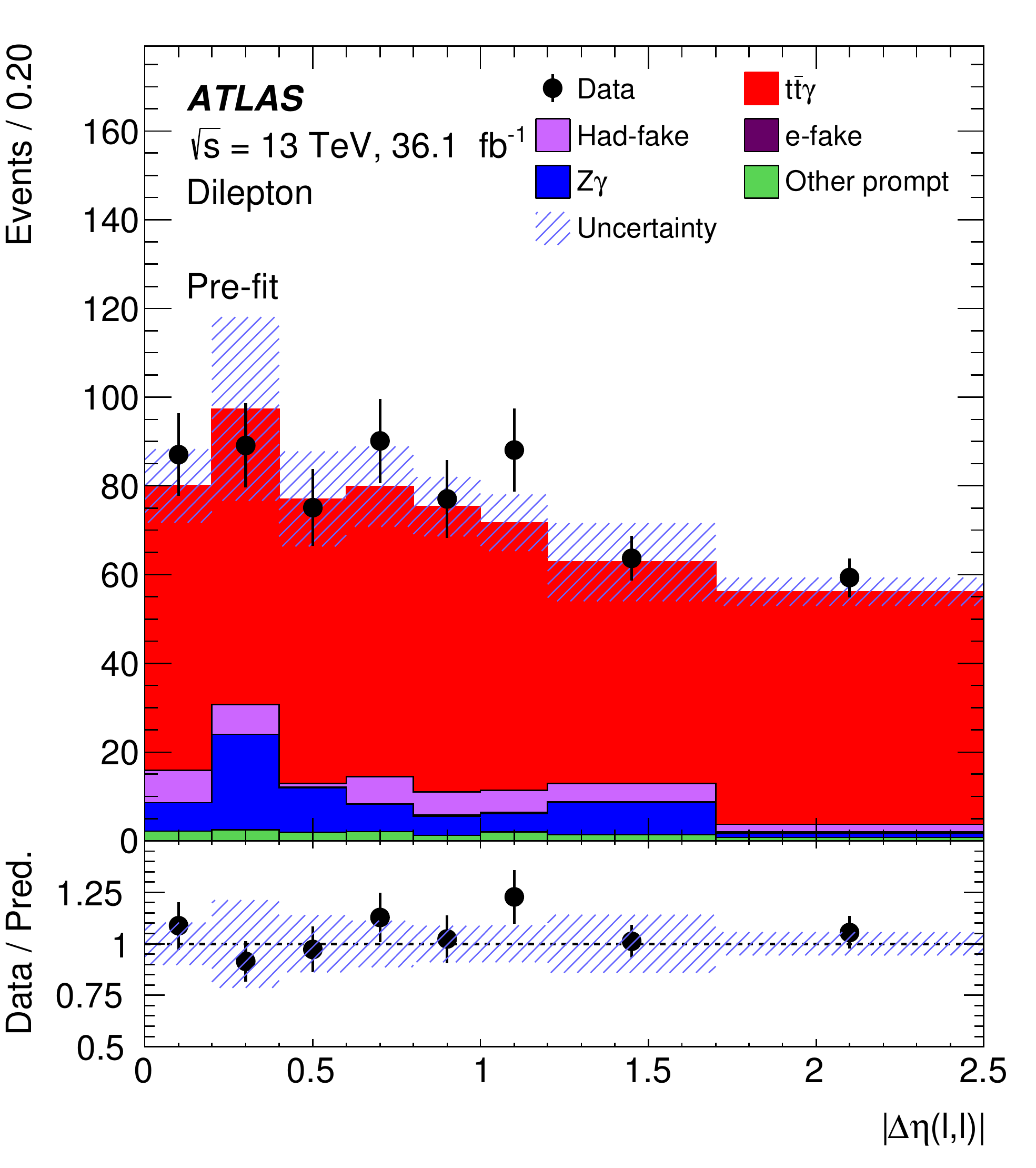}
}
 
\subfloat[]{
\includegraphics[width=0.34\linewidth]{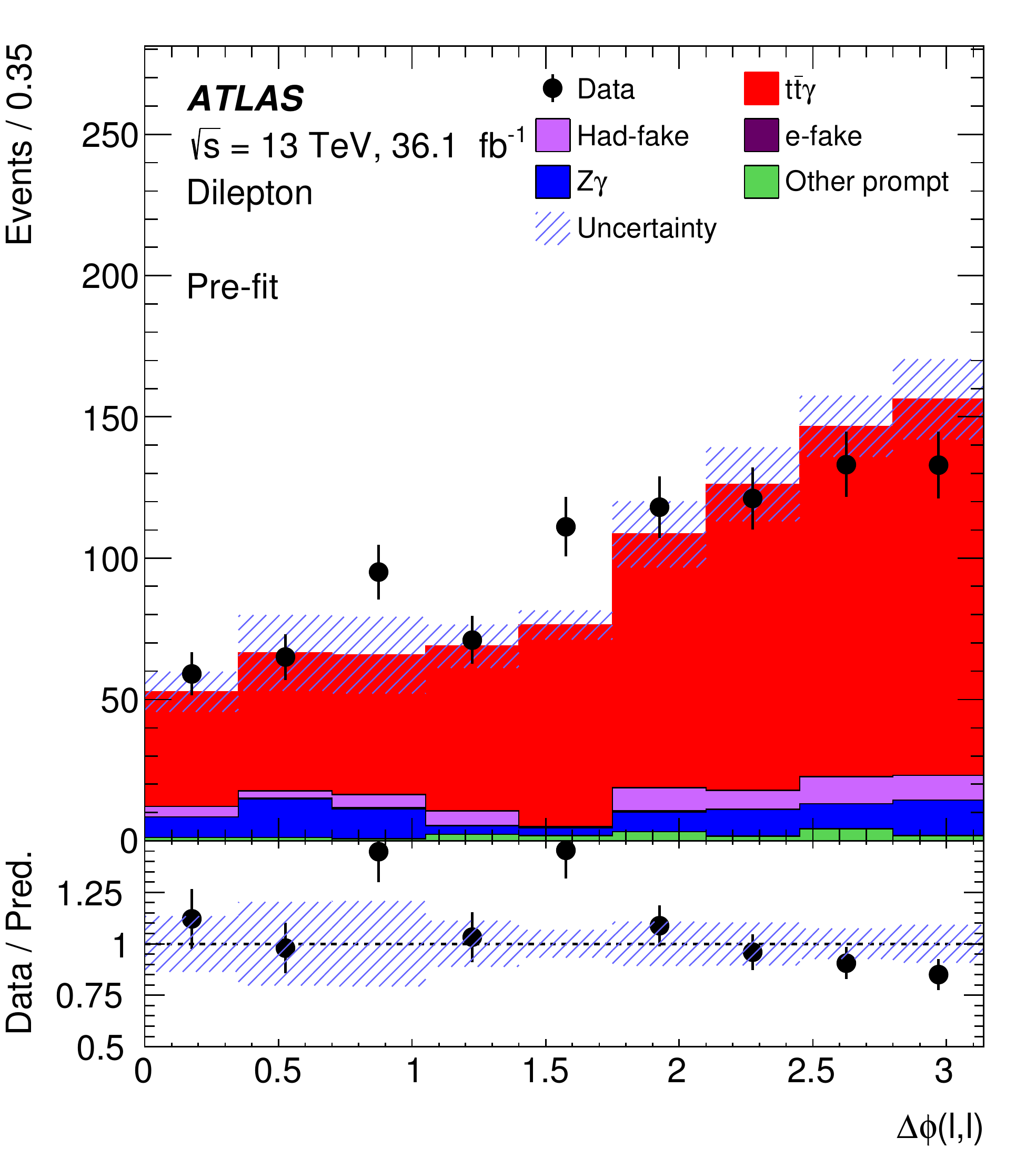}
}
\caption [] {Distributions of the (a) photon \pT, (b) photon $|\eta|$, (c) minimum $\Delta R (\gamma,\ell)$, (d) $|\Delta\eta(\ell, \ell)|$, and (e) $\Delta\phi(\ell,\ell)$ in the \chll channel after event selection and before likelihood fit. All data-driven corrections and systematic uncertainties are included. Overflow events are included in the last bin. In particular, events with $|\Delta\eta(\ell, \ell)|>2.5$ are included in the last bin of (d).}
\label{fig:prefitPlotsDL}
\end{figure}
 
\subsection{Multivariate analysis}
\label{sec:nn}
 
To discriminate the \ttg signal from backgrounds, a neural-network algorithm, called the event-level discriminator (ELD), is trained separately for the \chljets and \chll channels. Given the significant contribution of hadronic-fake photons in the \chljets channel, a dedicated neural network, referred to as the prompt-photon tagger (PPT) in the following, is trained to discriminate between prompt photons and hadronic-fake photons. The PPT is used as one of the inputs to the ELD in the \chljets channel.
 
Both neural-network algorithms are feedforward binary classifiers that have been trained using Keras~\cite{keras} and evaluated using \textsc{lwtnn}~\cite{lwtnn}. Theano~\cite{DBLP:journals/corr/Al-RfouAAa16} is used as backend. The input variables are normalized to have a standard deviation of 1 and a mean of 0. To reduce the risk of over-training, regularization methods such as \emph{dropout}~\cite{JMLR:v15:srivastava14a} and \emph{batch normalization}~\cite{DBLP:journals/corr/IoffeS15} layers are used. Additionally, $k$-fold cross-validation is performed.
 
Five variables which characterize the photon candidate shower shape in the transverse and lateral directions utilizing the energy deposits in the first and second layer of the ECAL, $R_\eta$, $R_\phi$, $w_{\eta_{2}}$, $w_{s3}$, and $F_\mathrm{side}$, and one variable which characterizes the energy leakage fraction into the HCAL, $R_\mathrm{had}$, are used in the PPT.
These are the standard discriminating variables used in ATLAS for photon identification~\cite{PERF-2017-02} and their definitions are given in the Appendix.
Prompt photons from simulated QCD-Compton processes and hadronic-fake photons from simulated dijet events are used as signal and background photons in the training and testing of the PPT. Photons are required to pass the Tight identification and have $\pt > 25~\GeV$ and $|\eta_{\textrm{clu}}| < 2.37$, excluding the calorimeter transition region. The PPT shape of the prompt photons in simulation is corrected to match data in photon \pT-$\eta$ bins. The correction factors for each bin are extracted from the ratio between the PPT output distribution in data and that of simulation, using photons in a $Z\to\ell\ell\gamma$ control region. The control region is defined by requiring exactly one photon and two opposite-sign leptons, with the invariant mass of the lepton pair between $60$ and $\SI{100}{~GeV}$. The resulting correction factors range from 0.5 to 2.0 and are in general around unity. PPT systematic uncertainties are evaluated separately for prompt photons and fake
photons and are discussed in Section~\ref{sec:expsys}. The PPT output distribution after event selection in the \chljets channel is shown in Figure~\ref{fig:prefitPlotsSLPPT}. The shape difference between data and prediction of the PPT is caused by the shape difference between data and simulation of the input discriminating variables and is covered by the assigned systematic uncertainties.
 
\begin{figure}[!htbp]
\centering
\includegraphics[width=0.45\linewidth]{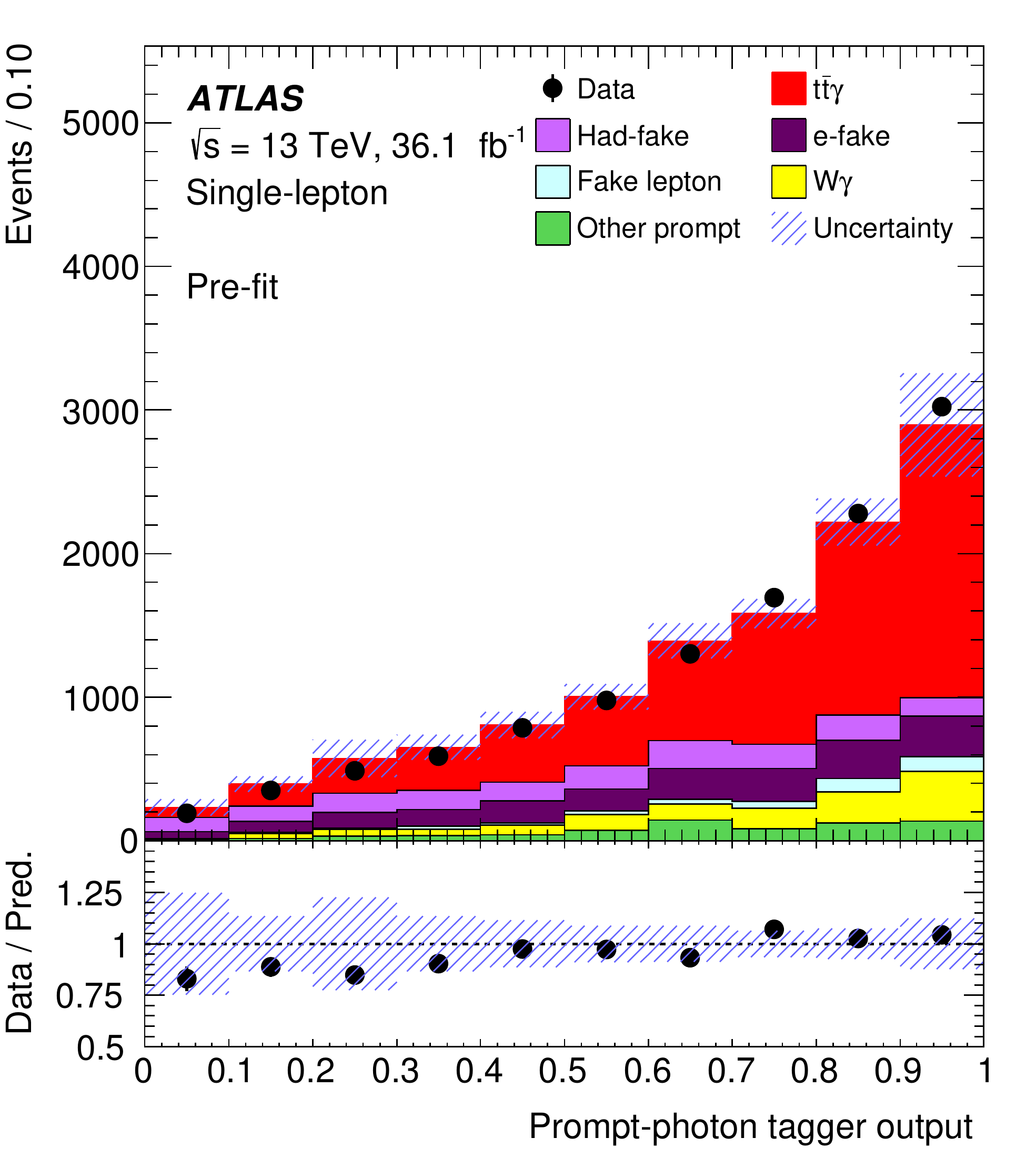}
\caption [] {Distributions of the output of the prompt-photon tagger in the \chljets channel after event selection and before likelihood fit. All data-driven corrections and systematic uncertainties are included.}
\label{fig:prefitPlotsSLPPT}
\end{figure}
 
Simulated signal and background events passing the event selection are used for the training and testing of the ELD, except for the fake-lepton background in the \chljets channel which is taken from data as described in Section~\ref{sec:nonprompt}. In the \chll channel, the selection criteria on the \met, the invariant masses, and the jet multiplicity are removed to increase the sample size for training. The training takes 15 (7) variables as input for the \chljets (\chll) channel, which are summarized in Table~\ref{tab:eldinputs}. The $b$-tagging related variables are important for the ELD training in both channels, because of their discriminating power against background without real heavy flavour jets. which have significant contributions. The use of the PPT as input to ELD improves the discrimination power against hadronic-fake background.
 
Variables like the dilepton invariant mass and missing transverse energy are useful for the ELD training in the dilepton channel, due to the dominant background of \Zgamma. The distributions of the ELD after event selection are shown in Figure~\ref{fig:prefitELD} for the \chljets and \chll channels. The shapes of the ELD are compared between signal and total background in Figure~\ref{fig:prefitELDshape} for the \chljets and \chll channels. In the \chljets channel, the kinematic properties and jet flavour compositions are similar between the \ttg signal and the dominating background, which is \ttbar production with a hadronic-fake or electron-fake photon. In the \chll channel, this is not the case since \Zgamma production is dominant. Thus the ELD is more discriminating in the \chll channel than in the \chljets channel. The ELD is used in the likelihood fit to data to extract the fiducial cross-sections.
 
\begin{table}[htbp]
\centering
\caption{Input variables for the event-level discriminator for the \chljets and \chll channels. For events without the 5th jet, the $\pt(j_5)$ is set to zero.}
\scalebox{0.93}{
\begin{tabular}{llcc}
\toprule
Variable & Description & Single lepton & Dilepton\\
\hline
PPT & Prompt-photon tagger output &  \checkmark &  \\
$H_\text{T}$ & Scalar sum of the \pt of the leptons and jets & \checkmark &  \\
$m(\gamma,\ell)$ & Invariant mass of the system of the photon and the lepton & \checkmark &  \\
\met & Missing transverse energy & \checkmark & \checkmark \\
$m_{W}^\text{T}$ & Reconstructed transverse mass of the leptonically decaying $W$-boson & \checkmark &  \\
& $= \sqrt{2 \times \pt(\ell) \times \met \times (1-\cos(\Delta\phi(\ell,\met))) }$ &  &  \\
$N_{\textrm{jets}}$ & Jet multiplicity & \checkmark &  \\
$\pt(j_1)$ & \pt of the leading jet (ordered in \pt)  & \checkmark& \checkmark \\
$\pt(j_2)$ & \pt of the sub-leading jet  &  \checkmark & \checkmark \\
$\pt(j_3)$ & \pt of the third jet  &  \checkmark &  \\
$\pt(j_4)$ & \pt of the fourth jet & \checkmark &  \\
$\pt(j_5)$ & \pt of the fifth jet & \checkmark &  \\
$N_{b\textrm{-jets}}$ & $b$-jet multiplicity & \checkmark & \checkmark \\
$b_1(j)$ & highest $b$-tagging score of all jets &  \checkmark & \checkmark \\
$b_2(j)$ & second highest $b$-tagging score of all jets & \checkmark  & \checkmark\\
$b_3(j)$ & third highest $b$-tagging score of all jets &\checkmark  &  \\
$m(\ell,\ell)$ & Invariant mass of the system of the two leptons &  & \checkmark \\
\bottomrule
\end{tabular}}
\label{tab:eldinputs}
\end{table}
 
\begin{figure}[!htbp]
\centering
\subfloat[]{
\includegraphics[width=0.45\linewidth]{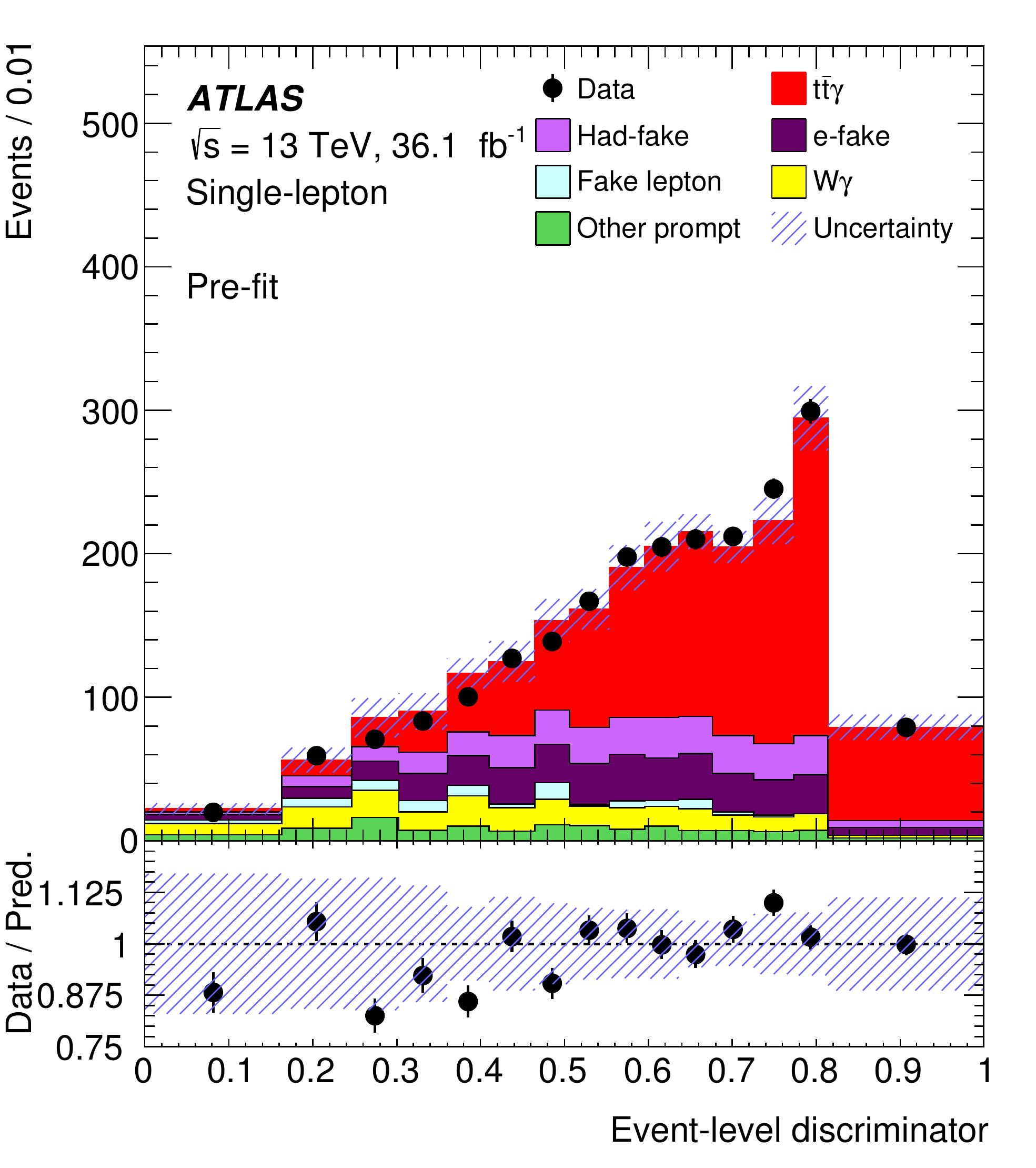}
}
\subfloat[]{
\includegraphics[width=0.45\linewidth]{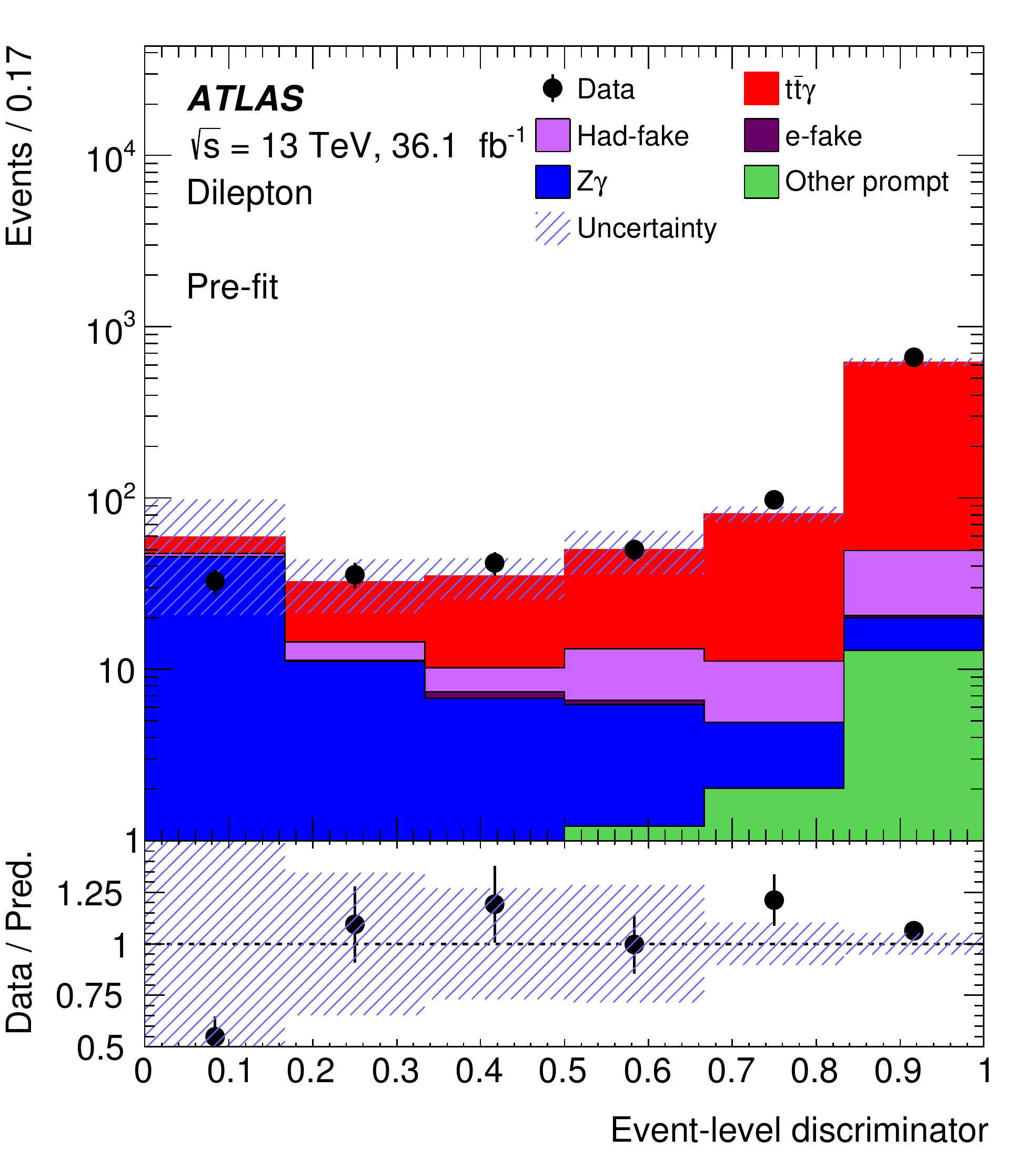}
}
\caption [] {Distributions of the ELD for the (a) \chljets and (b) \chll channels after event selection and before likelihood fit. All data-driven corrections and systematic uncertainties are included.}
\label{fig:prefitELD}
\end{figure}
 
\begin{figure}[!htbp]
\centering
\subfloat[]{
\includegraphics[width=0.45\linewidth]{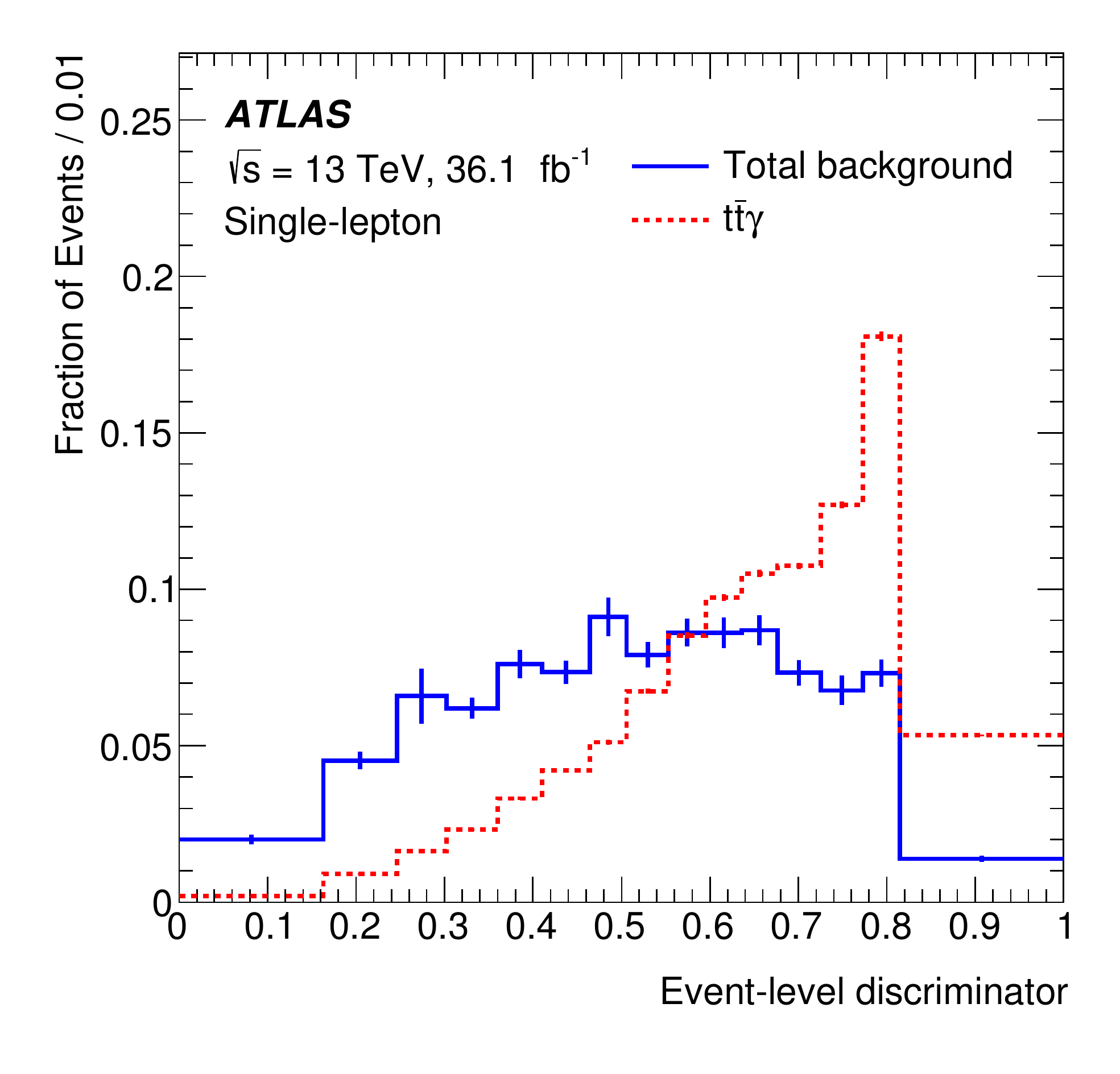}
}
\subfloat[]{
\includegraphics[width=0.45\linewidth]{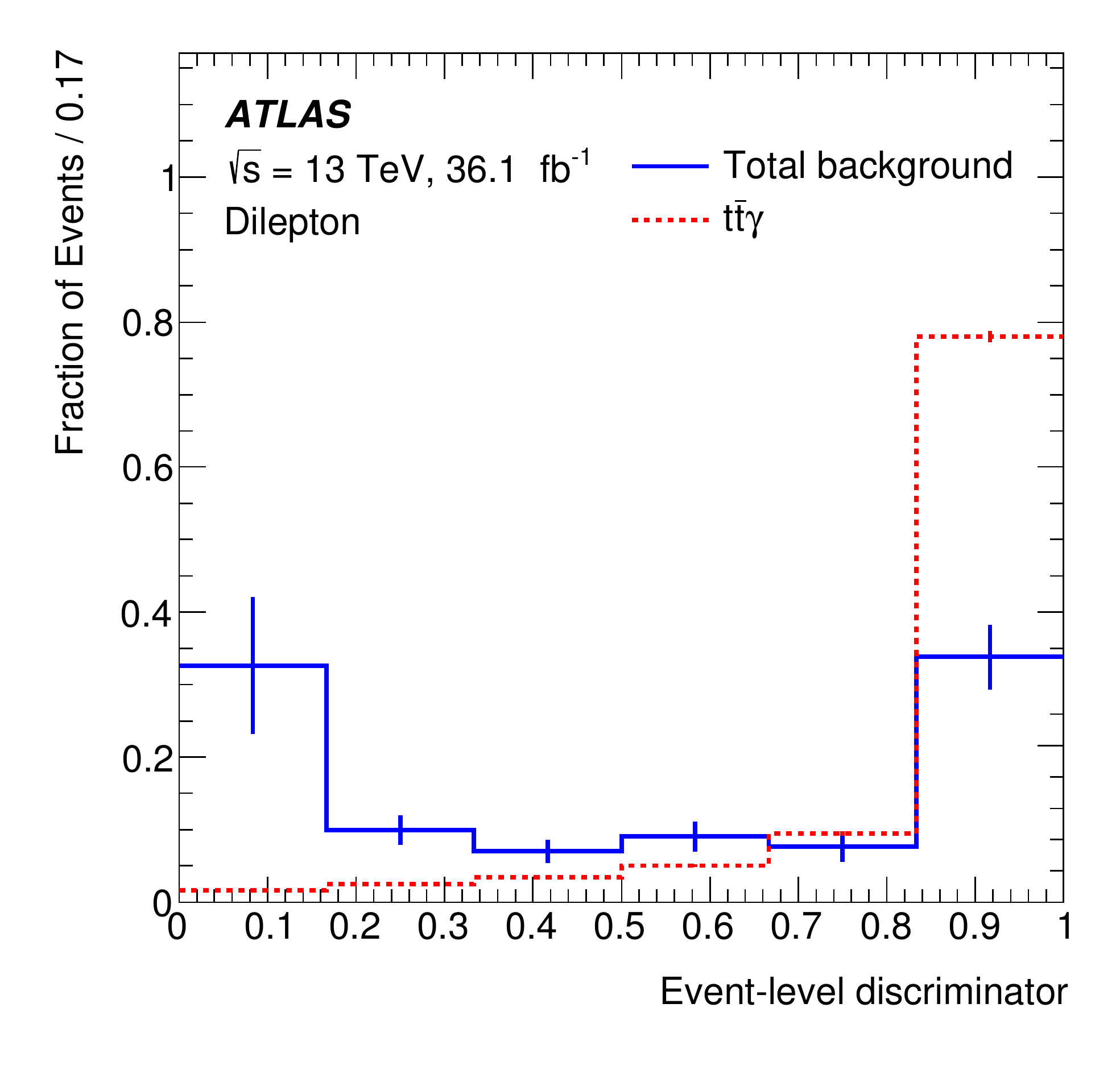}
}
\caption [] {Comparison of the shape of the ELD between signal and total background in the (a) \chljets and (b) \chll channels after event selection. All data-driven corrections are included.}
\label{fig:prefitELDshape}
\end{figure}
 
\section{Background estimation}
\label{sec:bkg}
 
\subsection{Hadronic-fake background}
\label{sec:hfake}
 
The hadronic-fake background is an important background in this analysis. Its main source is the \ttbar process, where one of the final state jets is reconstructed and identified as a photon. In addition, there are small contributions from \Wjets and single top processes to the \chljets channel and from \Zjets events to the \chll channel.
 
The hadronic-fake background is estimated using all the simulation samples by requiring that the selected photon is a hadron or from a hadronic decay at generator level. A data-driven method, called the ABCD method, is applied to derive a set of scale factors, based on the ratio of hadronic-fake background estimated by the method over the one from simulation. This set of scale factors is derived in the \chljets channel and applied to both the \chljets and \chll channels to calibrate the simulation to match data.
 
In the ABCD method, the isolation selection and part of the Tight identification criteria of the photon, which are assumed to be uncorrelated, are inverted to define three regions enriched with hadronic-fake photons. These regions are orthogonal to one another, and to the signal region. Region A uses photons that pass the isolation selection defined in Section~\ref{sec:objselection} but fail at least two out of the four identification requirements on the discriminating variables $F_\mathrm{side}$, $w_{s3}$, $\Delta E$, and $E_\mathrm{ratio}$ (defined in the Appendix), while passing all other Tight identification criteria. These four variables describe the shower shape in the first layer of ECAL and are chosen for their small correlation with the photon isolation but strong discrimination power between prompt and hadronic-fake photons. Region B uses photons that fail the identification criteria as in region A and do not pass the isolation selection. Additionally, the sum of the \pT of all tracks within $\Delta R = 0.2$ around the photon is required to be larger than 3~\GeV\ to further suppress the prompt-photon contribution. Region C selects photons that fail the isolation requirements as in region B but pass the Tight identification. Region D is the signal region.
 
The hadronic-fake background in the signal region can be expressed as:
\begin{eqnarray*}
N_{\text{D, est.}}^{\text{h-fake}}&=&\frac{N_{\text{A, data}}^{\text{h-fake}}~\times~N_{\text{C, data}}^{\text{h-fake}}}{N_{\text{B, data}}^{\text{h-fake}}}~\times~\theta_{\text{MC}}\,, \nonumber \\
\theta_{\text{MC}}&=&\frac{{}^{N_\text{D, MC}^{\text{h-fake}}}/_{N_\text{C, MC}^{\text{h-fake}}}}{{}^{N_\text{A, MC}^{\text{h-fake}}}/_{N_\text{B, MC}^{\text{h-fake}}}}\,,
\end{eqnarray*}
where $N_{\text{A, data}}^{\text{h-fake}}$, $N_{\text{B, data}}^{\text{h-fake}}$ and $N_{\text{C, data}}^{\text{h-fake}}$ are the numbers of hadronic-fake events in regions A, B, and C, estimated by subtracting the events with prompt photons and other backgrounds from the number of data events in these regions, and $N_{\text{A, MC}}^{\text{h-fake}}$, $N_{\text{B, MC}}^{\text{h-fake}}$, $N_{\text{C, MC}}^{\text{h-fake}}$ and $N_{\text{D, MC}}^{\text{h-fake}}$ are the numbers of hadronic-fake events predicted by simulation in regions A, B, C, and D. The factor $\theta_{\text{MC}}$ corrects for possible bias caused by residual correlation between the isolation variables and the four discriminating variables used to define the regions.
 
The ABCD method is applied to photons in different \pT-$\eta$ bins, separately for converted and unconverted photons. The resulting scale factors range from 0.8 to 3.2, with typical values around 1.5 and large statistical and systematic uncertainties of more than 0.5. These scale factors are applied to the MC-based hadronic-fake background prediction.
 
\subsection{Electron-fake background}
\label{sec:efake}
 
Dilepton events where an electron is mis-identified as a photon contribute to the electron-fake background in the \chljets channel. Its main source is the \ttbar \chllic decay process. When the lepton is an electron, there is also some contribution from the $Z\to ee$ process. A data-driven method is applied to derive a set of scale factors to correct the electron-to-photon fake rate in simulation to match data.
 
The electron-to-photon fake rate is measured with a tag-and-probe method using two control regions (CR), exploiting the $Z\to ee$ process. For the first CR, $Z\to ee$ event candidates are selected in which one of the electrons fulfills the photon selection criteria, i.e. contributes to the electron-fake background, and the electron-photon pair should have an invariant mass in the range [40, 140]~$\GeV$ and an opening angle greater than 2.62 rad. The electron is called the tag electron and the electron-fake photon candidate is referred to as probe photon. Non-$Z$-boson backgrounds are subtracted with a sideband fit of the invariant mass distribution. The fitted signal contains $Z\to ee\gamma$ contributions with one of the electrons not reconstructed or identified which is subtracted using simulation. For the second CR, events with an electron-positron pair satisfying the same requirements as in the first CR are selected and the same procedure is applied. The fake rate is calculated as the ratio of the number of probe photons over the number of probe electrons. To avoid a trigger bias, the tag electron in both CRs must match the \chljets
trigger. A set of \pT-$\eta$ binned fake-rate scale factors is determined by taking the ratio between the fake rate in data and in the simulation. The values of the scale factors range from 0.8 to 2.1 and are in general consistent with unity within their uncertainties.
 
The electron-fake background in the \chljets channel is validated in a control region selected by replacing the photon in the signal region event selection with an electron. This region is dominated by \ttbar events and \Zjets events in the single electron channel with negligible contribution from other processes. The ratio of data over prediction in this region is 0.98 $\pm$ 0.01, where the uncertainty is due to the sample size. This overall correction is applied to the electron-fake background predicted by simulation in the \chljets channel signal region, in addition to the fake-rate scale factors.
 
The electron-fake background in the \chll channel is very small. Simulation is used to predict its contribution. No dedicated control region is selected to validate this background.
 
\subsection{Fake-lepton background}
\label{sec:nonprompt}
 
In the \chljets channel, the fake-lepton background is dominated by multi-jet processes with an additional photon which could either be a prompt or a fake photon. It is estimated directly from data by using a matrix method~\cite{ATLAS-CONF-2014-058}. The number of background events in the signal region is evaluated by applying efficiency factors (fake lepton and real lepton efficiencies) to the number of events satisfying a tight (identical to the signal selection) as well as a looser lepton selection. The fake-lepton efficiency is measured using data in control regions dominated by multi-jet background with the real lepton contribution subtracted using simulation. The real lepton efficiency is extracted from a tag-and-probe technique using leptons from $Z$ boson decays. The efficiencies are parametrized as a function of the lepton $\eta$ and $m_{W}^{T}$ (the lepton $\pt$ and $m_{W}^{T}$) when the lepton is an electron (muon).
 
In the \chll channel, the contamination of background processes with at least one fake lepton is estimated by selecting same-sign \chll events in data, after subtracting events with two same-sign prompt leptons, using simulation. The fake-lepton background in the \chll channel is found to be negligible.
 
\subsection{Prompt-photon background}
\label{sec:prompt}
 
All background processes to \ttbar production are also background to \ttg production when accompanied by prompt-photon radiation. These processes include \Wgamma, \Zgamma, and associated production of a photon in single top, diboson, and \ttV productions. In the \chljets channel, \Wgamma is the dominant prompt-photon background, and \Zgamma and the others are grouped together as ``Other prompt''. In the \chll channel, \Zgamma is dominant, with all others grouped as ``Other prompt''. Background from \ttbar production with a photon produced in an additional $pp$ interaction in the same bunch crossing has been studied and is found to be negligible.
 
Validation regions are selected to check the modelling of \Wgamma in the single-lepton channel and \Zgamma in the dilepton channel. The \Zgamma validation region is selected by requiring exactly one $b$-tagged jet and the invariant mass of the system of the two leptons in a mass window of [60, 100]~\GeV. The \Wgamma validation region is selected with the same event selection as for the signal region of the single-lepton channel, with the following modifications: the number of jets must be either two or three; exactly one $b$-jet is required; the \MET is required to be larger than 40~\GeV; the ELD value must be smaller than 0.04; and the invariant mass of the system of the lepton and photon is required to be smaller than 80~$\GeV$ if the lepton is an electron. The modelling of \Wgamma is also checked in a light-flavour validation region requiring zero $b$-jet and without the ELD cut.
 
The normalization of the \Wgamma background is treated as a free parameter of the likelihood fit in the \chljets channel, since this background is well separated from the \ttg signal by the ELD and the uncertainty of its theoretical prediction is large. The shape of \Wgamma is taken from simulation and checked in the validation region to ensure good modelling. The normalization and shape of the \Zgamma background in the \chll channel as well as other prompt backgrounds in both channels are predicted by simulation.
 
\section{Analysis strategy}
\label{sec:strategy}
 
The analysis is performed in two parts, one being the measurement of the fiducial cross-section and the other the measurement of normalized differential cross-sections in the same fiducial region. Both parts share the same strategy for the estimation of backgrounds and systematic uncertainties. In the fiducial cross-section measurement, the ELD is fitted and the post-fit background yields and systematic uncertainties are used. In the normalized differential cross-section measurements, no fit is performed, except for the determination of the \Wgamma contribution in the \chljets channel, where a systematics-free ELD fit is performed.
 
\subsection{Fiducial region definition}
\label{sec:fiducial}
 
The fiducial region of the analysis is defined at particle level in a way that mimics the event selection in Section~\ref{sec:evtselection}. Leptons must have $\pT > 25~\GeV$ and $|\eta| < 2.5$ and must not originate from hadron decays. Photons not from hadron decays and in a $\Delta R = 0.1$ cone around a lepton are added to the lepton before the lepton selection. Photons are required to have $\pT > 20~\GeV$ and $|\eta| < 2.37$ and must not originate from hadron decays or be used for lepton dressing. The photon isolation computed from the ratio of the scalar sum of the transverse momentum of all stable\footnote{A stable particle is a particle with $c\times\tau$ > 10 mm, where $c$ is the speed of light and $\tau$ is the lifetime of the particle.} charged particles around the photon over its transverse momentum must be smaller than 0.1. Jets are clustered using the anti-$k_{t}$ algorithm with $R=0.4$ using all final state particles excluding non-interacting particles and
muons that are not from hadron decays. Jets must have $\pT > 25~\GeV$ and $|\eta| < 2.5$. A ghost matching method~\cite{Cacciari:2008gn} is used to determine the flavour of the jets, with those matched to $b$-hadrons tagged as $b$-jets. A simple overlap removal is performed: jets within $\Delta R < 0.4$ of a selected lepton or photon are removed. For events in the \chljets (\chll) channel, exactly one photon and exactly one lepton (two leptons) are required. At least four (two) jets are required with at least one of them $b$-tagged. Events are rejected if there is any lepton and photon pair satisfying $\Delta R(\gamma,\ell) < 1.0$. The acceptance for the generated signal events to pass the fiducial selection of the \chljets (\chll) channel is 8.2\% (0.96\%).

\subsection{Fiducial cross-section}
\label{sec:likelihood}
 
The fiducial cross-section is extracted using a profile likelihood fit to the ELD distribution. The parameter of interest, the fiducial cross-section $\sigma_{\text{fid}}$, is related to the number of signal events in bin $i$ of the ELD as
\begin{equation*}
N_i^s = L \times \sigma_{\text{fid}} \times C \times f_i^{\text{ELD}}\,,
\end{equation*}
where $L$ is the integrated luminosity, $C$ is the correction factor for the signal efficiency and for migration into the fiducial region, and $f_i^{\text{ELD}}$ is the fraction of signal events falling into bin $i$ of the ELD. The correction factor $C$ is defined as $N_{\text{MC}}^{s, \text{sel.}}/N_{\text{MC}}^{s, \text{fid}}$, where $N_{\text{MC}}^{s, \text{sel.}}$ is the simulated number of signal events passing the event selection described in Section~\ref{sec:evtselection} and $N_{\text{MC}}^{s, \text{fid}}$ is the corresponding number of signal events generated in the fiducial region defined in Section~\ref{sec:fiducial}. The value of $C$ is 0.36 (0.30) for the \chljets (\chll) channel, with negligible statistical uncertainty.
 
A likelihood function is defined from the product over all bins of the ELD distribution:
\begin{equation*}
\mathcal{L} = \prod_i P(N_i^{\text{obs}} | N_i^s(\vec{\theta}) + \sum_b N_i^b(\vec{\theta})) \times \prod_t G(0|\theta_t,1)\,,
\end{equation*}
where $N_i^{\text{obs}}$, $N_i^s$, and $N_i^b$ are the observed number of events in data, the predicted number of signal events, and the estimated number of background events in bin $i$ of the ELD, which form a Poisson term $P$ in that bin. Nuisance parameter $\theta_t$ is to parameterize a systematic uncertainty $t$, which is constrained by a Gaussian $G(0|\theta_t,1)$, so that when it changes from zero to $\pm$1, the quantities affected by this systematics in the likelihood change by $\pm$1 standard deviation. The collection of all the systematic uncertainties is denoted as $\vec{\theta}$. For systematic uncertainties related to the finite number of MC events, the Gaussian terms in the likelihood are replaced by Poisson terms. Each systematic uncertainty affects $N^s_i$ and $N^b_i$ in each bin of the ELD. The cross-section is measured by profiling the nuisance parameters and maximizing this likelihood.
 
\subsection{Normalized differential cross-sections}
\label{sec:unfold}
 
An unfolding procedure is applied to the observed detector-level distribution of a given observable, with backgrounds subtracted, to derive the true distribution of the signal at particle level, from which the differential cross-section as a function of the observable is calculated. The differential cross-section is normalized to unity.
 
The differential cross-section is given by
\begin{equation*}
\sigma_{k} = \frac{1}{L} \times \frac{1}{\epsilon_k} \times \sum_j M_{jk}^{-1} \times (N^{\mathrm{obs}}_{j} - N^{b}_{j}) \times (1-f_{\mathrm{out},j})\,.
\end{equation*}
The indices $j$ and $k$ indicate the bin of the observable at detector and particle levels, respectively. The variables $N^{\textrm{obs}}_{j}$ and $N^{b}_{j}$ are the number of observed events and of estimated background events in bin $j$ at detector level, respectively. The efficiency $\epsilon_k$ is the fraction of signal events generated at particle level in bin $k$ of the fiducial region to be reconstructed and selected at detector level and have the objects, that are used to define the observable to be unfolded, matched between reconstruction and particle-levels with $\Delta R$ < 0.1. The migration matrix $M_{kj}$ expresses the probability for an event in bin $k$ at particle level to end up in bin $j$ at detector level, calculated from events passing both the fiducial region selection and the event selection, as well as the above matching procedure. The outside-migration fraction $f_{\mathrm{out},j}$ is the fraction of signal events generated outside the fiducial region but reconstructed and selected in bin $j$ at detector level or events failing the above matching. The signal MC sample is used to determine $\epsilon_k$, $f_{\mathrm{out},j}$, and $M_{kj}$, the values of which are illustrated in Figure~\ref{fig:effmigsl_pt}, using the photon \pT in the \chljets channel as an example. The normalization and the corresponding uncertainty of the \Wgamma contribution in the \chljets channel are taken from the likelihood fit introduced in Section~\ref{sec:likelihood} but without systematic uncertainties included. The normalized differential cross-section is
\begin{equation*}
\sigma_{k}^{\textrm{norm}} = \frac{\sigma_{k}}{\sum_k \sigma_{k}} \,,
\end{equation*}
where the sum is over all the bins of the observable.
 
\begin{figure}[!htbp]
\centering
\subfloat[]{
\includegraphics[width=0.45\linewidth]{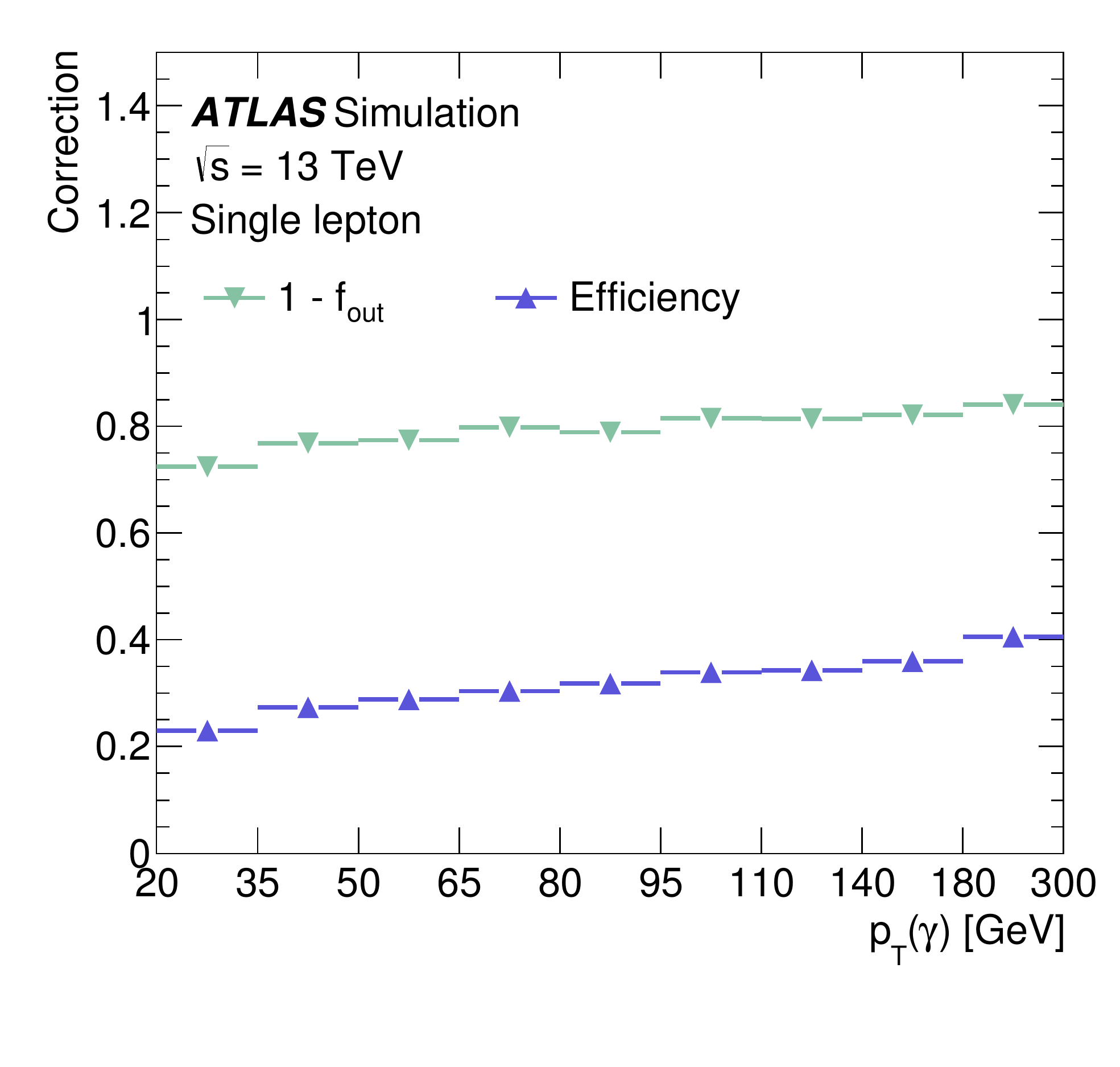}
}
\subfloat[]{
\includegraphics[width=0.45\linewidth]{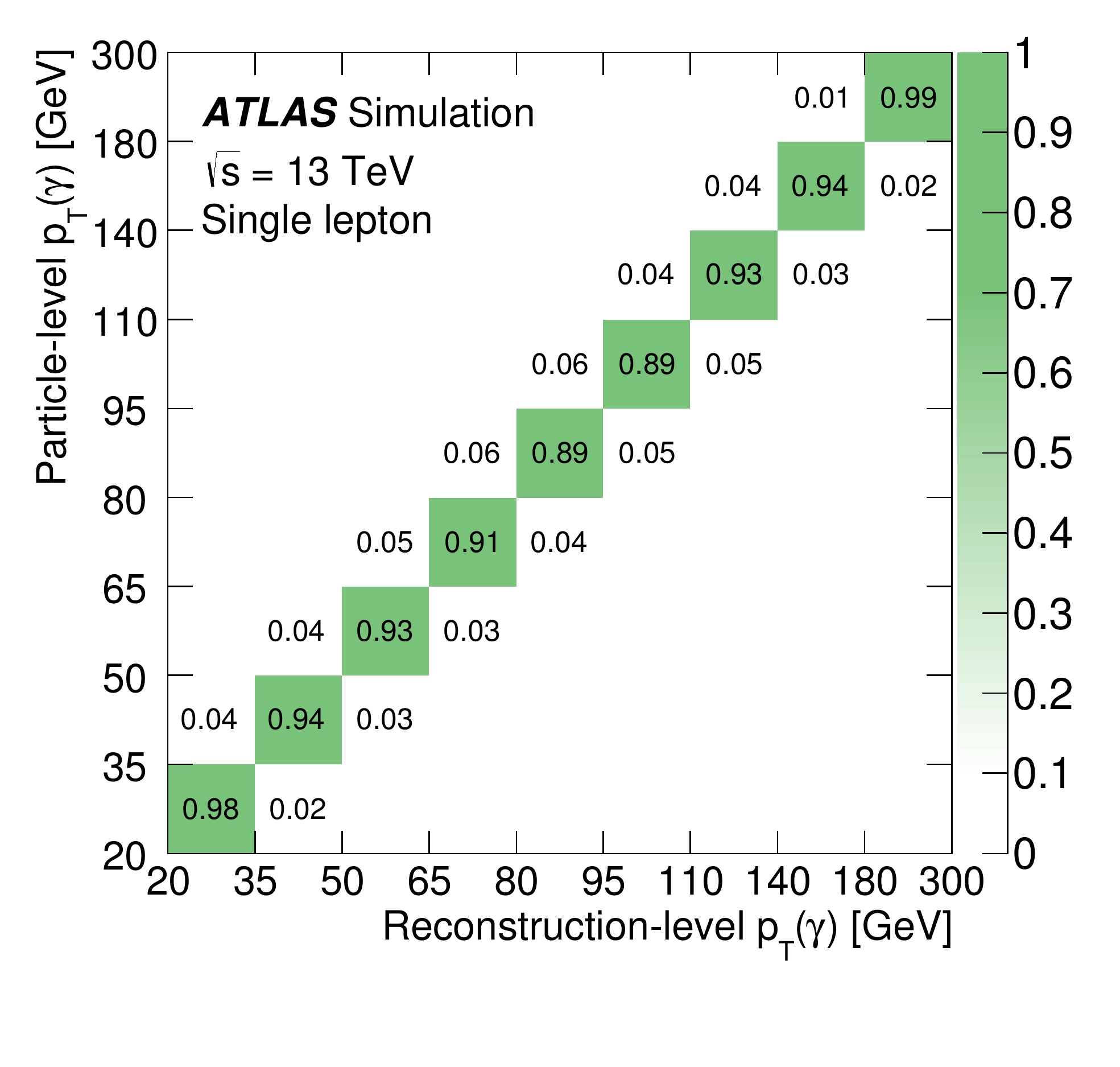}
}
\caption [] {The (a) efficiency and outside fraction and (b) migration matrix for the photon \pT in the \chljets channel.}
\label{fig:effmigsl_pt}
\end{figure}
 
The inversion of the migration matrix $M_{kj}$ is approximated using the iterative Bayesian method~\cite{DAgostini:1994fjx} implemented in the \textsc{RooUnfold} package~\cite{Adye:2011gm}. The method relies on the Bayesian probability formula to invert the migration matrix, starting from a given prior of the particle-level distribution and iteratively updating it with the posterior distribution. The binning choices of the unfolded observables take into account the detector resolution and the expected statistical uncertainty, with the latter being the dominating factor. Three iterations are chosen which give a good convergence of the unfolded distribution and a statistically stable result. Tests are performed, using simulation, to verify that the unfolding procedure does not bias the results while the estimated uncertainties are still reasonable. The results are cross-checked with other unfolding methods, which give consistent results.
 
The chosen observables to unfold are the photon \pT and $|\eta|$ and the $\Delta R$ between the photon and the closest lepton for both \chljets and \chll channels and the $\Delta \phi$ and $|\Delta \eta|$ between the two leptons for the \chll channel. These are all lepton or photon observables, therefore the migration matrices are almost diagonal, making the unfolding simple and converging fast. The kinematic properties of the photon are sensitive to the $t\gamma$ coupling, while the \chll $\Delta \phi$ is sensitive to the \ttbar spin correlation. The normalized differential cross-sections are measured, since the overall signal normalization is given by the measured fiducial cross-section.
 
\section{Systematic uncertainties}
\label{sec:syst}
 
Signal and background modelling and experimental uncertainties in the analysis are described in this section, as well as the PPT systematic uncertainty. They affect the normalization of signal and background and/or the shape of their corresponding distributions, such as the ELD and the observables to be unfolded. Each of the signal and background modelling uncertainties is correlated between different channels for the relevant signal or background process. Each of the experimental uncertainties is correlated between signal and simulated backgrounds and between different channels. The PPT systematic uncertainty is separately studied for the prompt, electron-fake, and hadronic-fake photons. Table~\ref{tab:systSummary} gives a summary of these uncertainties and their impact to the fiducial cross-section measurements.
 
\subsection{Signal modelling uncertainties}
 
The signal modelling uncertainties include the uncertainties due to the choice of the QCD scales, the parton shower, the amount of ISR and FSR, and the PDF set. Their effects on the corrections defined in Section~\ref{sec:strategy} (both for the fiducial and normalized differential cross-section measurements) as well as on the shape of the ELD distributions are evaluated.
 
To study the QCD scale uncertainty, the renormalization and factorization scales are varied up and down by a factor of two from their nominal choices independently or simultaneously. The largest variation of the corrections or the shapes is assigned as the uncertainty. To evaluate the parton shower uncertainty, \PYTHIA8 and \HERWIG7 both interfaced to \MGaMC are compared. The ISR/FSR uncertainty is studied by comparing the variations of the A14 tune parameters of \PYTHIA8 with its nominal values. The PDF uncertainty is evaluated using the standard deviation of the distribution formed by the 100 eigenvector set of the NNPDF set~\cite{Pumplin:2002vw}.
 
\subsection{Background modelling uncertainties}
 
The systematic uncertainties on the hadronic-fake background due to background subtraction in the hadronic-fake control regions A, B, and C are estimated by varying up and down the signal by 100\%, the other MC-based backgrounds by 50\%, and the other data-driven backgrounds by their estimated uncertainties, separately. The statistical uncertainties in the three data control regions are also considered. Systematic uncertainties arising from the correction factor $\theta_{\text{MC}}$, the extrapolation of the hadronic-fake scale factors, and the shapes of the distributions of the ELD and the observables to be unfolded are estimated using \ttbar samples rather than all of the simulated hadronic-fake samples, since \ttbar is the dominant source of the hadronic-fake background. The uncertainty due to the rate of additional QCD radiation is estimated by comparing the samples with enhanced/reduced parton shower radiation as described in Section~\ref{sec:samples} with the nominal sample, and the uncertainty due to the modelling of the generator and parton shower is estimated by comparing \textsc{Powheg}+\PYTHIA8 with \SHERPA.
 
The systematic uncertainties on the electron-fake background mainly come from the sideband fit when measuring the fake rate, which is estimated by varying the fit parameters within their uncertainties. The uncertainty due to $Z\to ee\gamma$ subtraction is also considered by replacing the \Zgamma MC sample by the \Zjets sample, where the photon radiation is described by the parton shower. Uncertainties of the shapes of the distributions are evaluated using \ttbar systematic variations MC samples as for the hadronic-fake background. In the \chll channel where the electron-fake background is very small, a 50\% uncertainty is assumed to cover a possible mis-modelling in the estimate.
 
For evaluating the systematic uncertainty of the fake-lepton background in the single-lepton channel, several alternative parameterizations of the real and fake efficiencies of the matrix method are studied and two predicting larger and smaller yields are selected as up and down variations. The lepton $\eta$, $b$-jet multiplicity, and $m_{W}^{T}$ (the lepton $\eta$, minimum $\Delta R$ between the lepton and the closest jet, and the jet \pT) parameterization is used as up (down) variation when the lepton is an electron. The jet \pT and $b$-jet multiplicity (the lepton \pT, $\eta$, and minimum $\Delta R$ between the lepton and the closest jet) parameterization is used as up (down) variation when the lepton is a muon. The resulting uncertainties are around 50\%, and given that this background contribution is relatively small, no additional systematic uncertainties are considered.
 
The uncertainty on the \Vgamma background shape is studied by varying the renormalization and factorization scales up and down by a factor of two from their nominal values independently or simultaneously and then choosing the maximum shape distortions as the QCD scale uncertainties. For the \Zgamma background in the \chll channel, \SHERPA is compared with \textsc{MG5}\_\textsc{aMC}+\PYTHIA8 to evaluate the shape uncertainty due to the choice of generator and parton shower. No shape modelling uncertainty is assigned to the other small prompt backgrounds. Apart from the \Wgamma background in the \chljets channel whose normalization is a free parameter in the likelihood fit, a normalization uncertainty of 50\% is assigned to each source of the prompt-photon background, included in Table~\ref{tab:prefiteventYieldswSFs} as ``\Zgamma'' and ``Other prompt''.
 
\subsection{Experimental uncertainties}
\label{sec:expsys}
 
Experimental systematic uncertainties affect the normalization and shape of the simulated signal and background samples. For MC-based backgrounds calibrated to data using data-driven techniques, only the shape variations are considered.
 
The photon identification and isolation efficiencies as well as the efficiencies of the lepton reconstruction, identification, isolation, and trigger in the MC samples are all corrected as mentioned in Section~\ref{sec:objselection}. These corrections, which are \pT and $\eta$ dependent, are varied to study their impact on the final results. Similarly, the corrections to the lepton and photon momentum scale and resolution in simulation are varied within their uncertainties~\cite{PERF-2017-03,PERF-2015-10}.
 
The PPT systematic uncertainty is evaluated separately for prompt photons and hadronic-fake photons. The data-driven PPT scale factors as mentioned in Section~\ref{sec:objselection} are turned on and off to assign a PPT systematic uncertainty for the prompt photon. The resulting uncertainty is also assigned to the electron-fake photon PPT output distribution, since its shape and shape difference between data and simulation are similar to that of the prompt photons. The maximum PPT shape difference between data and prediction in the hadronic-fake control region C of Section~\ref{sec:hfake}, with the expected signal contamination in this region varied by $\pm 50\%$,  is used to estimate the hadronic-fake PPT uncertainty. The hadronic-fake photons in region C are non-isolated while those of the signal region are isolated. To account for a possible underestimation of the systematic uncertainty caused by this difference, the shape differences between data and prediction in the isolated hadronic-fake control region A of Section~\ref{sec:hfake} are considered as an additional PPT systematic uncertainty. The PPT shape uncertainties are estimated in photon $\pt$ and $\eta$ bins.
 
The jet energy scale (JES) uncertainty is derived using a combination of simulations, test beam data and \textit{in situ} measurements~\cite{PERF-2012-01,PERF-2011-03,PERF-2011-05}. Additional contributions from jet flavour composition, $\eta$-intercalibration, punch-through, single-particle response, calorimeter response to different jet flavours, and pile-up are taken into account, resulting in 21 uncorrelated JES uncertainty subcomponents. The jet energy resolution (JER) in simulation is smeared up by the measured JER uncertainty~\cite{PERF-2011-04}. The uncertainty associated with the JVT cut is obtained by varying the efficiency correction factors. The $b$-tagging weights used for jet flavour tagging are corrected by data, separately for $b$-jets, $c$-jets, and light-flavour jets~\cite{ATL-PHYS-PUB-2015-022,ATL-PHYS-PUB-2016-012}. The corrections are varied by their measured uncertainties.
 
The uncertainties associated with energy scales and resolutions of photons, leptons and jets are propagated to the \MET. Additional uncertainties originate from the modelling of its soft term~\cite{ATLAS-CONF-2013-082}.
 
The uncertainty in the combined 2015+2016 integrated luminosity is 2.1\%. It is derived, following a methodology similar to that detailed in Ref.~\cite{DAPR-2013-01}, and using the LUCID-2 detector for the baseline luminosity measurements~\cite{Avoni_2018}, from calibration of the luminosity scale using x-y beam-separation scans.
 
The uncertainty associated to the modelling of pile-up in the simulation is assessed by varying the reweighting of the pile-up in the simulation within its uncertainties.
 
\subsection{Systematic uncertainties of the measured differential cross-section}
 
Systematic uncertainties for unfolding arise from the detector response description, signal modelling, and background modelling. The systematic uncertainties due to background modelling and the detector response are evaluated by varying the input detector-level pre-fit distributions, unfolding them with corrections based on the nominal signal sample, and calculating the difference of the resulting unfolded distributions with respect to the nominal one. The systematic uncertainties due to signal modelling are evaluated by varying the signal corrections, i.e. the migration matrix $M_{kj}$, the efficiency $\epsilon_k$ and the fraction $f_{\mathrm{out},j}$ as defined in Section~\ref{sec:unfold}, with which the nominal input detector-level pre-fit distributions are unfolded, and calculating the difference of the resulting unfolded distributions with respect to the nominal one. The statistical uncertainties of the signal and background MC samples are also considered. The covariance matrix $C_{ij}$ for each of these systematic uncertainties is estimated as $\sigma_i \times \sigma_j$, where $\sigma_i$ and $\sigma_j$ are the symmetrized uncertainties for bin $i$ and bin $j$ of the unfolded distribution. The covariance matrix for the statistical uncertainty of data is calculated by the unfolding algorithm~\cite{Adye:2011gm}.
 
\section{Results}
\label{sec:result}
 
\subsection{Fiducial cross-sections}
 
The fiducial cross-section is extracted via a binned maximum likelihood fit to the ELD distribution in data as described in section~\ref{sec:likelihood}. The measured cross-sections are
\begin{align*}
\sigma^{\text{SL}}_{\text{fid}} & =  521 \pm 9\text{(stat.)} \pm 41\text{(sys.)}~\text{fb}~\text{and}  \nonumber \\
\sigma^{\text{DL}}_{\text{fid}} & =  69  \pm 3\text{(stat.)} \pm 4\text{(sys.)}~\text{fb}\,,
\end{align*}
for the \chljets and \chll channels, respectively, and agree well within uncertainties with the corresponding predicted cross-sections of $495 \pm 99$~fb and $63 \pm 9$~fb. The ELD distributions after the fit (post-fit) are shown in Figure~\ref{fig:ELDpostfit} for the \chljets and \chll channels. The corresponding event yields are summarized in Table~\ref{tab:postFitYields}, including all the systematic uncertainties. Compared to pre-fit (Table~\ref{tab:prefiteventYieldswSFs}), the signal event yields are higher, while some of the background event yields are lower. Some of the systematic uncertainties are moderately constrained, e.g. the parton shower uncertainty of the \ttg and \ttbar modelling and the PPT shape uncertainty of prompt photons.
 
The fiducial cross-sections in each of the individual channels ($e$+jets, $\mu$+jets, $ee$, $e\mu$, and $\mu\mu$) as well as a combined \chljets and \chll cross-section are also measured. The former are measured by fitting the ELD distribution in each individual channel separately, while the latter is measured by fitting them simultaneously, sharing the same signal strength parameter $\mu = \sigma_{t\bar{t}\gamma}/\sigma_{t\bar{t}\gamma}^{\textrm{NLO}}$, which scales coherently the fiducial cross-sections of each channel. A comparison of all measurements with the predictions is shown in Figure~\ref{fig:crossSectionMu}.
 
\begin{figure}[!htbp]
\centering
\subfloat[]{
\includegraphics[width=0.45\linewidth]{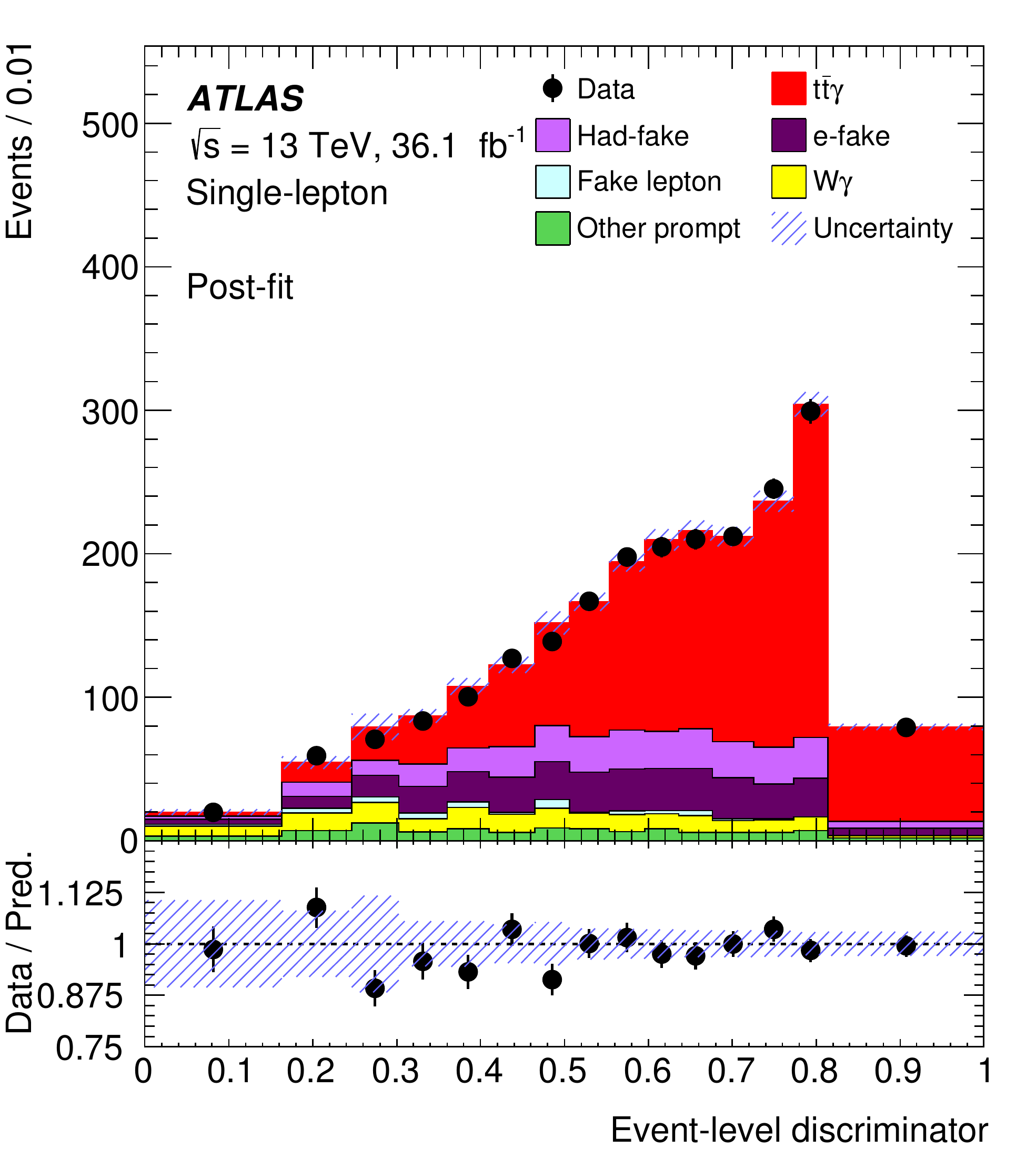}
}
\subfloat[]{
\includegraphics[width=0.45\linewidth]{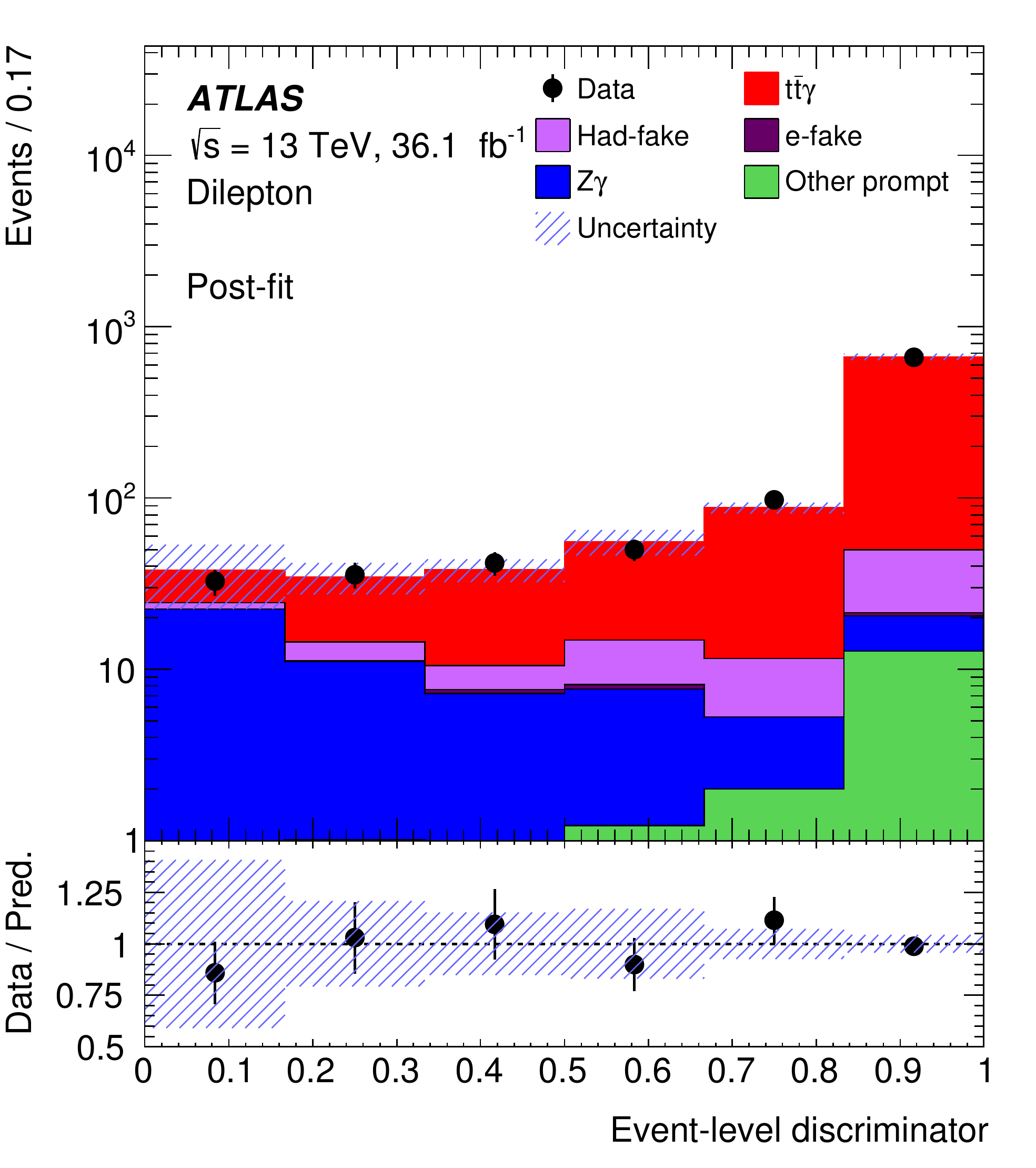}
}
\caption [] {The post-fit ELD distributions for the (a) \chljets and (b) \chll channels. All the systematic uncertainties are included.}
\label{fig:ELDpostfit}
\end{figure}
 
\begin{table}[!h]
\begin{center}
\caption{The observed data and post-fit event yields for the signal and backgrounds in the \chljets and \chll channels. All data-driven corrections and systematic uncertainties are included. The fake-lepton background in the dilepton channel is negligible, represented by a ``-''. The \Zgamma (\Wgamma) background in the single-lepton (dilepton) channel is included in ``\Other.''}
\scalebox{1.0}{
\begin{tabular}{l|r@{$\,\pm\,$}r|r@{$\,\pm\,$}r}
\toprule
Channel & \multicolumn{2}{c|}{Single lepton} & \multicolumn{2}{c}{Dilepton}  \\ \hline
\ttg & $7\,040$ & 350 &  780 & 44 \\ \hline
Hadronic-fake & $1\,470$ & 180 &  49 & 26 \\ \hline
Electron-fake & $1\,620$ & 160 &  2 & 1 \\ \hline
Fake lepton & 186 & 68 &  \multicolumn{2}{c}{-} \\ \hline
\Wgamma & 900 & 370 &  \multicolumn{2}{c}{} \\ \hline
\Zgamma & \multicolumn{2}{c|}{} & 55 & 29 \\ \hline
\Other & 570 & 180 &  18 & 7 \\ \hline
\hline
Total & $11\,790$ & 180 &  906 & 38 \\ \hline
Data & \multicolumn{2}{l|}{$11\,662$} &  \multicolumn{2}{l}{902} \\
\bottomrule
\end{tabular}
}
\label{tab:postFitYields}
\end{center}
\end{table}
 
\begin{figure}[!htbp]
\centering
\includegraphics[width=0.7\textwidth]{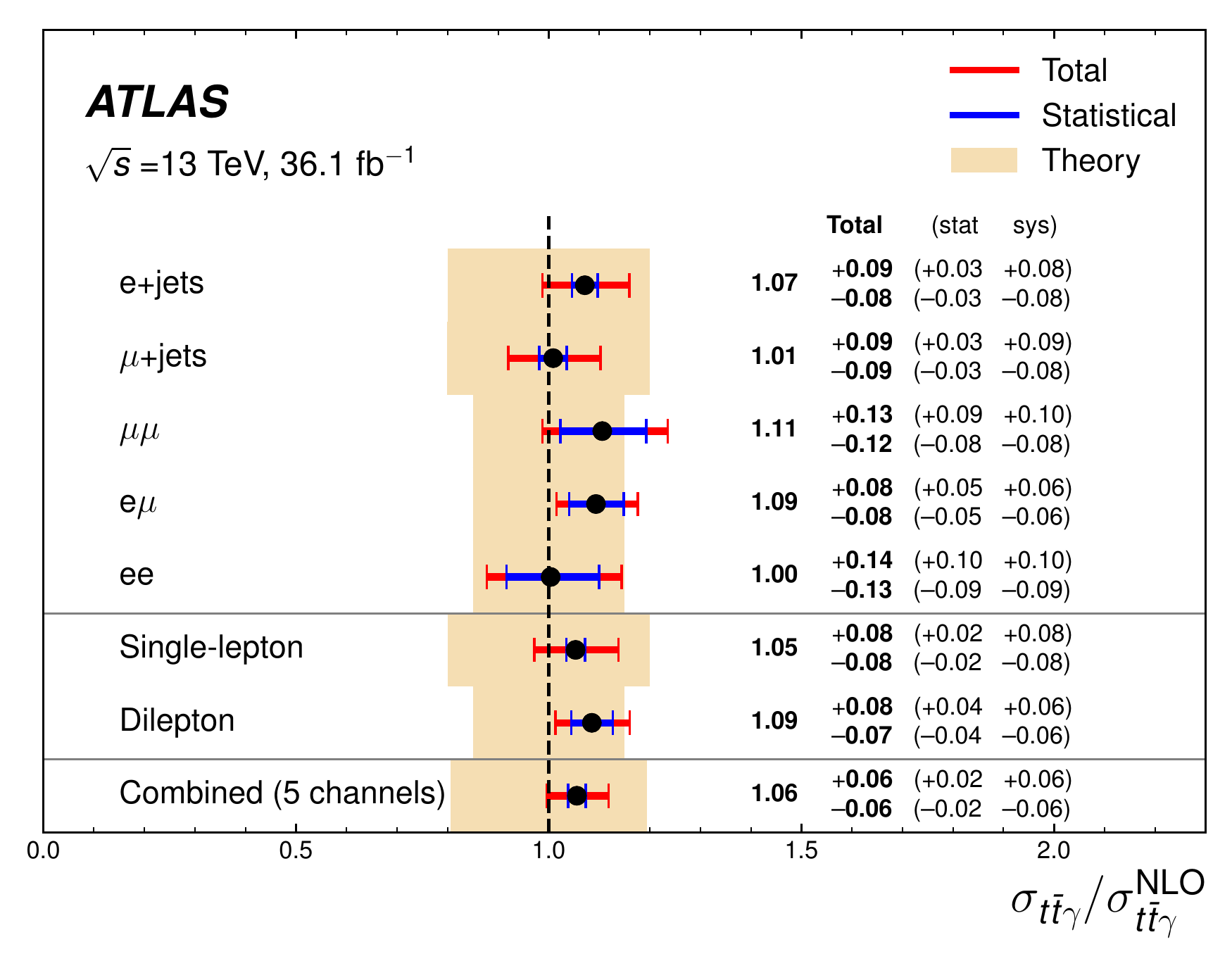}
\caption [] {
The measured fiducial cross-sections normalized to their corresponding NLO SM predictions~\cite{melnikov} for the five individual channels and for the \chljets and \chll channels, as well as the combination of all channels.  The statistical uncertainties are the inner error bars, while the total uncertainties are the outer error bars. The NLO prediction for the inclusive fiducial  cross-section is represented by the dashed vertical line, and the theoretical uncertainties are represented by the shaded bands.}
\label{fig:crossSectionMu}
\end{figure}
 
All the systematic uncertainties introduced in Section~\ref{sec:syst} are grouped into a smaller set of classes and summarized in Table~\ref{tab:systSummary} for the \chljets and \chll channels. The effect of each group of uncertainties is calculated from the quadratic difference between the relative uncertainty in the measured fiducial cross-section with this group of uncertainties included or excluded from the fit with corresponding nuisance parameters fixed to their fitted values. In the \chljets channel, the jet-related and background modelling systematic uncertainties are dominant, followed by the PPT and signal modelling systematic uncertainties. In the \chll channel, the data statistical uncertainty is the leading contribution, followed by the signal and background modelling systematic uncertainties. The luminosity and pile-up uncertainties are also important in this channel.
 
\begin{table}[h]
\centering
\caption{
Summary of the effects of the groups of systematic uncertainties on the fiducial cross-section in the \chljets and \chll channels. Due to rounding effects and small correlations between the different sources of uncertainty, the total systematic uncertainty is different from the sum in quadrature of the individual sources.}
\scalebox{1.0}{
\begin{tabular}{l|r|r}
\toprule
Source &  Single lepton (\%) &  Dilepton (\%) \\
\hline
Signal modelling 	& $\pm$ 1.6 	& $\pm$ 2.9   \\
Background modelling	& $\pm$ 4.8 	& $\pm$ 2.9   \\
Photon			& $\pm$ 1.1 	& $\pm$ 1.1   \\
Prompt-photon tagger 	& $\pm$ 4.0 	&   -    \\
Leptons			& $\pm$ 0.3 	& $\pm$ 1.3   \\
Jets			& $\pm$ 5.4 	& $\pm$ 2.0   \\
$b$-tagging		& $\pm$ 0.9 	& $\pm$ 0.4   \\
Pile-up			& $\pm$ 2.0 	& $\pm$ 2.3   \\
Luminosity		& $\pm$ 2.3 	& $\pm$ 2.3   \\
MC sample size		& $\pm$ 1.9 	& $\pm$ 1.7   \\
\hline
Total systematic uncertainty & $\pm$ 7.9 & $\pm$ 5.8 \\
Data sample size   & $\pm$ 1.5 & $\pm$ 3.8 \\
\hline
Total uncertainty & $\pm$ 8.1 & $\pm$ 7.0 \\
\bottomrule
\end{tabular}
}
\label{tab:systSummary}
\end{table}
 
\FloatBarrier
\subsection{Normalized differential cross-sections}
 
The normalized differential cross-sections are shown in Figures~\ref{fig:unfolded_sl} and \ref{fig:unfolded_dl}, for the \chljets and \chll channels, respectively. They are compared to the nominal \ttg sample (\textsc{MG5}\_\textsc{aMC}+\PYTHIA~8) and the samples with variations of the \PYTHIA8 A14 tune parameters and the alternative parton shower model of \textsc{MG5}\_\textsc{aMC}+\Herwig7. In addition, a comparison with the nominal \ttbar \textsc{Powheg}+\PYTHIA8 MC sample where prompt-photon radiation is modelled in the parton shower is included. All \ttg samples predict very similar shapes and describe the data well. A small deviation from the prediction is observed in the \chll $\Delta\phi$ distribution, where the leptons in the prediction are more back-to-back than in data. The deviation of data from the prediction is 1.5 standard deviations, based on the $\chi^2$ calculated according to the procedure described in what follows. The \textsc{Powheg}+\PYTHIA8 \ttbar sample gives an improved agreement with data compared to the nominal and varied \ttg samples, although the overall agreement is still poor. It can also be seen from Figures~\ref{fig:unfolded_sl_pt} and \ref{fig:unfolded_dl_pt} that the photons generated by \PYTHIA8 have a softer \pT spectrum than in data.
 
The systematic uncertainties of the unfolded distributions are decomposed into the signal modelling uncertainty, experimental uncertainty, and background modelling uncertainty in both channels. In the \chljets (\chll) channel, the background modelling uncertainty is split into \ttbar (\Zgamma) and the others. These decomposed uncertainties are illustrated in Figures~\ref{fig:sys_unfolded_sl} and \ref{fig:sys_unfolded_dl} for the \chljets and \chll channels, respectively. For the \chljets channel, the systematic uncertainty is dominated by the \ttbar modelling, which is used to model the shapes of the hadronic-fake and electron-fake backgrounds. For the \chll channel, the systematic uncertainty is dominated by the \Zgamma modelling, mostly from the comparison between \SHERPA and \textsc{MG5}\_\textsc{aMC}+\PYTHIA8 generators. Because the unfolding is performed with the distributions before the fit, the background modelling uncertainties are not constrained as in the case of fiducial cross-section measurement where a fit to ELD distribution is performed, and thus they have a much larger impact on the result.
 
The differences between the unfolded and the predicted distributions are quantified by the chi-squared per degree of freedom $\chi^2$/ndf, where the $\chi^2$ is
\begin{equation*}
\chi^2 = (\sigma_{j,\textrm{data}}^{\textrm{norm}} - \sigma_{j,\textrm{pred.}}^{\textrm{norm}})
\cdot C_{jk}^{-1} \cdot (\sigma_{k,\textrm{data}}^{\textrm{norm}} - \sigma_{k,\textrm{pred.}}^{\textrm{norm}})\,,
\end{equation*}
where $\sigma_{\textrm{data}}^{\textrm{norm}}$ and $\sigma_{\textrm{pred.}}^{\textrm{norm}}$ are the unfolded and predicted normalized differential cross-sections, $C_{jk}$ is the covariance matrix of $\sigma_{\textrm{data}}^{\textrm{norm}}$, and $j$ and $k$ are the binning indices of the distribution. For normalized differential cross-sections, the last bin of the above formula is removed from the $\chi^2$ calculation and ndf is reduced by one since this bin gives redundant information. The total correlation matrix is shown in Table~\ref{tab:correlation_dPhi_lep_dilepton}, taking the $\Delta\phi(\ell,\ell)$ in the \chll channel as an example. There is moderate correlation, either positive or negative, between different bins of the unfolded $\Delta\phi(\ell,\ell)$ distribution. The calculated $\chi^2$/ndf values and their corresponding $p$-values are summarized in Tables~\ref{tab:chi2_sl} and \ref{tab:chi2_dl}, quantifying the compatibility between data and each of the predictions.
 
\begin{figure}[!htbp]
\centering
\subfloat[]{
\includegraphics[width=0.45\textwidth]{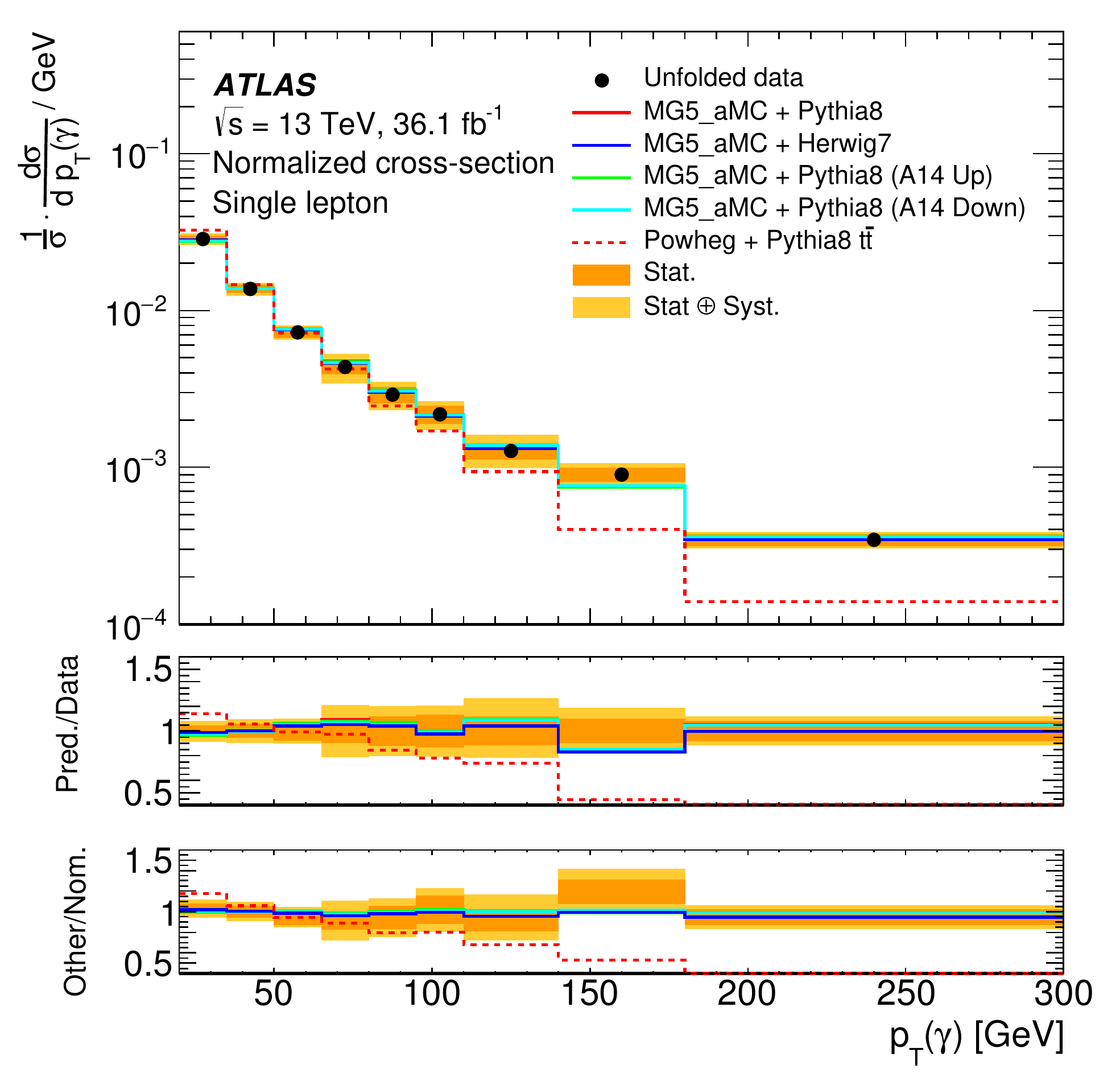}
\label{fig:unfolded_sl_pt}
}
\subfloat[]{
\includegraphics[width=0.45\textwidth]{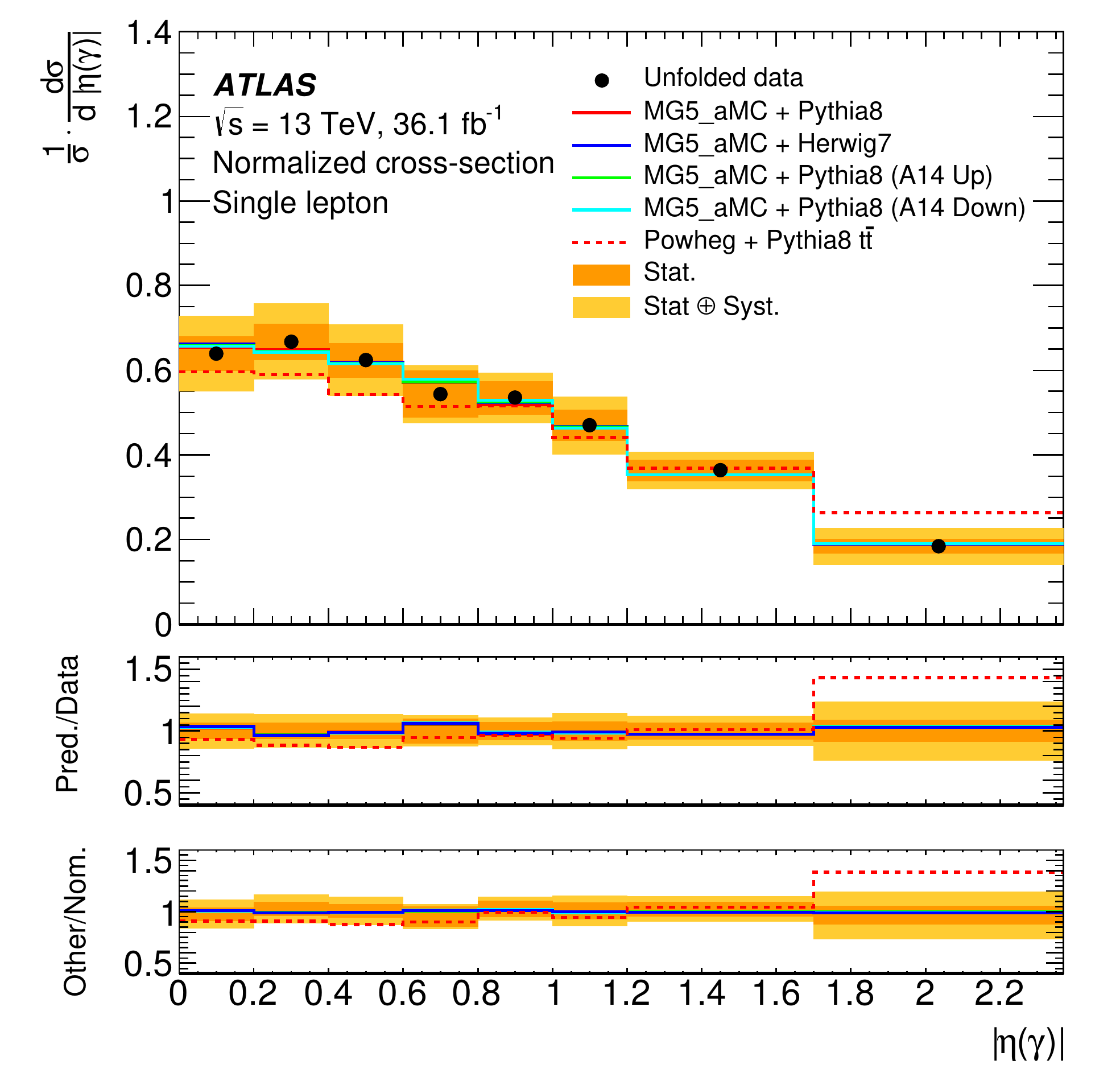}
}
 
\subfloat[]{
\includegraphics[width=0.45\textwidth]{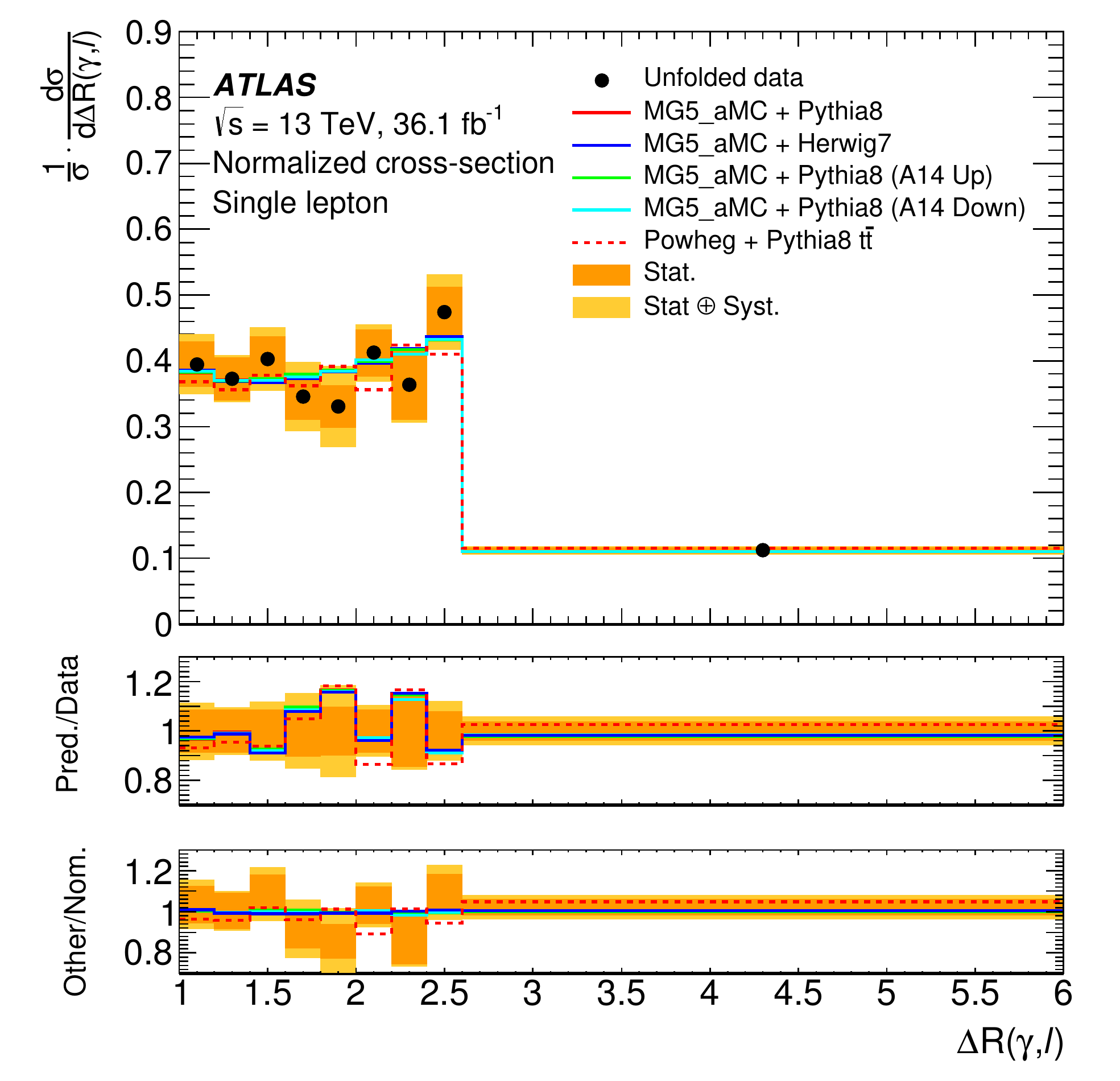}
}
 
\caption{The normalized differential cross-sections as a function of the (a) photon \pt, (b) photon $|\eta|$, and (c) $\Delta R (\gamma,\ell)$ in the \chljets channel. The unfolded distributions are compared to the predictions of the \textsc{MG5\_aMC}+\PYTHIA~8 together with the up and down variations of the \PYTHIA~8 A14 tune parameters, the \textsc{MG5\_aMC}+\Herwig7, and the \textsc{Powheg}+\PYTHIA8 \ttbar where photon radiation is modelled in the parton shower. The top ratio-panel shows the ratios of all the predictions over data. The bottom ratio-panel shows the ratios of the alternative predictions and data over the nominal prediction. Overflows are included in the last bin.}
\label{fig:unfolded_sl}
\end{figure}
 
\begin{figure}[!htbp]
\centering
\subfloat[]{
\includegraphics[width=0.34\textwidth]{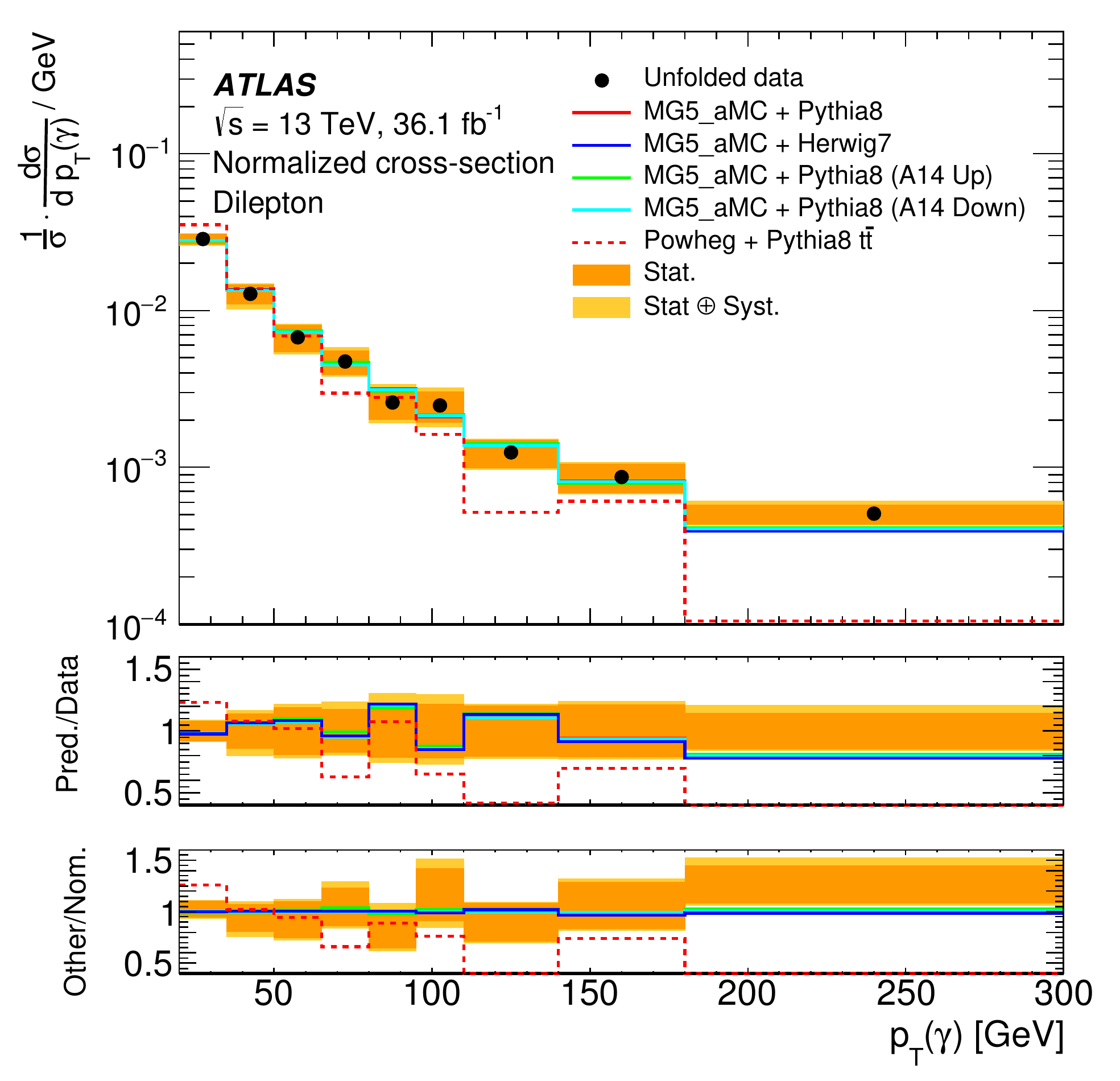}
\label{fig:unfolded_dl_pt}
}
\subfloat[]{
\includegraphics[width=0.34\textwidth]{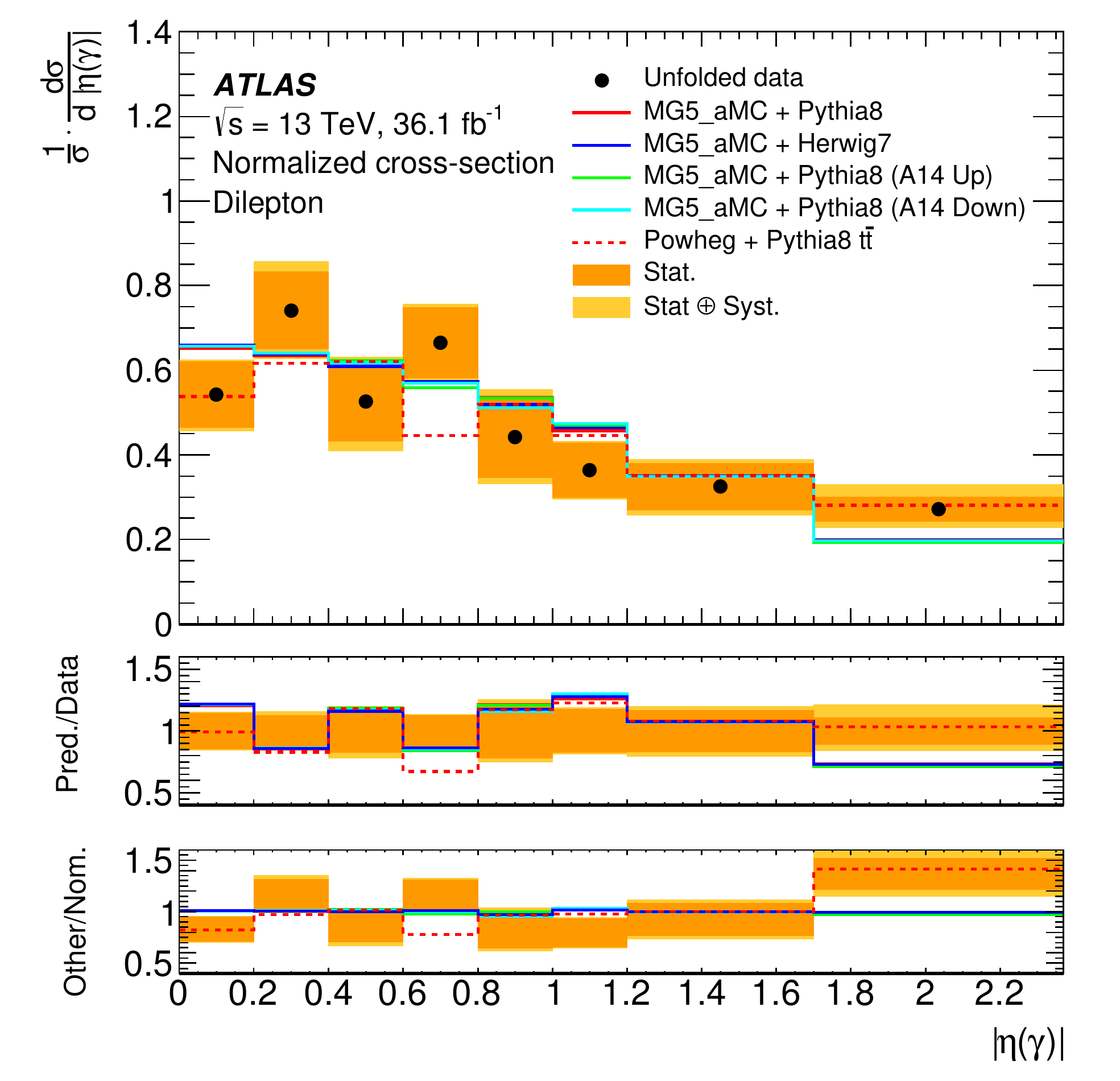}
}
 
\subfloat[]{
\includegraphics[width=0.34\textwidth]{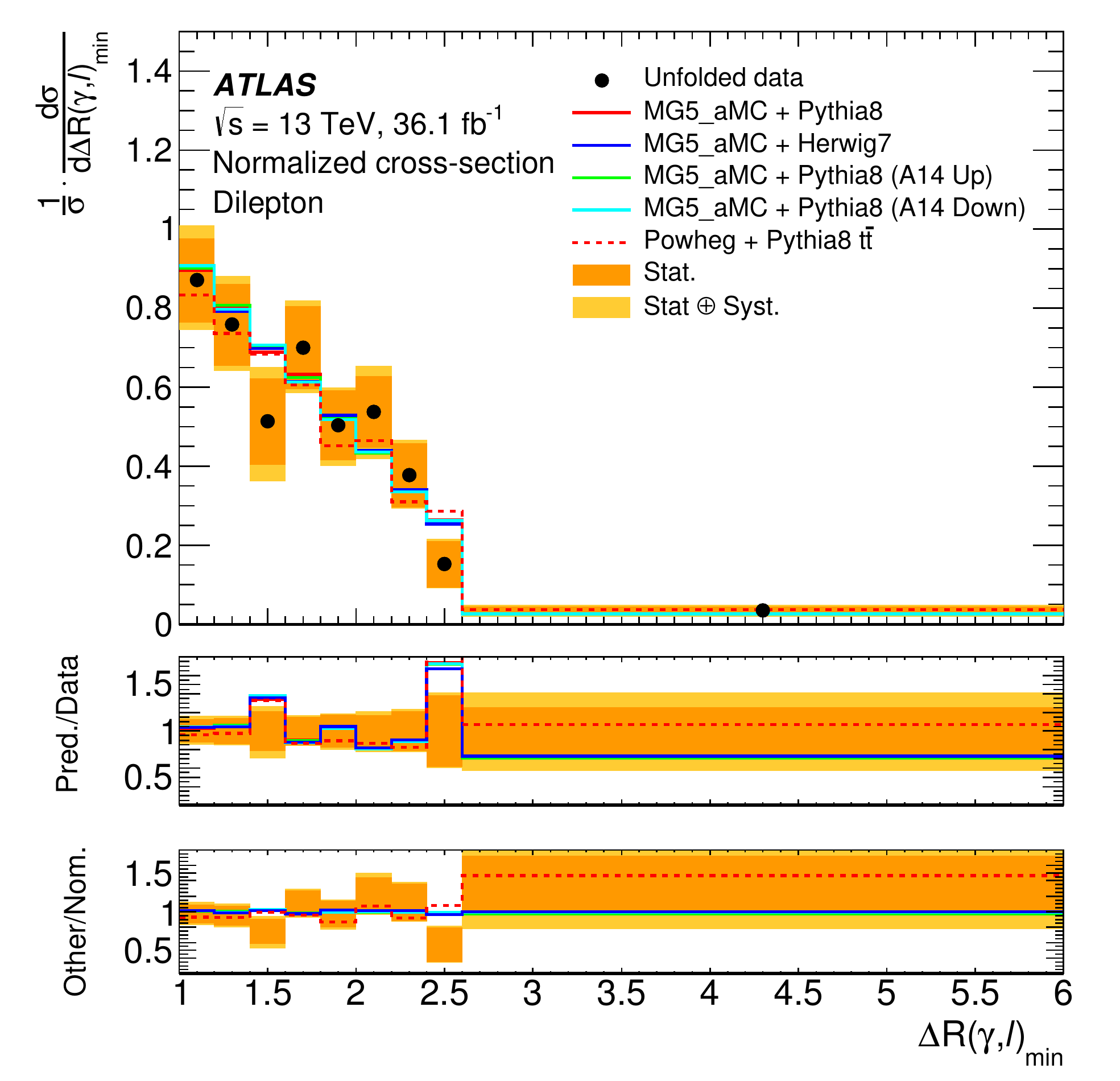}
}
\subfloat[]{
\includegraphics[width=0.34\textwidth]{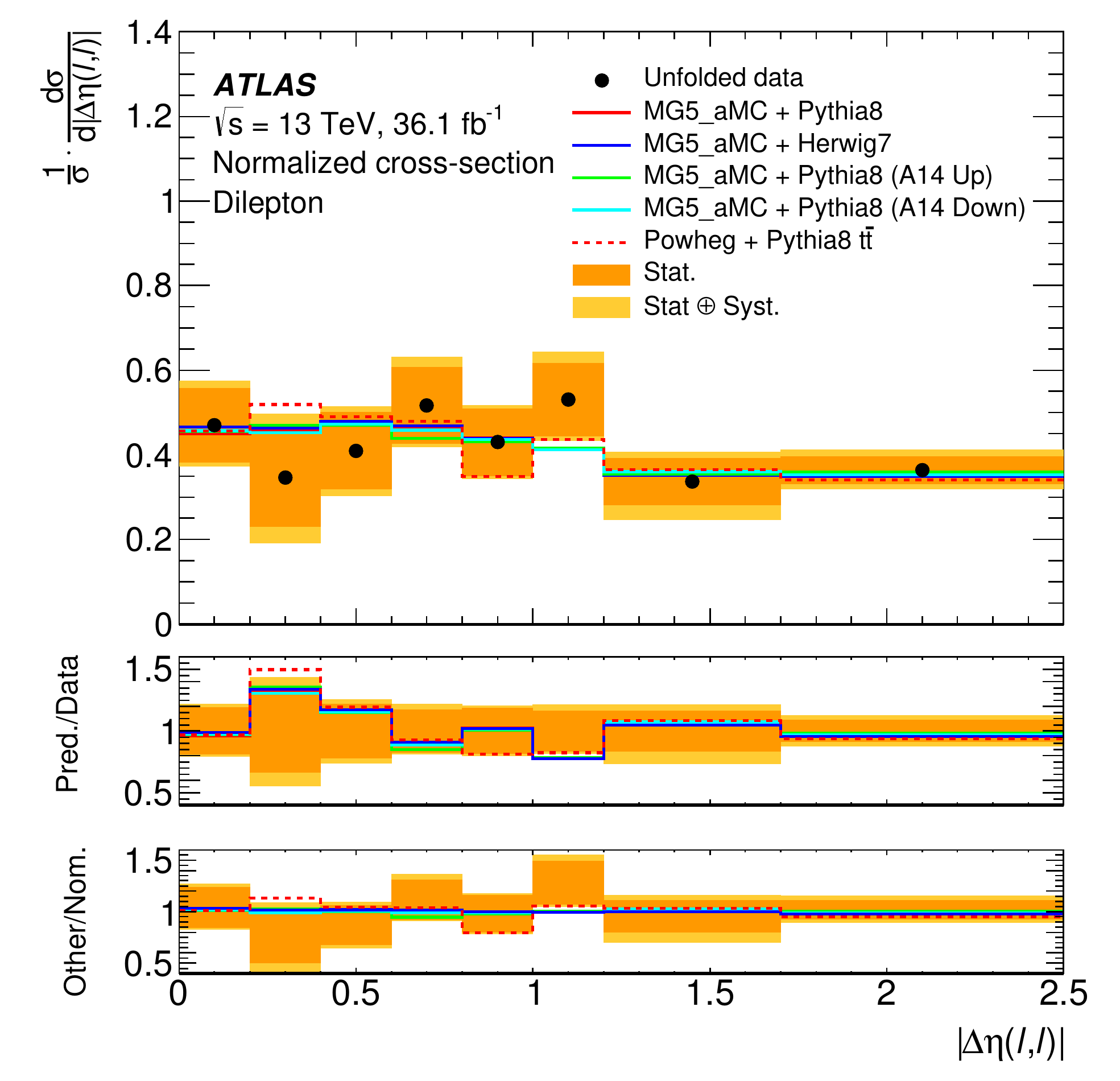}
}
 
\subfloat[]{
\includegraphics[width=0.34\textwidth]{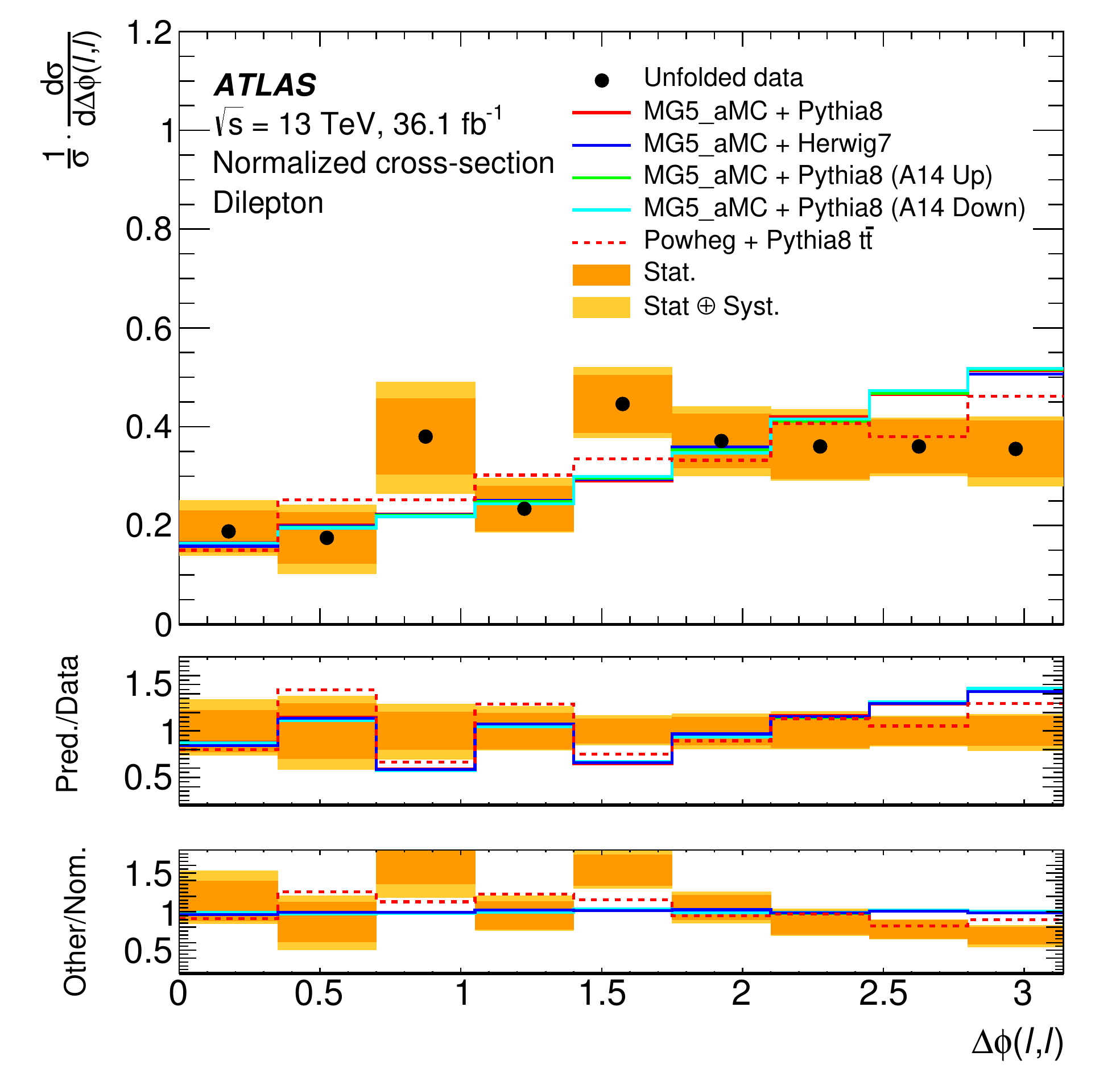}
}
\caption{The normalized differential cross-sections as a function of the (a) photon \pt, (b) photon $|\eta|$, (c) minimum $\Delta R (\gamma,\ell)$, (d) $|\Delta \eta (\ell,\ell)|$, and (e) $\Delta \phi (\ell,\ell)$ in the \chll channel. The unfolded distributions are compared to the predictions of the \textsc{MG5\_aMC}+\PYTHIA8 together with the up and down variations of the \PYTHIA8 A14 tune parameters, the \textsc{MG5\_aMC}+\Herwig7, and the \textsc{Powheg}+\PYTHIA8 \ttbar where photon radiation is modelled in the parton shower. The top ratio-panel shows the ratios of all the predictions over data. The bottom ratio-panel shows the ratios of the alternative predictions and data over the nominal prediction. Overflows are included in the last bin.}
\label{fig:unfolded_dl}
\end{figure}
\FloatBarrier
 
\begin{figure}[!htbp]
\centering
\subfloat[]{
\includegraphics[width=0.45\textwidth]{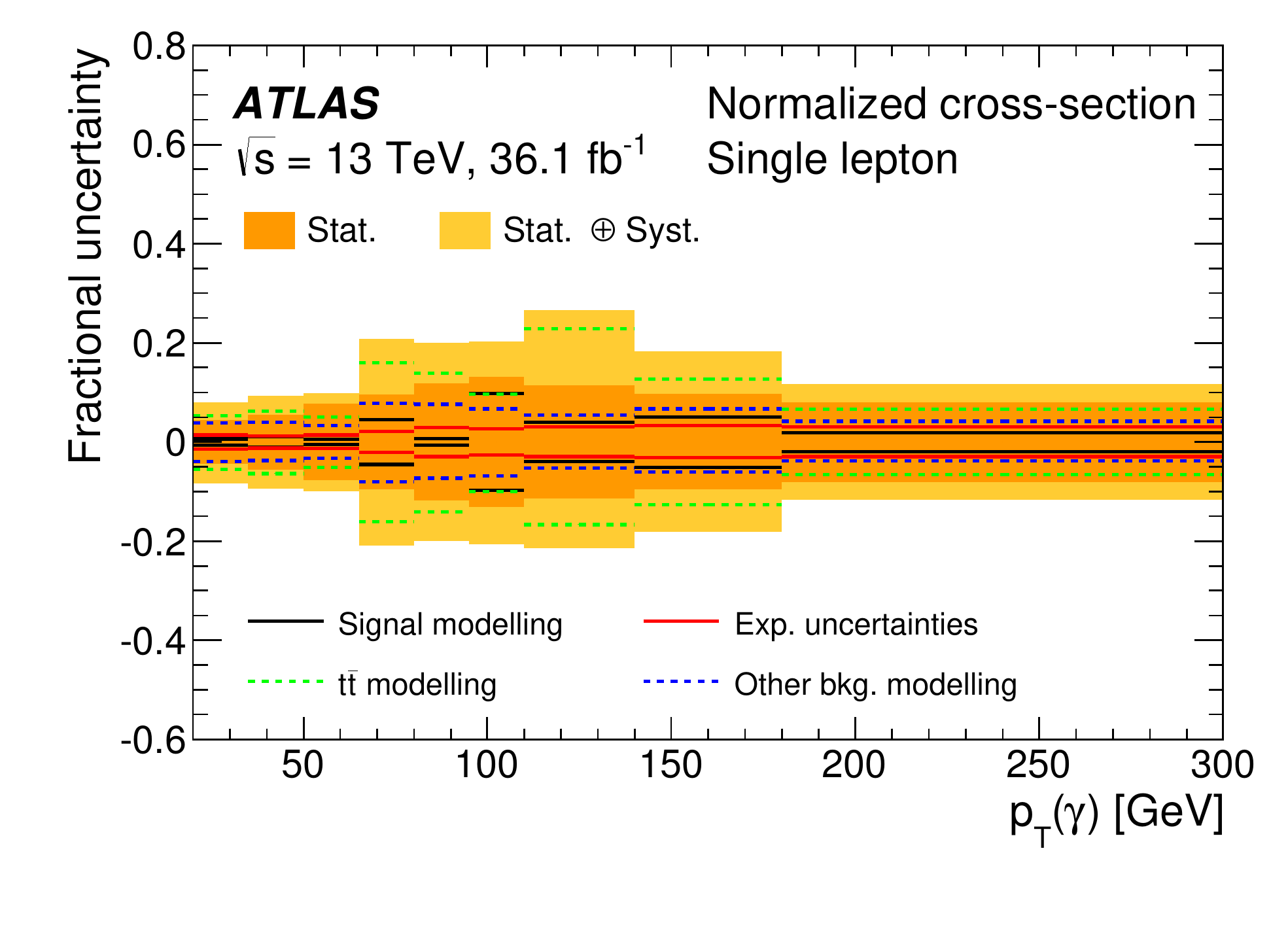}
}
\subfloat[]{
\includegraphics[width=0.45\textwidth]{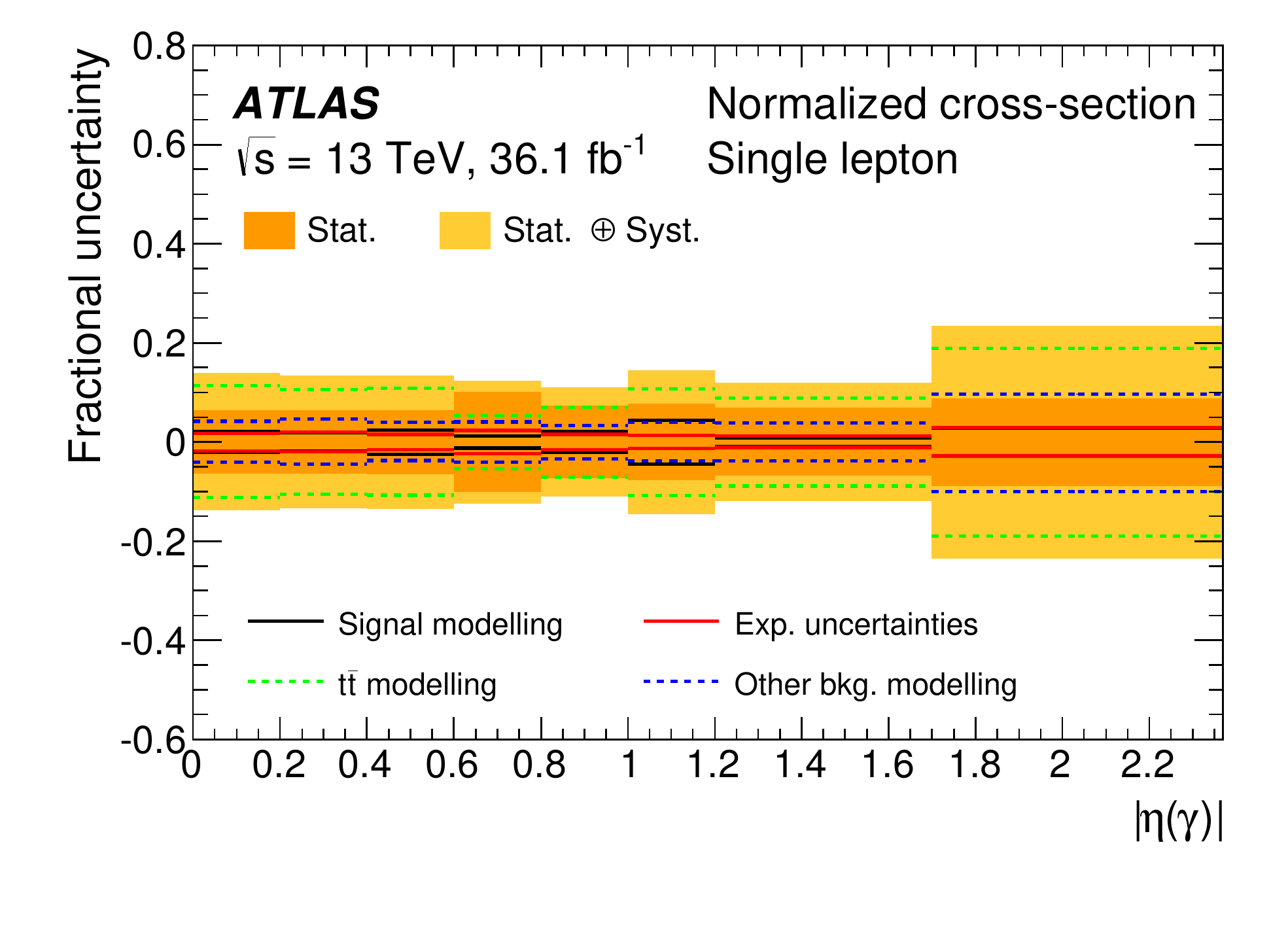}
}
 
\subfloat[]{
\includegraphics[width=0.45\textwidth]{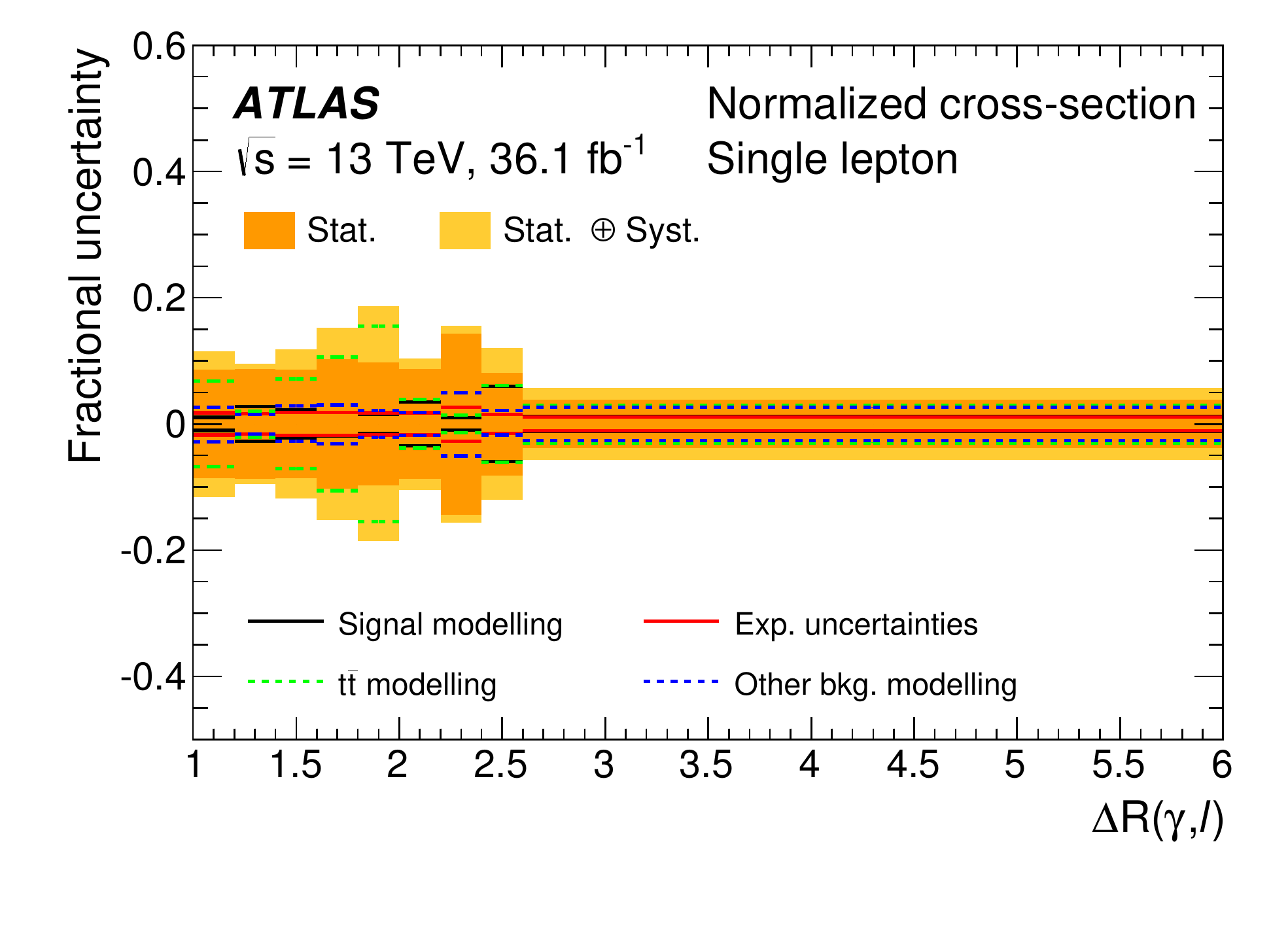}
}
 
\caption{The decomposed systematic uncertainties for the normalized differential cross-sections as a function of the (a) photon \pt, (b) photon $|\eta|$, and (c) $\Delta R (\gamma,\ell)$ in the \chljets channel.}
\label{fig:sys_unfolded_sl}
\end{figure}
 
\begin{figure}[!htbp]
\centering
\subfloat[]{
\includegraphics[width=0.45\textwidth]{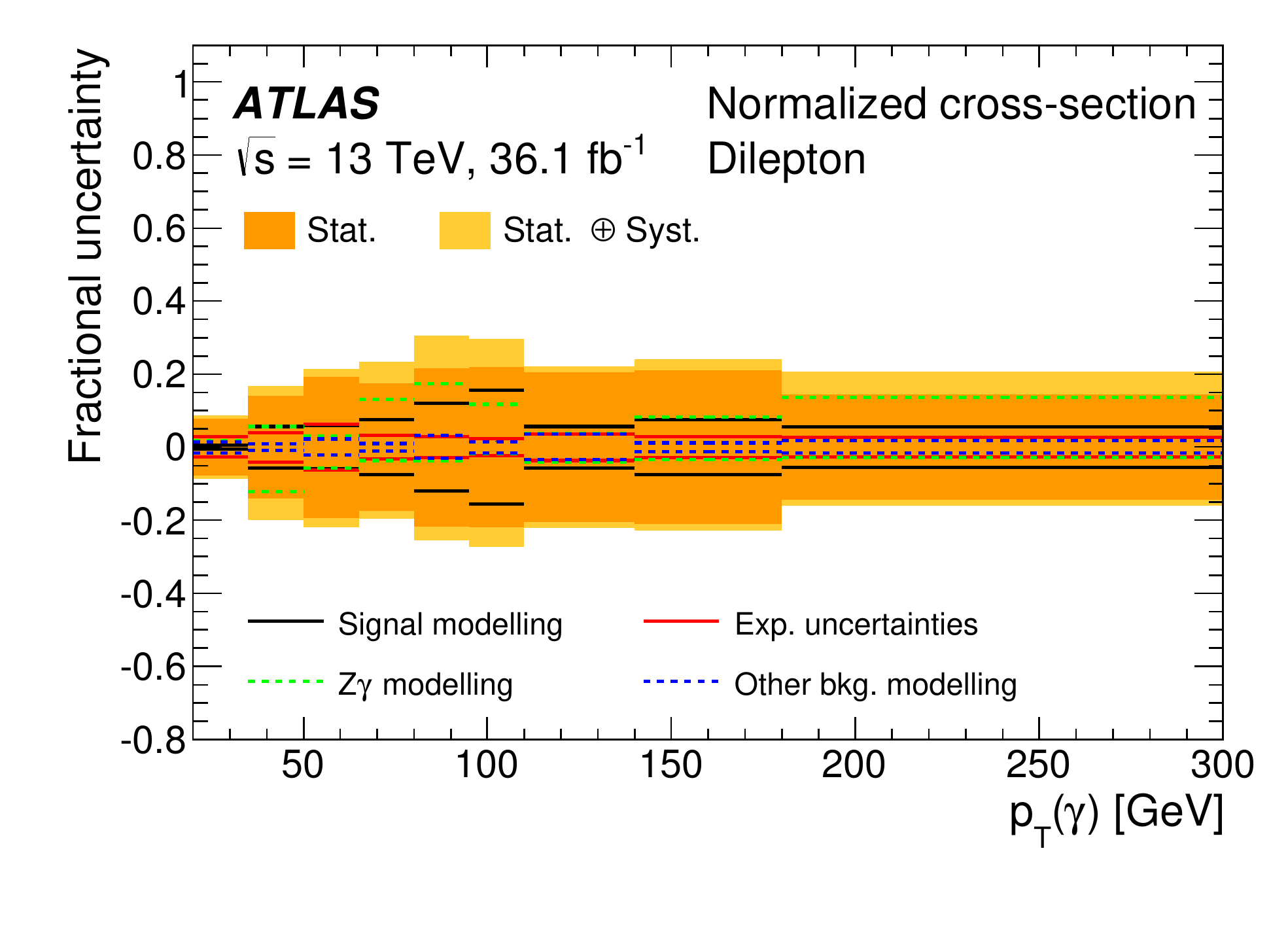}
}
\subfloat[]{
\includegraphics[width=0.45\textwidth]{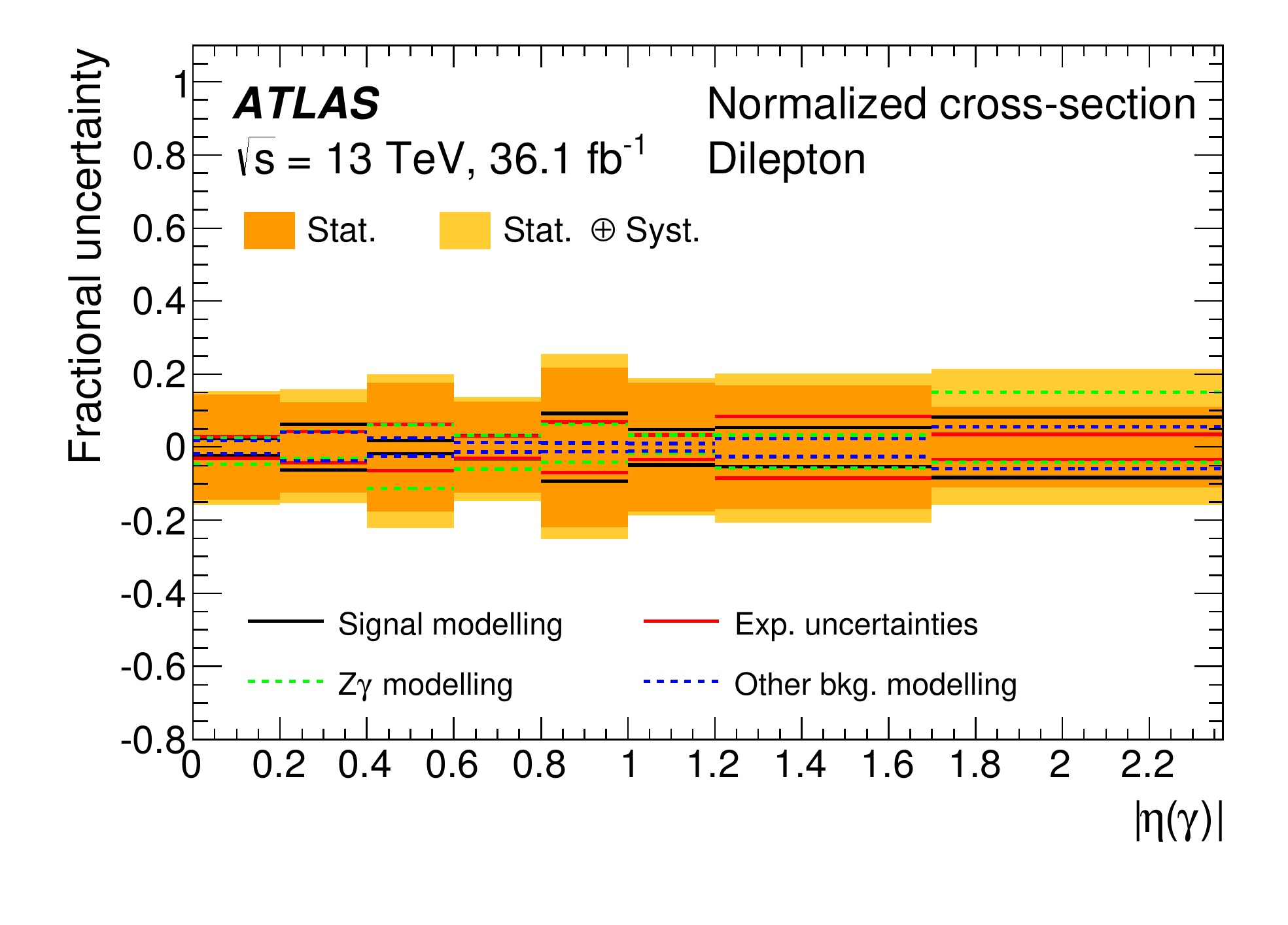}
}
 
\subfloat[]{
\includegraphics[width=0.45\textwidth]{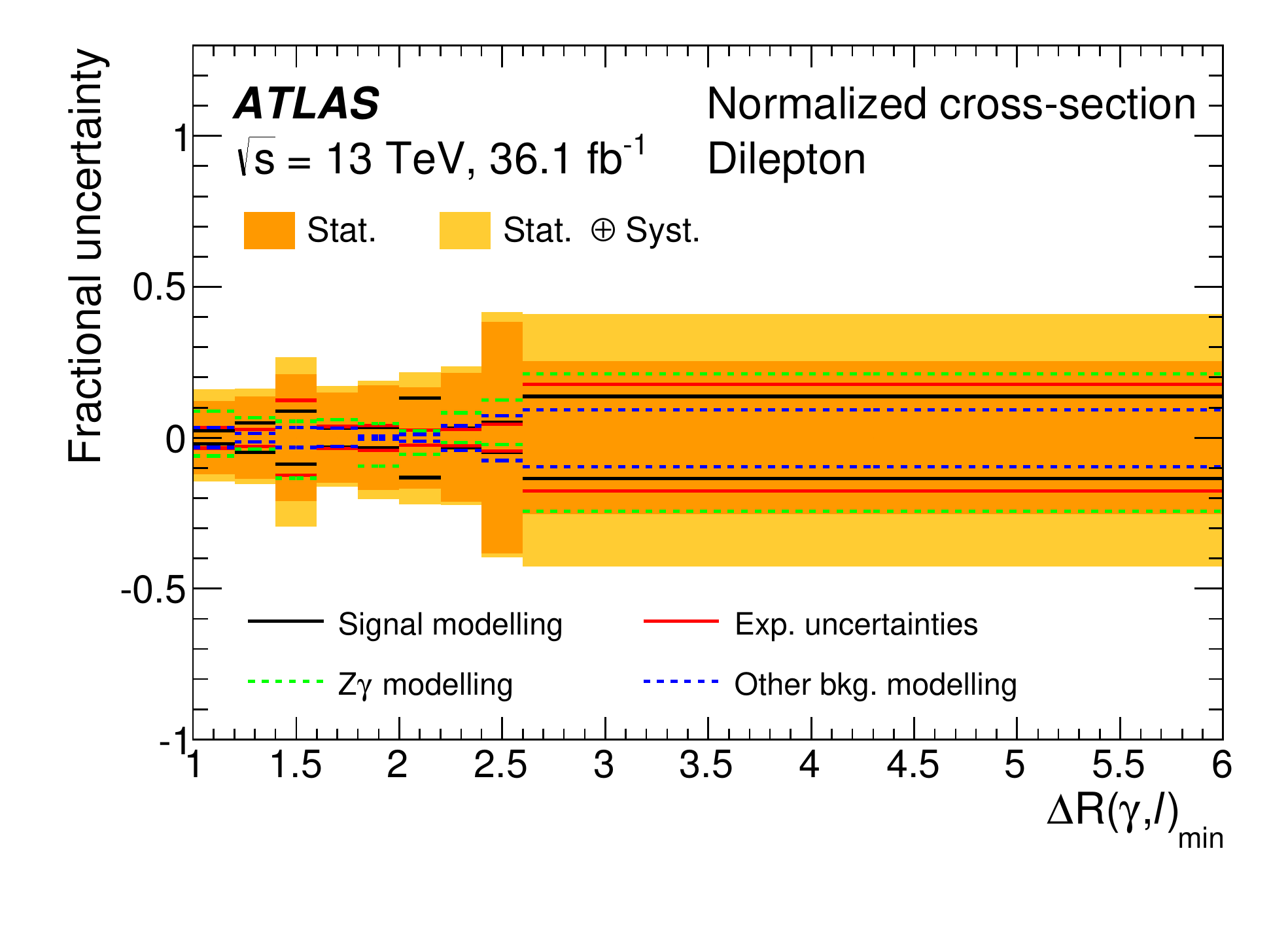}
}
\subfloat[]{
\includegraphics[width=0.45\textwidth]{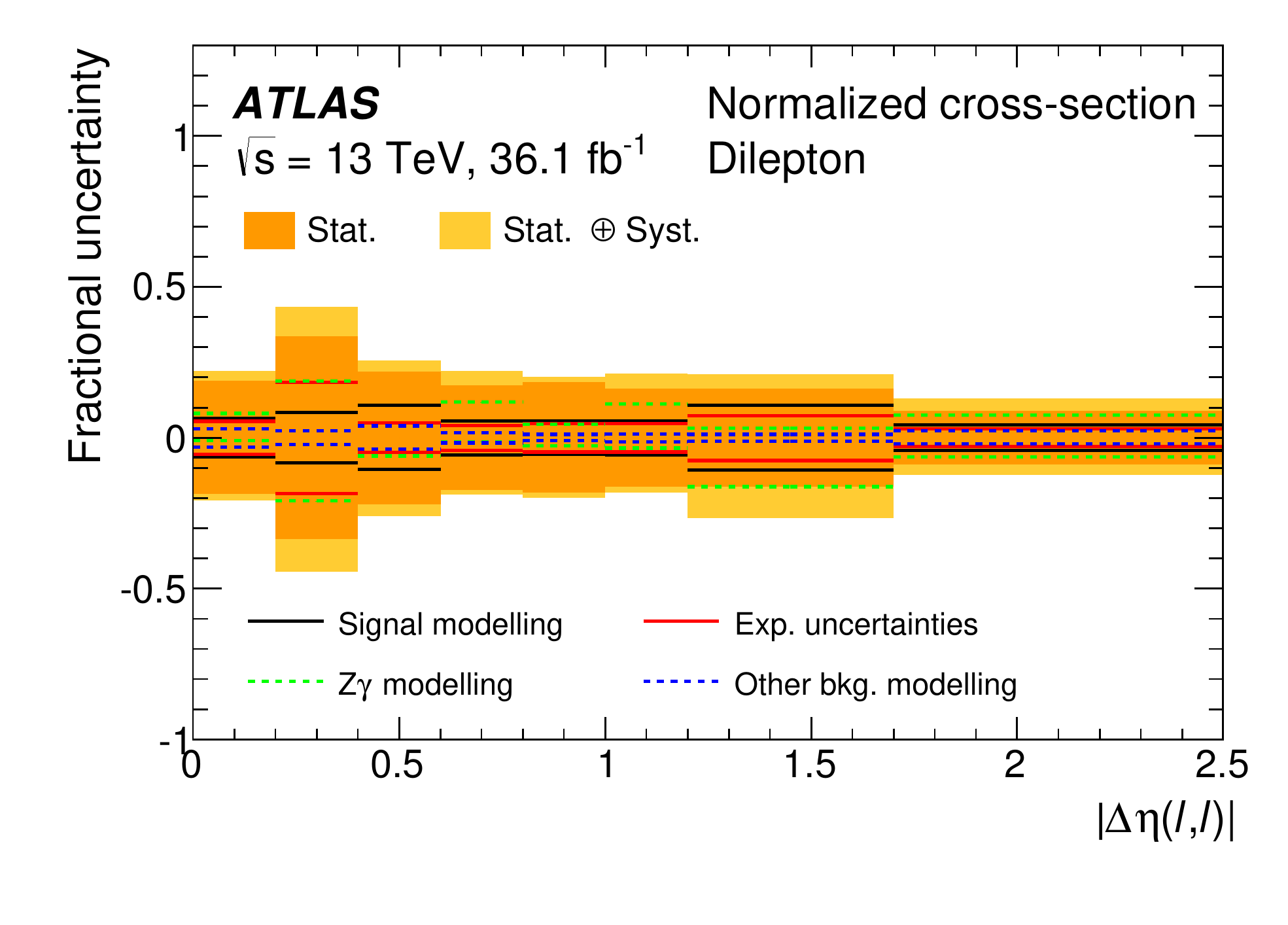}
}
 
\subfloat[]{
\includegraphics[width=0.45\textwidth]{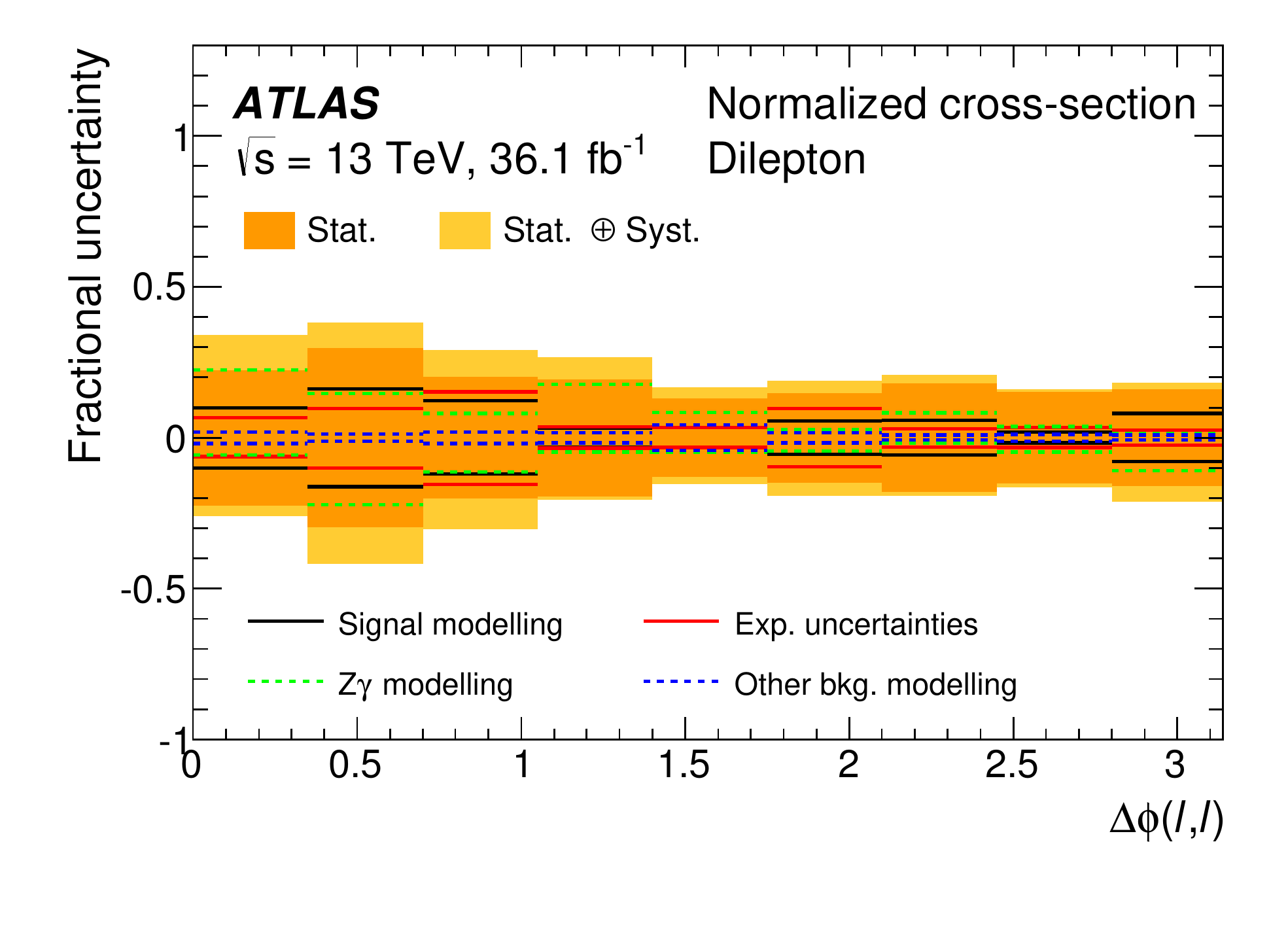}
}
 
\caption{The decomposed systematic uncertainties for the normalized differential cross-sections as a function of the (a) photon \pt, (b) photon $|\eta|$, (c) minimum $\Delta R (\gamma,\ell)$, (d) $|\Delta \eta (\ell,\ell)|$, and (e) $\Delta \phi (\ell,\ell)$ in the \chll channel.}
\label{fig:sys_unfolded_dl}
\end{figure}
 
\begin{table}
\centering
\caption{The correlation matrix for the normalzed differential cross-section as a function of $\Delta\phi(\ell,\ell)$ in the \chll channel, accounting for the statistical and systematic uncertainties.}
\scalebox{0.7}{
\begin{tabular}{ c | r r r r r r r r r}
\toprule
Bin & 0.0 - 0.35 & 0.35 - 0.7 & 0.7 - 1.05 & 1.05 - 1.4 & 1.4 - 1.75 & 1.75 - 2.1 & 2.1 - 2.45 & 2.45 - 2.8 & 2.8 - 3.14\\
\hline
0.0 - 0.35 & 1.00 & $-0.07$ & 0.05 & $-0.36$ & $-0.18$ & $-0.08$ & $-0.19$ & $-0.13$ & $-0.24$\\
0.35 - 0.7 & & 1.00 & 0.29 & 0.21 & 0.09 & 0.11 & 0.21 & 0.02 & 0.32\\
0.7 - 1.05 & & & 1.00 & 0.12 & 0.05 & 0.10 & 0.14 & 0.05 & 0.23\\
1.05 - 1.4 & & & & 1.00 & 0.26 & 0.18 & 0.29 & 0.20 & 0.41\\
1.4 - 1.75 & & & & & 1.00 & 0.15 & 0.16 & 0.14 & 0.21\\
1.75 - 2.1 & & & & & & 1.00 & 0.17 & 0.15 & 0.23\\
2.1 - 2.45 & & & & & & & 1.00 & 0.12 & 0.33\\
2.45 - 2.8 & & & & & & & & 1.00 & 0.18\\
2.8 - 3.14 & & & & & & & & & 1.00\\
\bottomrule
\end{tabular}
\label{tab:correlation_dPhi_lep_dilepton}
}
\end{table}
 
\begin{table}
\centering
\caption{$\chi^2$/ndf values and $p$-values between the measured normalized differential cross-sections and predictions from several generators in the \chljets channel.}
\scalebox{0.7}{
\begin{tabular}{l | c c  | c c  | c c }
\toprule
& \multicolumn{2}{c}{$p_{T}(\gamma)$}& \multicolumn{2}{c}{$|\eta(\gamma)|$}& \multicolumn{2}{c}{$\Delta R(\gamma,\ell)$}\\
Predictions & $\chi^2$/ndf & $p$-value & $\chi^2$/ndf & $p$-value & $\chi^2$/ndf & $p$-value \\
\midrule
\MGaMC + \textsc{Pythia}8		  & 3.2/8 &  $ \enspace 0.92$ & 0.7/7 & 1.0 & 5.0/8 & 0.76\\
\MGaMC + \textsc{Herwig}7	     	  & 2.3/8 &  $ \enspace 0.97$ & 0.9/7 & 1.0 & 4.8/8 & 0.78\\
\MGaMC + \textsc{Pythia}8 (A14 Up)   	  & 3.3/8 &  $ \enspace 0.91$ & 0.8/7 & 1.0 & 4.9/8 & 0.77\\
\MGaMC + \textsc{Pythia}8 (A14 Down)	  & 2.6/8 & $ \enspace 0.96 $ & 0.9/7 & 1.0 & 4.6/8 & 0.80\\
\textsc{POWHEG} + \textsc{Pythia}8 \ttbar& 25.4/8 & <0.01 & 2.8/7 & 0.9 & 8.7/8 & 0.37\\
\bottomrule
\end{tabular}}
\label{tab:chi2_sl}
\end{table}
 
\begin{table}
\centering
\caption{$\chi^2$/ndf values and $p$-values between the measured normalized differential cross-sections and predictions from several generators in the \chll channel.}
\scalebox{0.7}{
\begin{tabular}{l | r c  | r c  | r c  | r c  | r c }
\toprule
& \multicolumn{2}{c}{$p_{T}(\gamma)$}& \multicolumn{2}{c}{$\eta(\gamma)$}& \multicolumn{2}{c}{$\Delta R(\gamma,\ell)$}& \multicolumn{2}{c}{$|\Delta\eta(\ell,\ell)|$}& \multicolumn{2}{c}{$\Delta\phi(\ell,\ell)$}\\
Predictions & $\chi^2$/ndf & $p$-value & $\chi^2$/ndf & $p$-value & $\chi^2$/ndf & $p$-value & $\chi^2$/ndf & $p$-value & $\chi^2$/ndf & $p$-value \\
\midrule
\MGaMC + \textsc{Pythia}8		  & 1.7/8 & 0.99 & 7.4/7 & 0.39 & 6.9/8 & 0.55 & 3.0/7 & 0.89 & 14.4/8 & 0.07\\
\MGaMC + \textsc{Herwig}7 		  & 2.0/8 & 0.98 & 7.4/7 & 0.39 & 6.6/8 & 0.58 & 3.1/7 & 0.88 & 14.4/8 & 0.07\\
\MGaMC + \textsc{Pythia}8 (A14 Up)	  & 1.6/8 & 0.99 & 8.4/7 & 0.30 & 7.4/8 & 0.49 & 3.4/7 & 0.85 & 14.0/8 & 0.08\\
\MGaMC + \textsc{Pythia}8 (A14 Down)	  & 1.6/8 & 0.99 & 7.9/7 & 0.34 & 7.5/8 & 0.48 & 3.2/7 & 0.87 & 14.4/8 & 0.07\\
\textsc{POWHEG} + \textsc{Pythia}8 \ttbar & 20.1/8 & 0.01 & 10.8/7 & 0.15 & 8.6/8 & 0.38 & 4.5/7 & 0.72 & 9.8/8 & 0.28\\
\bottomrule
\end{tabular}}
\label{tab:chi2_dl}
\end{table}
 
\FloatBarrier
\section{Conclusions}
\label{sec:conclusion}
 
Fiducial cross-sections of top-quark pair production in association with a photon are measured in the \chljets and \chll decay channels of the top-quark pair using 36.1~\ifb of 13~\TeV~$pp$ collision data collected in 2015 and 2016 by the ATLAS detector at the LHC. The normalized differential cross-sections are measured as a function of the photon \pT and $|\eta|$, and the $\Delta R$ between the photon and the closest lepton for both channels, and the $|\Delta\eta|$ and $\Delta\phi$ between the two leptons for the \chll channel.
 
In both channels, the measured fiducial cross-sections agree well with the NLO SM predictions within uncertainties. The measured normalized differential cross-sections also agree well with the LO \ttg prediction and the NLO \ttbar prediction, where the photon comes from the parton shower. The largest disagreement between data and LO \ttg prediction is observed in the distribution of the azimuthal opening angle between the two leptons in the \chll channel, which is sensitive to \ttbar spin correlation, while the NLO \ttbar sample provides an improved agreement with data in this variable.
 
\clearpage
\appendix
\part*{Appendix}
\addcontentsline{toc}{part}{Appendix}
 
The definitions of the discriminating variables used in ATLAS photon identification are given in Table~\ref{tab:showershapes}.
 
\begin{table}[htbp]
\centering
\caption{Definitions of photon discriminating variables.}
\begin{tabular}{c|p{13cm}}
\toprule
\textbf{Name} & \textbf{Description} \\
\hline
\multicolumn{2}{l}{\textbf{Hadronic leakage}}\\
\hline
$R_\mathrm{had}$ or $R_\mathrm{had_{1}}$ & Transverse energy leakage in the hadronic calorimeter normalized to transverse energy of the photon candidate in the ECAL. In the region $0.8\leq |\eta|\leq 1.37$, the entire energy of the photon candidate in the HCAL is used ($R_\mathrm{had}$), while in the region $|\eta| < 0.8$ and $|\eta| > 1.37$ the energy of the first layer of the HCAL is used ($R_\mathrm{had1}$).\\
\hline
\multicolumn{2}{l}{\textbf{Energy ratios and width in the second layer of ECAL}}\\
\hline
$R_\eta$ & Energy ratio of $3\times 7$ to $7\times 7$ cells in the $\eta\times\phi$ plane.\\
$R_\phi$ & Energy ratio of $3\times 3$ to $3\times 7$ cells in the $\eta\times\phi$ plane.\\
$w_{\eta_{2}}$ & Lateral width of the shower, using a window of $\eta\times\phi = 3\times 5$ cells.
\\
 
\hline
\multicolumn{2}{l}{\textbf{Energy ratios and widths in the first (strip) layer of ECAL}}\\
\hline
$w_{s3}$ &
Shower width along $\eta$, using 3 strips around the largest energy deposit.\\
$w_{s\,\mathrm{tot}}$ &
Shower width along $\eta$, using $20\times 2$ strip cells in the $\eta\times\phi$ plane.\\
$F_\mathrm{side}$ &
Energy outside the 3 central strips but within 7 strips, normalized to the energy within the 3 central strips.\\
$E_\mathrm{ratio}$ &
Ratio between difference of the first and second energy maximum divided by their sum ($E_\mathrm{ratio}=1$ if there is no second maximum).\\
$\Delta E$ & Difference between the second energy maximum and the minimum found between first and second maximum ($\Delta E = 0$ if there is no second maximum).\\
 
\bottomrule
\end{tabular}
\label{tab:showershapes}
\end{table}
 
The shapes of the PPT of the prompt, hadronic-fake, and electron-fake photons are compared in Figure~\ref{fig:pptoutput}.
 
\begin{figure}[!htbp]
\centering
\includegraphics[width=0.45\linewidth]{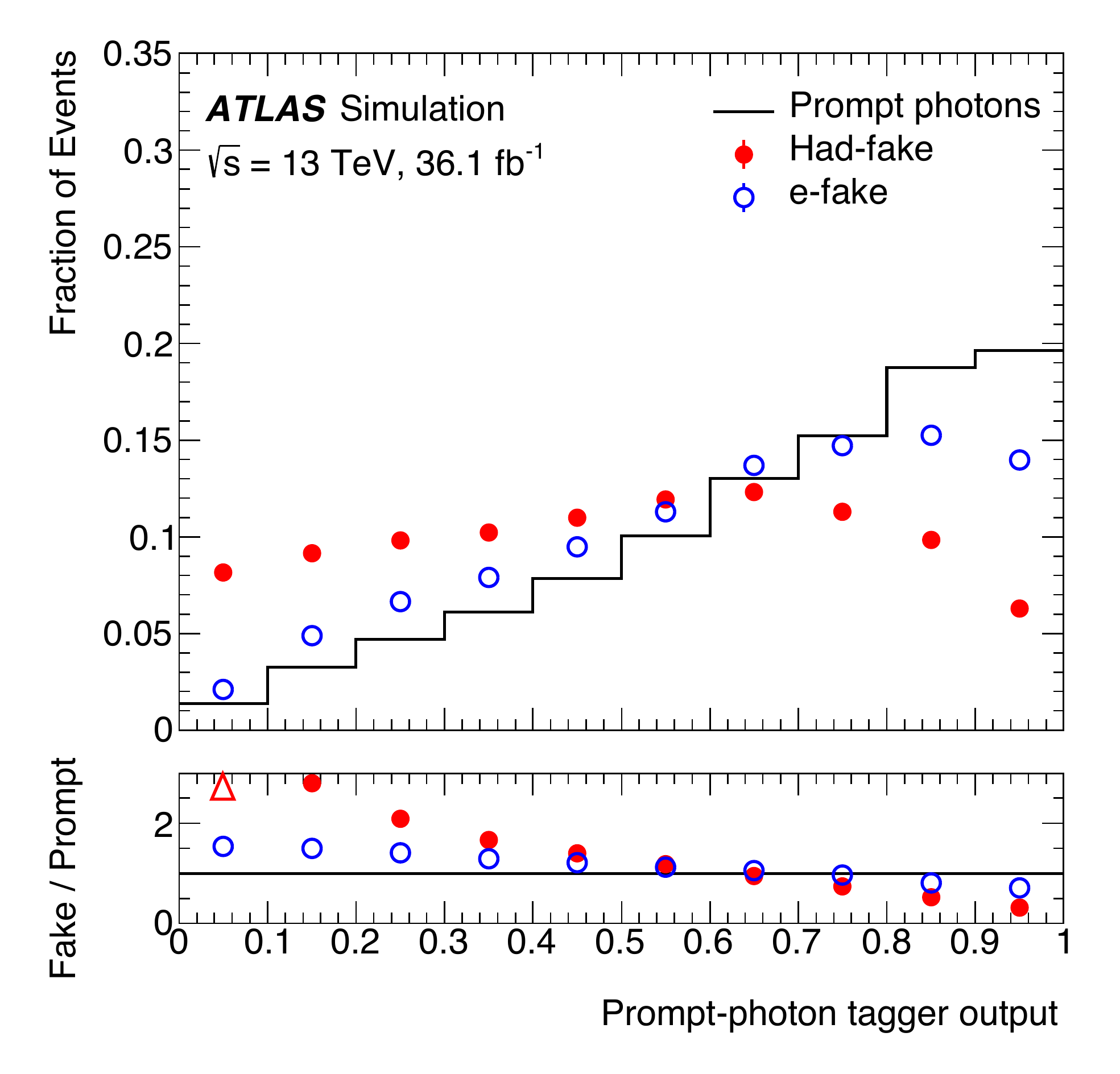}
\caption{The simulated shapes of the output of the prompt-photon tagger for the prompt, hadronic-fake, and electron-fake photons after applying Tight identification and isolation criteria.}
\label{fig:pptoutput}
\end{figure}
 
\FloatBarrier
 
Figure~\ref{fig:SLELDvars} shows the post-fit distributions of four important variables for the training of the ELD in the \chljets channel.
 
\begin{figure}[!htbp]
\centering
\subfloat[]{
\includegraphics[width=0.45\linewidth]{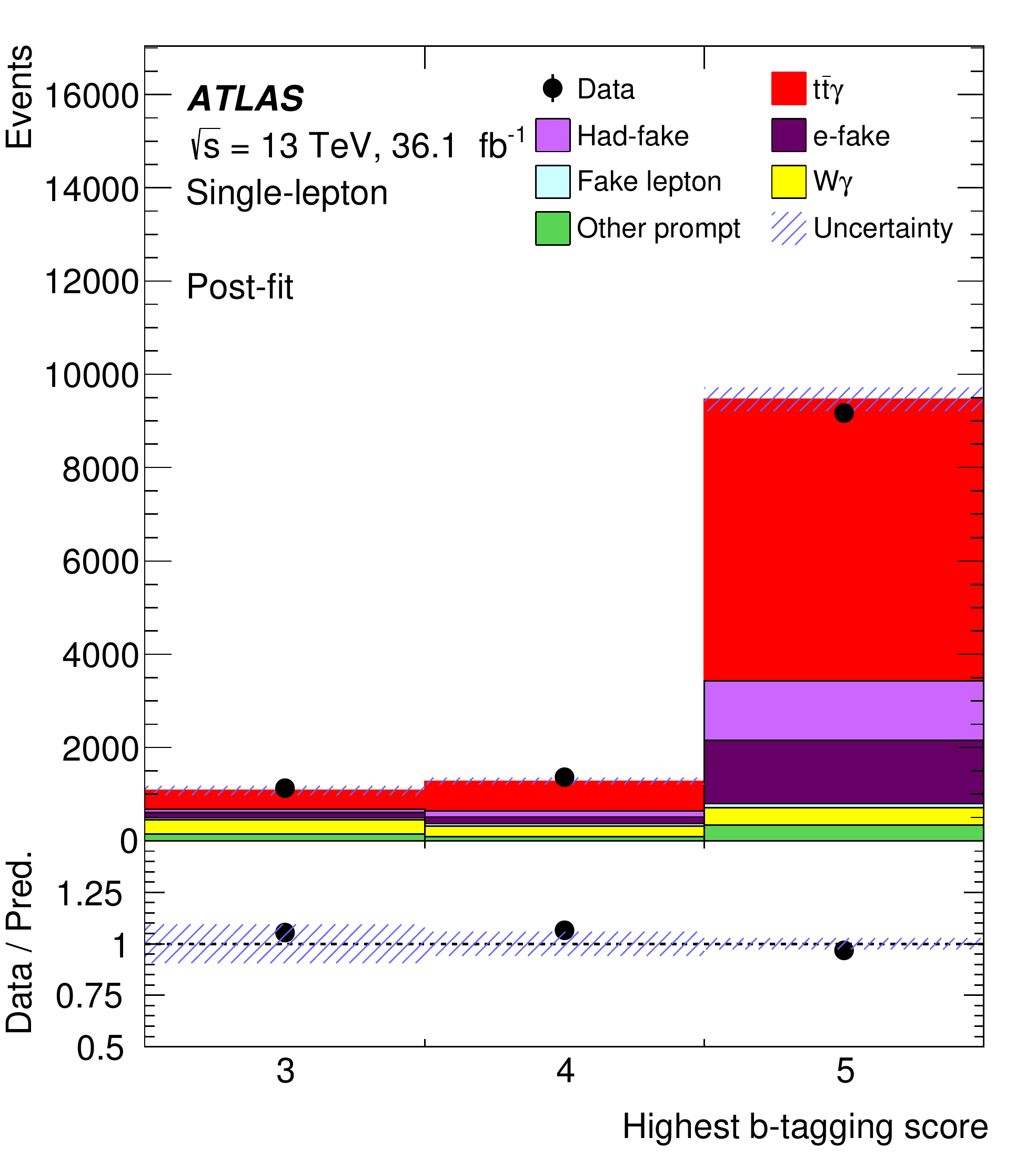}
}
\subfloat[]{
\includegraphics[width=0.45\linewidth]{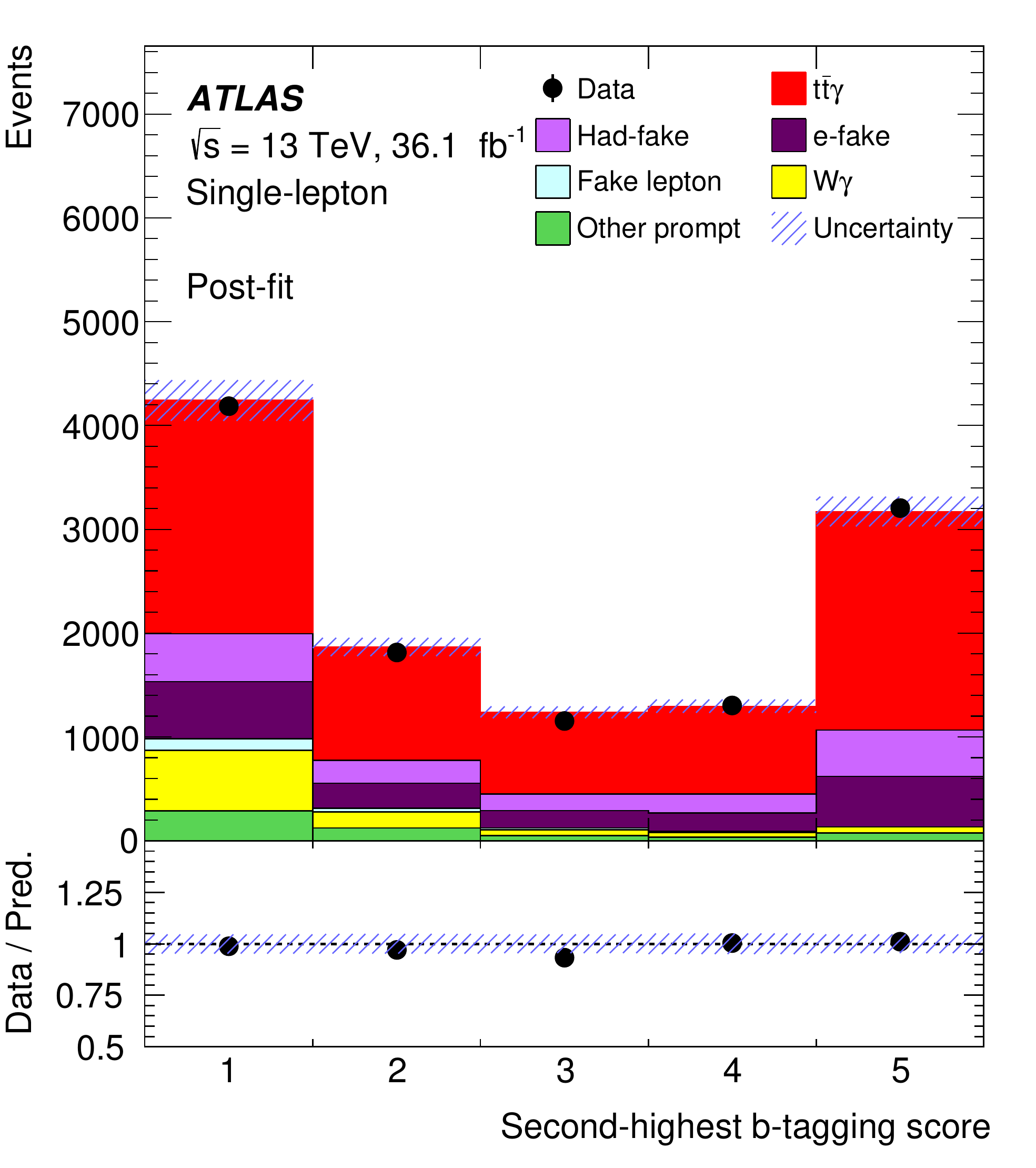}
}
 
\subfloat[]{
\includegraphics[width=0.45\linewidth]{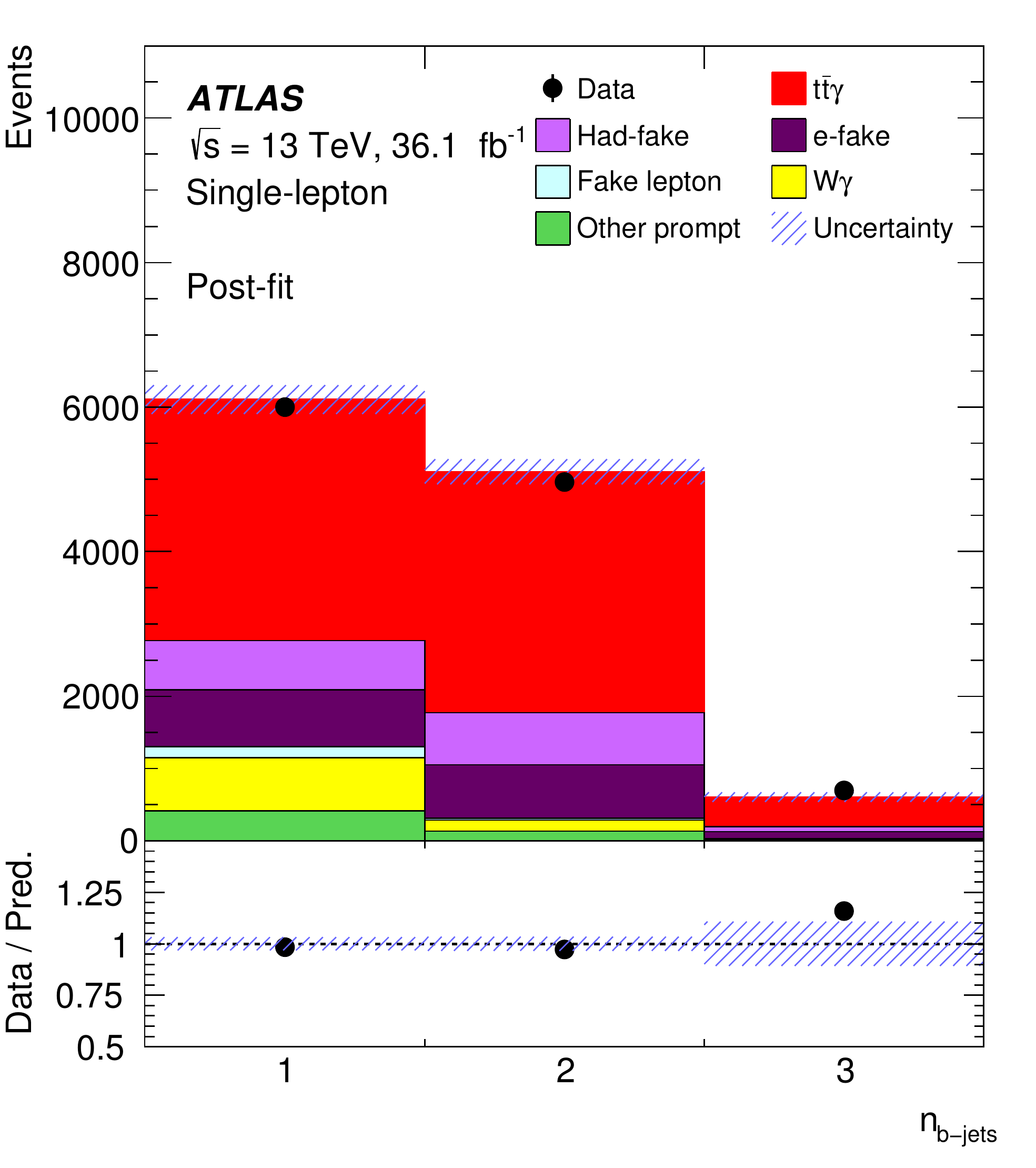}
}
\subfloat[]{
\includegraphics[width=0.45\linewidth]{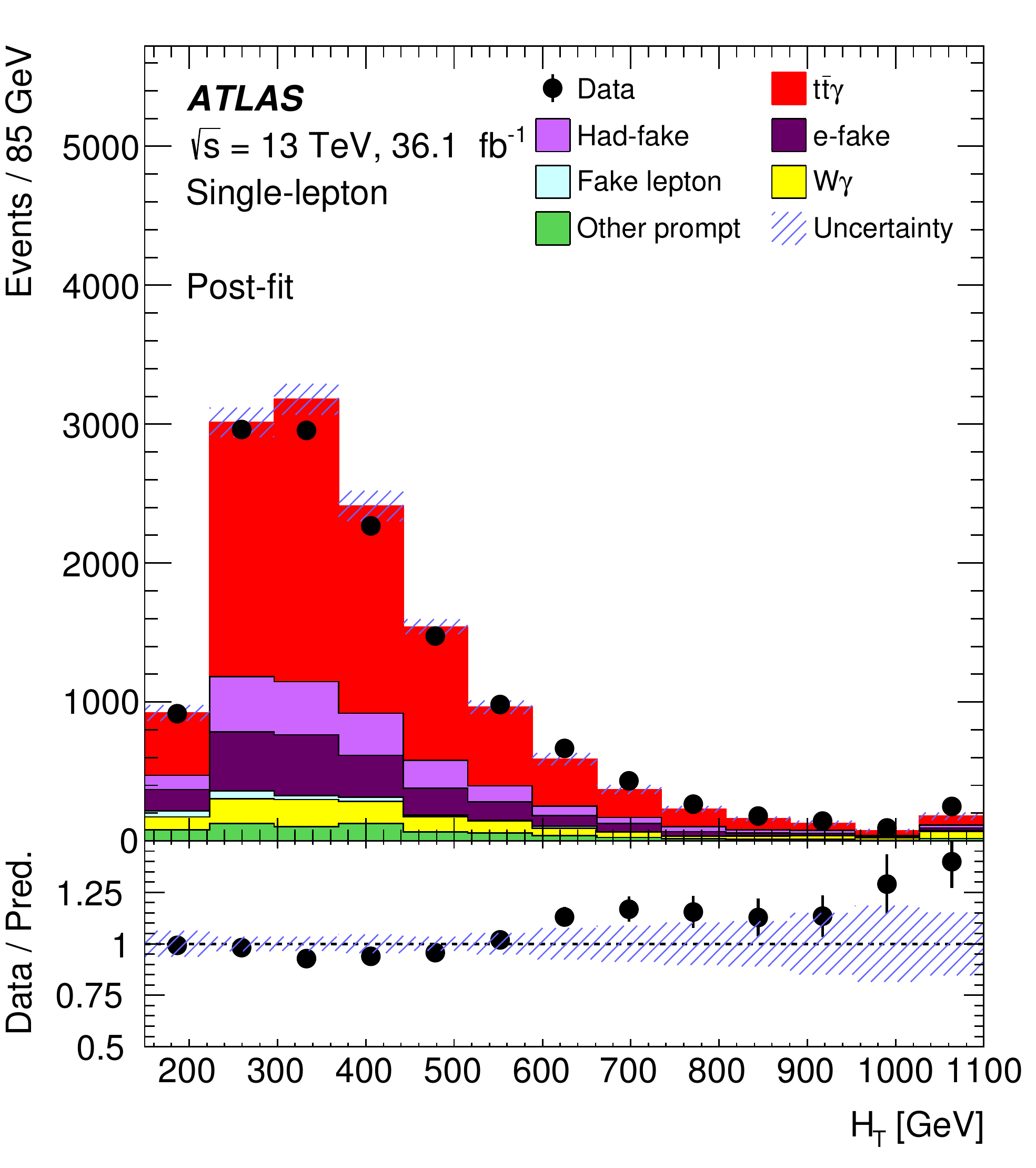}
}
\caption [] {Distributions of the (a) highest and (b) second highest $b$-tagging scores for all jets, (c) total number of $b$-tagged jets (at a 77\% efficiency), and (d) scalar sum of the transverse energies of all the selected physical objects in the \chljets channel after event selection and likelihood fit. Each bin of (a) and (b), from one to five, represents the $b$-tagging efficiencies of $>$ 85\%, 77--85\%, 70--77\%, 60--70\%, and $<$ 60\%. All data-driven corrections and systematic uncertainties are included.}
\label{fig:SLELDvars}
\end{figure}
 
\FloatBarrier
 
Figure~\ref{fig:DLELDvars} shows the post-fit distributions of four important variables for the training of the ELD in the \chll channel.
 
\begin{figure}[!htbp]
\centering
\subfloat[]{
\includegraphics[width=0.45\linewidth]{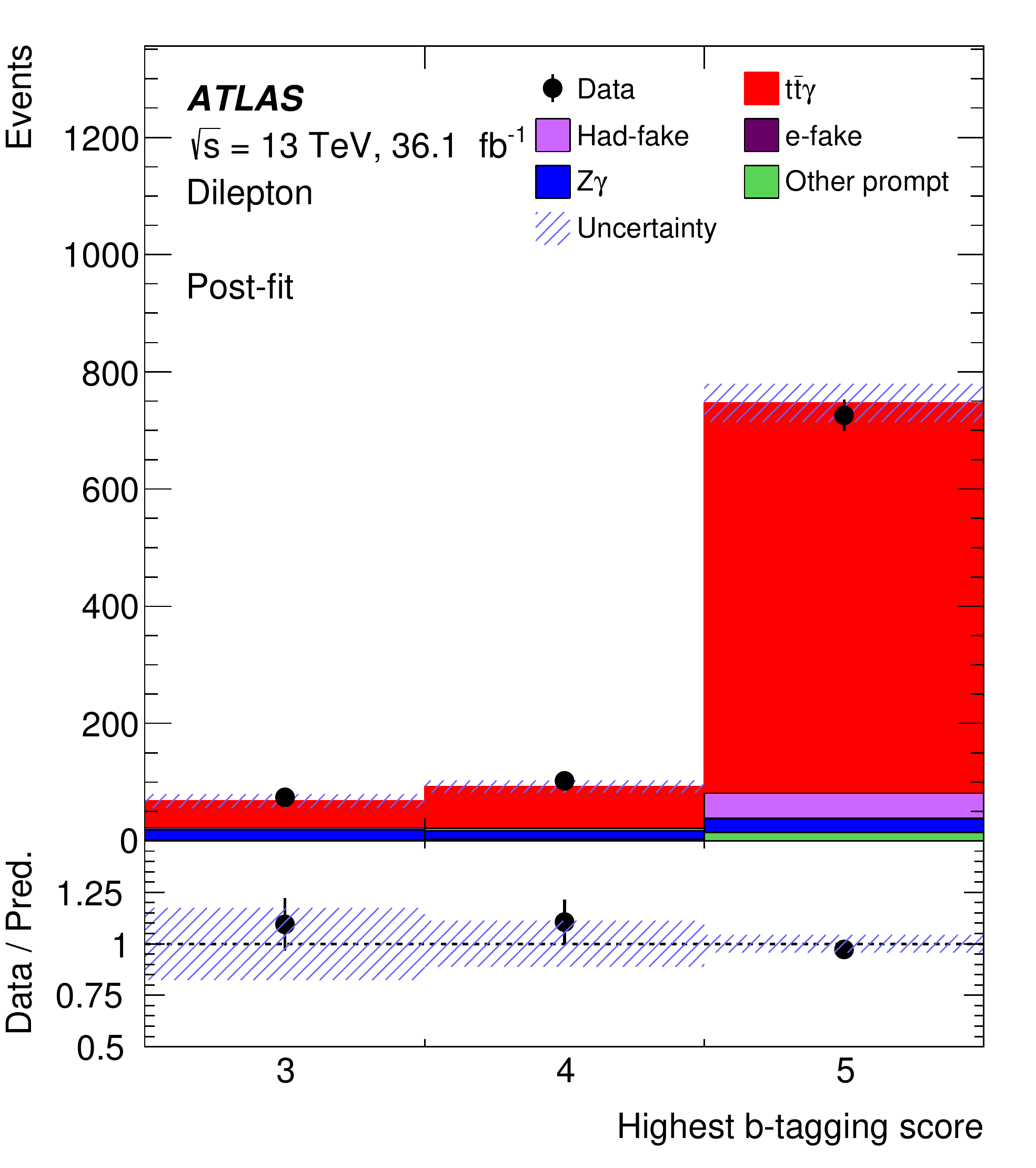}
}
\subfloat[]{
\includegraphics[width=0.45\linewidth]{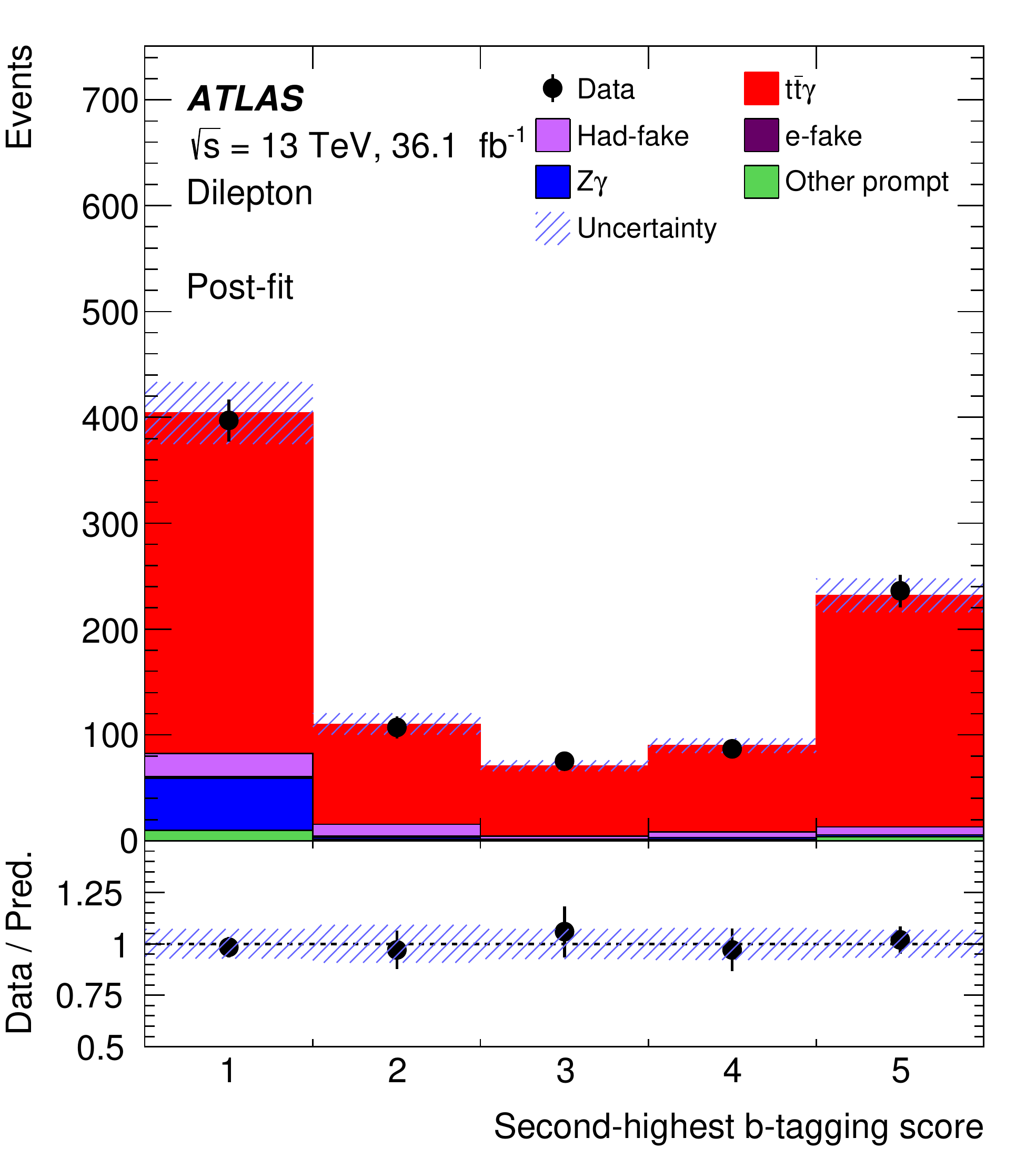}
}
 
\subfloat[]{
\includegraphics[width=0.45\linewidth]{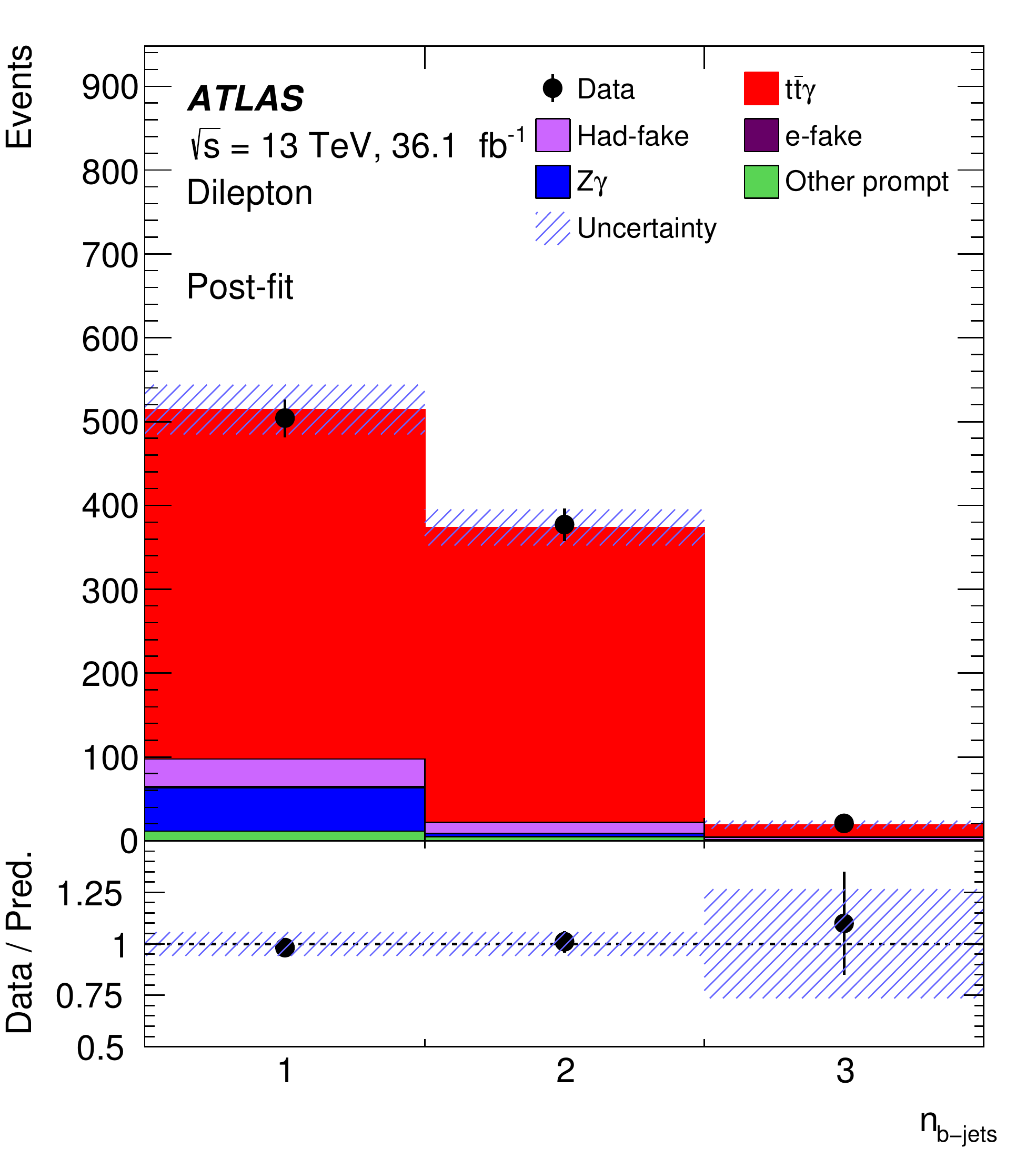}
}
\subfloat[]{
\includegraphics[width=0.45\linewidth]{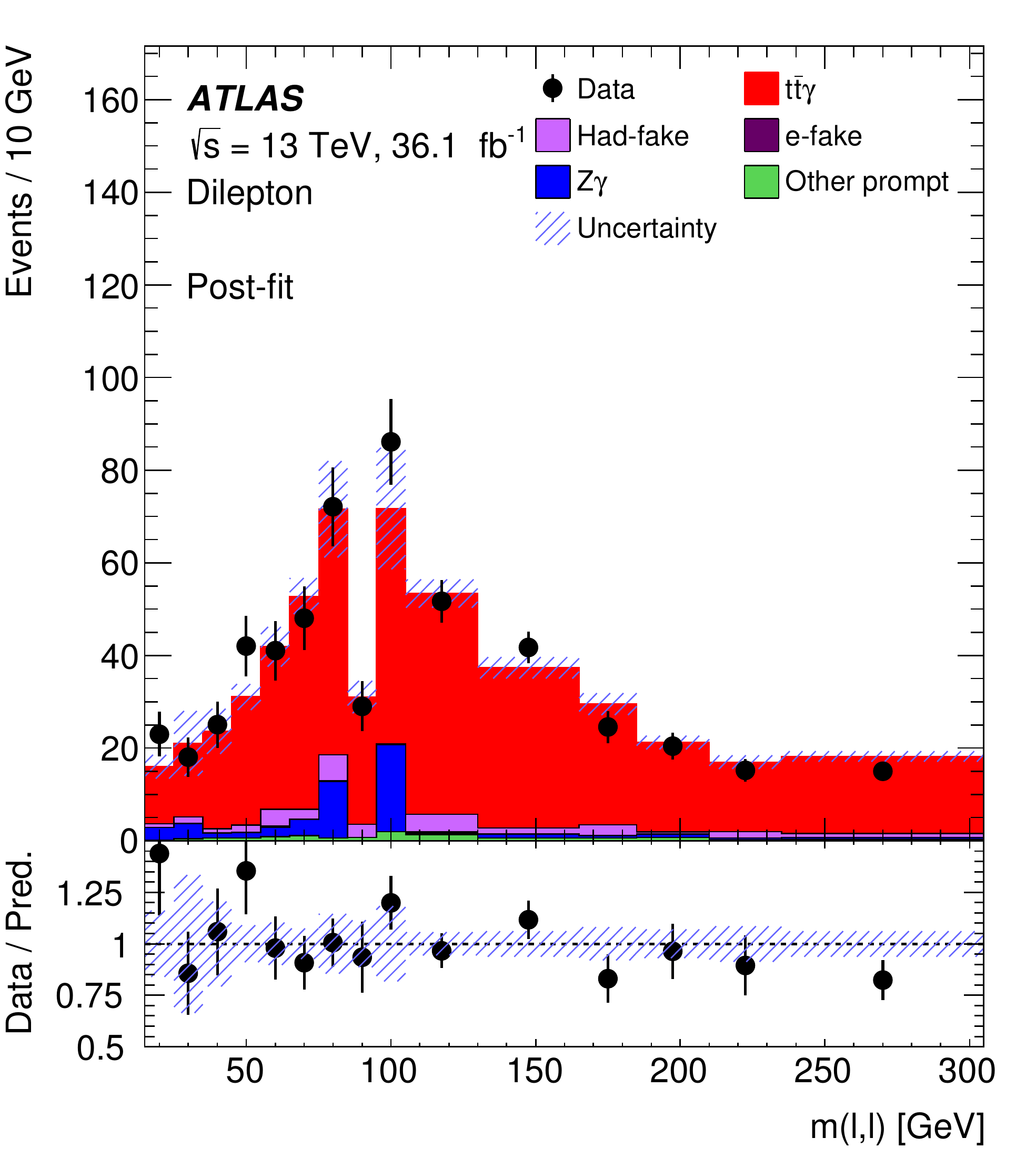}
}
\caption [] {Distributions of the (a) highest and (b) second highest $b$-tagging scores for all jets, (c) total number of $b$-tagged jets (at a 77\% efficiency), and (d) invariant mass between the two leptons in the \chll channel after event selection and likelihood fit. Each bin of (a) and (b), from one to five, represents the $b$-tagging efficiencies of $>$ 85\%, 77--85\%, 70--77\%, 60--70\%, and $<$ 60\%. All data-driven corrections and systematic uncertainties are included.}
\label{fig:DLELDvars}
\end{figure}
 
Figure~\ref{fig:hfakedd} shows the distributions of the ELD and $H_\text{T}$ in the hadronic-fake control region B, as defined in Section~\ref{sec:hfake}.
 
\begin{figure}[!htbp]
\centering
\subfloat[]{
\includegraphics[width=0.45\linewidth]{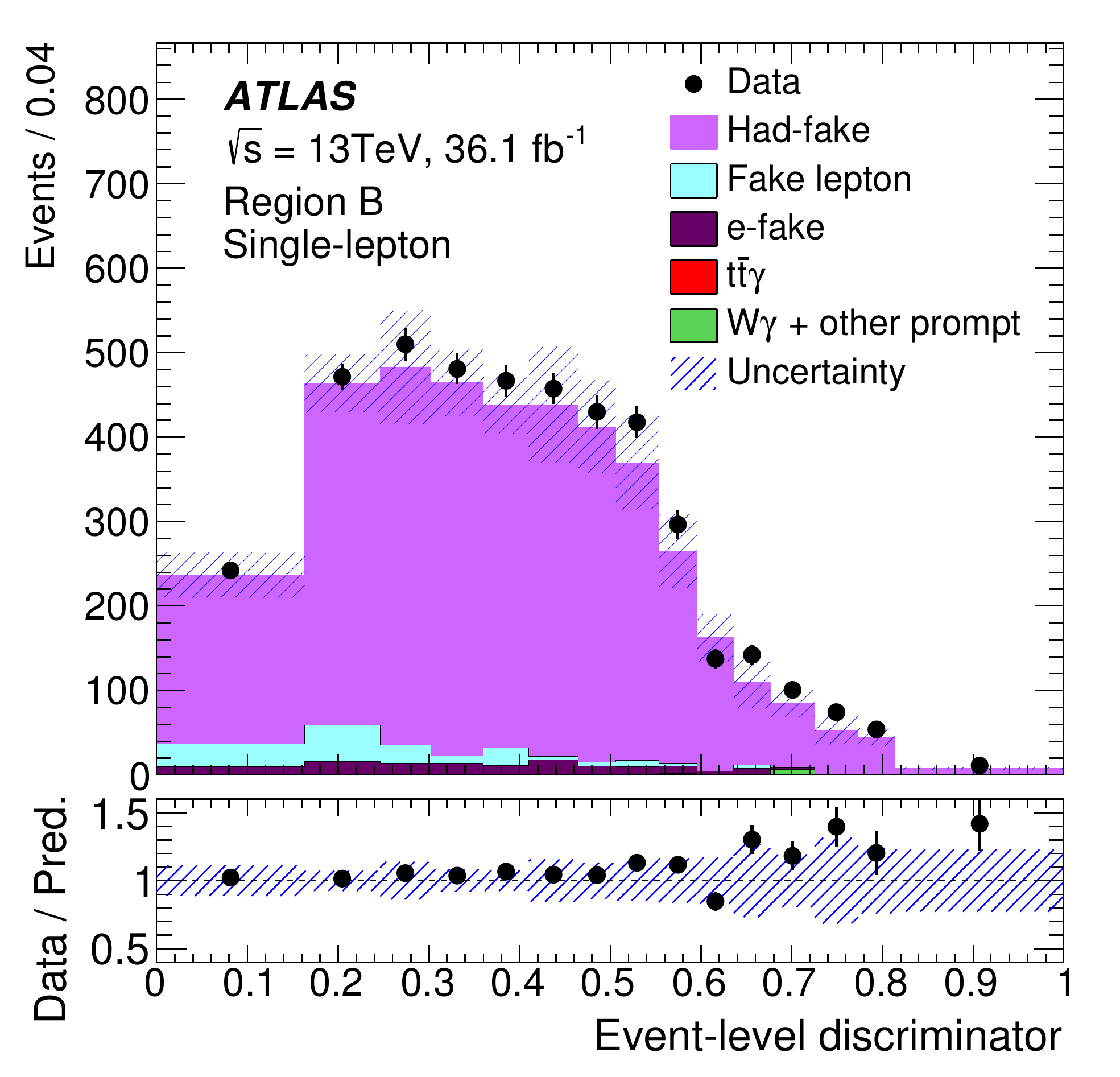}
}
\subfloat[]{
\includegraphics[width=0.45\linewidth]{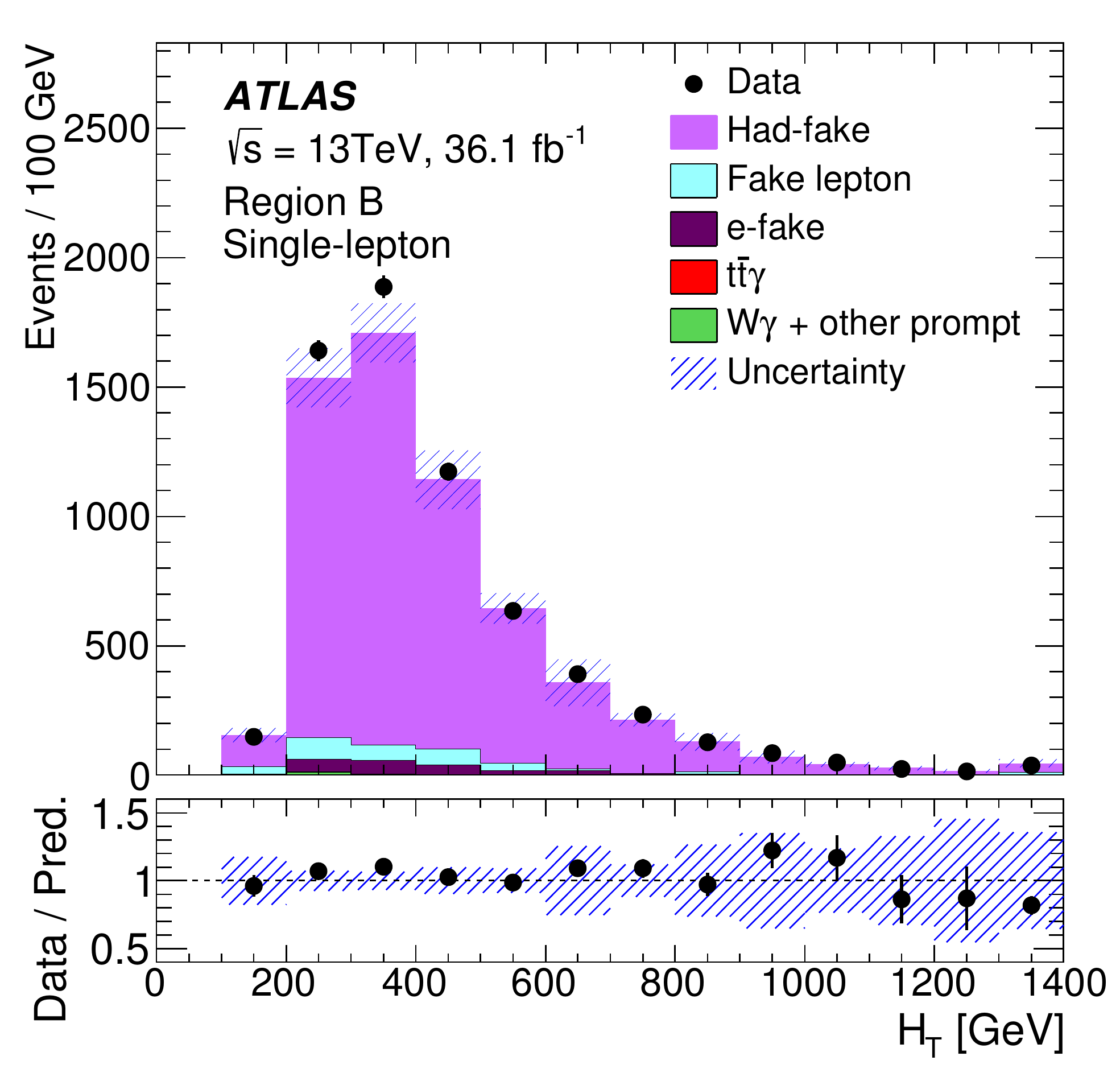}
}
\caption [] {Distributions of the (a) ELD and (b) $H_\text{T}$ in the hadronic-fake control region B of the \chljets channel. Data-driven correction to the electron-fake background and all systematic uncertainties are included. The contributions from $t\bar{t}\gamma$ and W$\gamma$ + other prompts are very small (0.2\% and 0.3\%, respectively), therefore not visible.}
\label{fig:hfakedd}
\end{figure}
 
Figure~\ref{fig:effmigdl_dphill} shows the distributions of the efficiency, outside fraction, and migration matrix for the \chll azimuthal angle in the \chll channel, which are the inputs for unfolding, as described in Section~\ref{sec:unfold}.
 
\begin{figure}[!htbp]
\centering
\subfloat[]{
\includegraphics[width=0.45\linewidth]{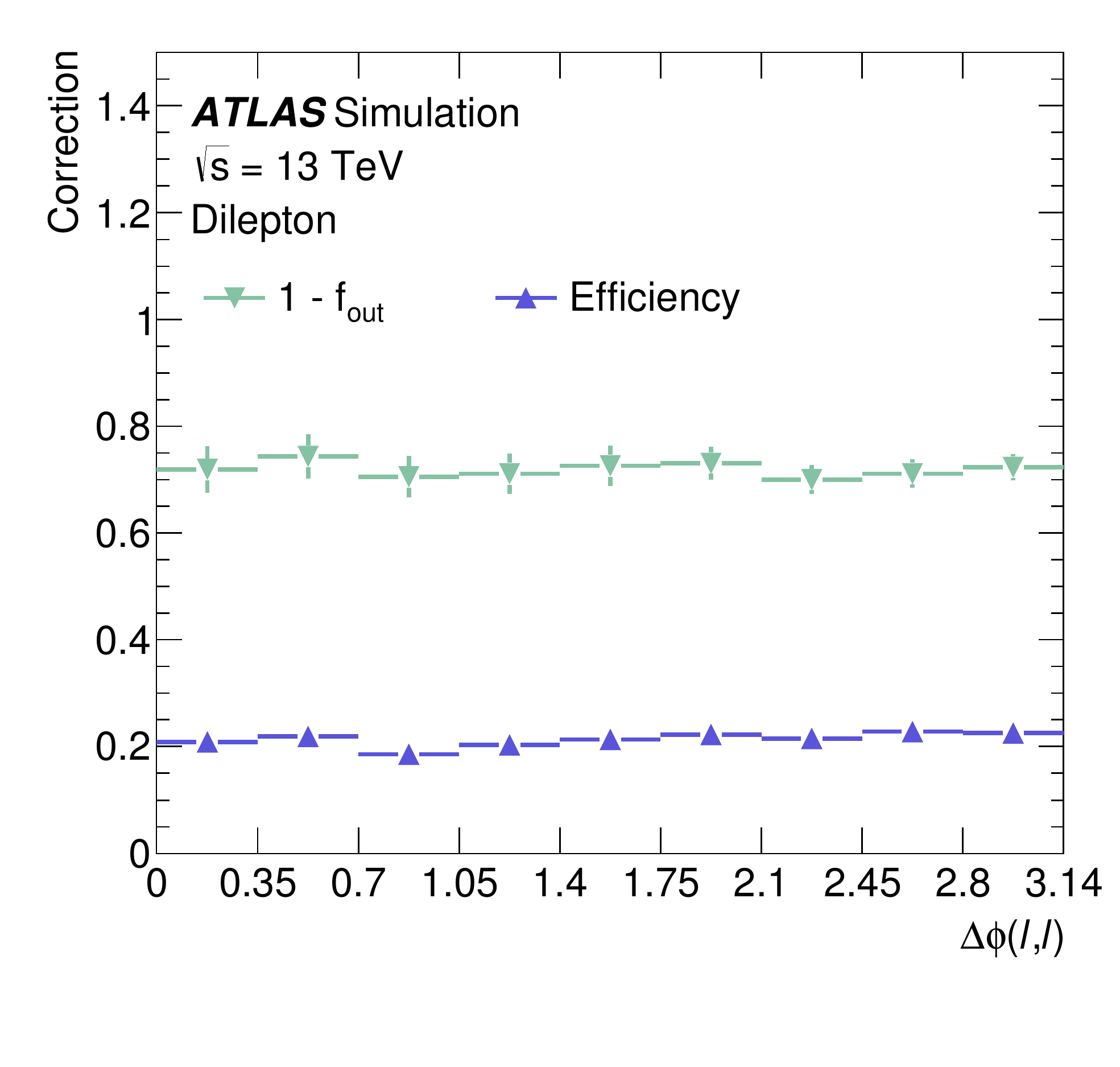}
}
\subfloat[]{
\includegraphics[width=0.45\linewidth]{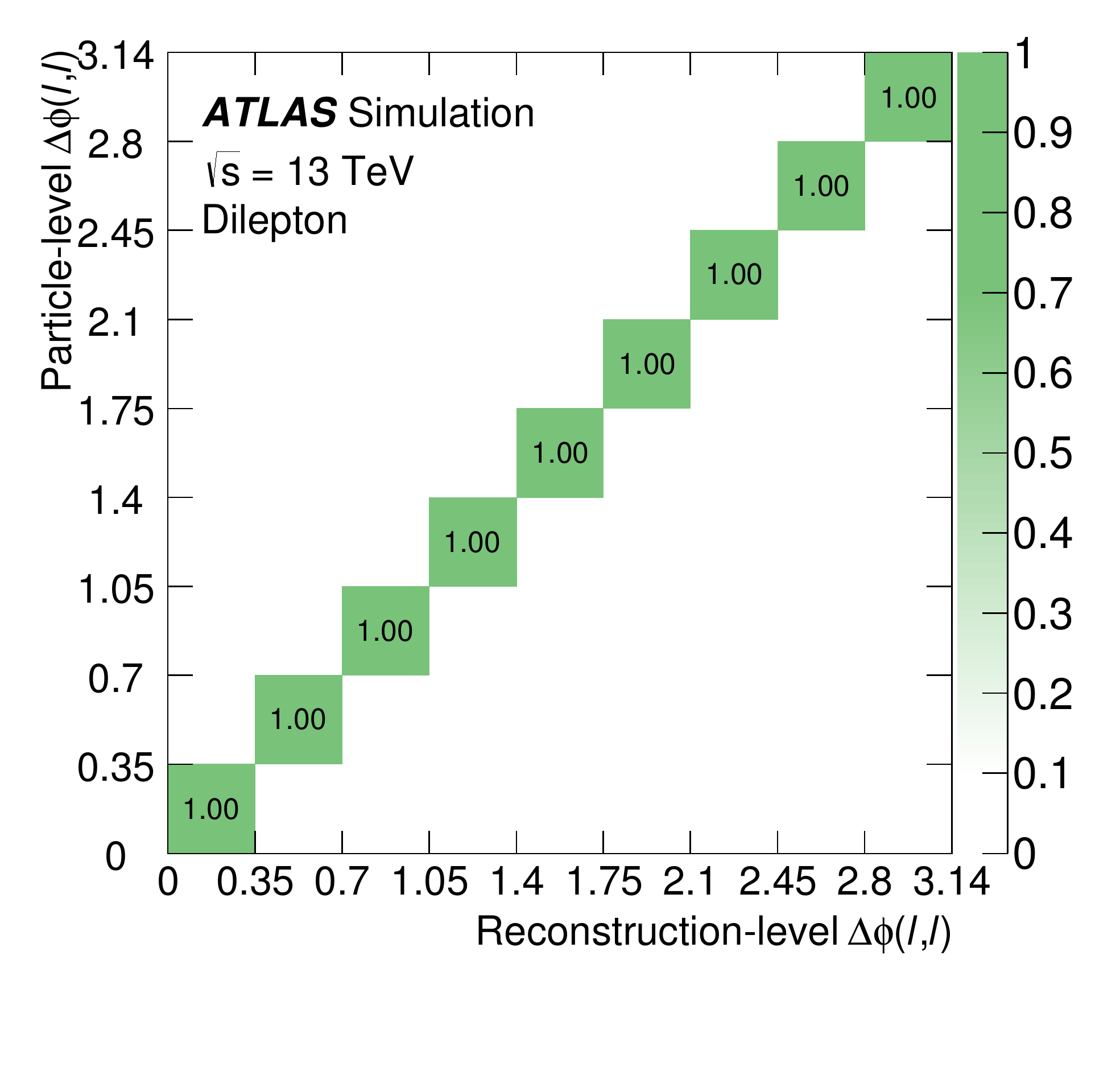}
}
\caption [] {The (a) efficiency and outside fraction and (b) migration matrix for the \chll azimuthal angle in the \chll channel.}
\label{fig:effmigdl_dphill}
\end{figure}
 
\section*{Acknowledgements}
 
 
We thank CERN for the very successful operation of the LHC, as well as the
support staff from our institutions without whom ATLAS could not be
operated efficiently.
 
We acknowledge the support of ANPCyT, Argentina; YerPhI, Armenia; ARC, Australia; BMWFW and FWF, Austria; ANAS, Azerbaijan; SSTC, Belarus; CNPq and FAPESP, Brazil; NSERC, NRC and CFI, Canada; CERN; CONICYT, Chile; CAS, MOST and NSFC, China; COLCIENCIAS, Colombia; MSMT CR, MPO CR and VSC CR, Czech Republic; DNRF and DNSRC, Denmark; IN2P3-CNRS, CEA-DRF/IRFU, France; SRNSFG, Georgia; BMBF, HGF, and MPG, Germany; GSRT, Greece; RGC, Hong Kong SAR, China; ISF and Benoziyo Center, Israel; INFN, Italy; MEXT and JSPS, Japan; CNRST, Morocco; NWO, Netherlands; RCN, Norway; MNiSW and NCN, Poland; FCT, Portugal; MNE/IFA, Romania; MES of Russia and NRC KI, Russian Federation; JINR; MESTD, Serbia; MSSR, Slovakia; ARRS and MIZ\v{S}, Slovenia; DST/NRF, South Africa; MINECO, Spain; SRC and Wallenberg Foundation, Sweden; SERI, SNSF and Cantons of Bern and Geneva, Switzerland; MOST, Taiwan; TAEK, Turkey; STFC, United Kingdom; DOE and NSF, United States of America. In addition, individual groups and members have received support from BCKDF, CANARIE, CRC and Compute Canada, Canada; COST, ERC, ERDF, Horizon 2020, and Marie Sk{\l}odowska-Curie Actions, European Union; Investissements d' Avenir Labex and Idex, ANR, France; DFG and AvH Foundation, Germany; Herakleitos, Thales and Aristeia programmes co-financed by EU-ESF and the Greek NSRF, Greece; BSF-NSF and GIF, Israel; CERCA Programme Generalitat de Catalunya, Spain; The Royal Society and Leverhulme Trust, United Kingdom.
 
The crucial computing support from all WLCG partners is acknowledged gratefully, in particular from CERN, the ATLAS Tier-1 facilities at TRIUMF (Canada), NDGF (Denmark, Norway, Sweden), CC-IN2P3 (France), KIT/GridKA (Germany), INFN-CNAF (Italy), NL-T1 (Netherlands), PIC (Spain), ASGC (Taiwan), RAL (UK) and BNL (USA), the Tier-2 facilities worldwide and large non-WLCG resource providers. Major contributors of computing resources are listed in Ref.~\cite{ATL-GEN-PUB-2016-002}.
 
 
\FloatBarrier
 
\printbibliography
 
\clearpage \input{atlas_authlist}

\clearpage
 
\end{document}

%% file: atlas_authlist.tex
 
\begin{flushleft}
{\Large The ATLAS Collaboration}

\bigskip

M.~Aaboud$^\textrm{\scriptsize 35d}$,    
G.~Aad$^\textrm{\scriptsize 100}$,    
B.~Abbott$^\textrm{\scriptsize 127}$,    
D.C.~Abbott$^\textrm{\scriptsize 101}$,    
O.~Abdinov$^\textrm{\scriptsize 13,*}$,    
B.~Abeloos$^\textrm{\scriptsize 131}$,    
D.K.~Abhayasinghe$^\textrm{\scriptsize 92}$,    
S.H.~Abidi$^\textrm{\scriptsize 166}$,    
O.S.~AbouZeid$^\textrm{\scriptsize 40}$,    
N.L.~Abraham$^\textrm{\scriptsize 155}$
% ...
\bigskip
\\

% ...
$^{13}$Institute of Physics, Azerbaijan Academy of Sciences, Baku; Azerbaijan.\\
% ...
$^{35}$$^{(a)}$Facult\'e des Sciences Ain Chock, R\'eseau Universitaire de Physique des Hautes Energies - Universit\'e Hassan II, Casablanca;$^{(b)}$Facult\'{e} des Sciences, Universit\'{e} Ibn-Tofail, K\'{e}nitra;$^{(c)}$Facult\'e des Sciences Semlalia, Universit\'e Cadi Ayyad, LPHEA-Marrakech;$^{(d)}$Facult\'e des Sciences, Universit\'e Mohamed Premier and LPTPM, Oujda;$^{(e)}$Facult\'e des sciences, Universit\'e Mohammed V, Rabat; Morocco.\\
% ...
$^{40}$Niels Bohr Institute, University of Copenhagen, Copenhagen; Denmark.\\
% ...
$^{92}$Department of Physics, Royal Holloway University of London, Egham; United Kingdom.\\
% ...
$^{100}$CPPM, Aix-Marseille Universit\'e, CNRS/IN2P3, Marseille; France.\\
$^{101}$Department of Physics, University of Massachusetts, Amherst MA; United States of America.\\
% ...
$^{127}$Homer L. Dodge Department of Physics and Astronomy, University of Oklahoma, Norman OK; United States of America.\\
% ...
$^{131}$LAL, Universit\'e Paris-Sud, CNRS/IN2P3, Universit\'e Paris-Saclay, Orsay; France.\\
% ...
$^{155}$Department of Physics and Astronomy, University of Sussex, Brighton; United Kingdom.\\
% ...
$^{166}$Department of Physics, University of Toronto, Toronto ON; Canada.\\
% ...
$^{a}$ Also at Borough of Manhattan Community College, City University of New York, New York NY; United States of America.\\
$^{b}$ Also at Centre for High Performance Computing, CSIR Campus, Rosebank, Cape Town; South Africa.\\
$^{c}$ Also at CERN, Geneva; Switzerland.\\
$^{d}$ Also at CPPM, Aix-Marseille Universit\'e, CNRS/IN2P3, Marseille; France.\\
$^{e}$ Also at D\'epartement de Physique Nucl\'eaire et Corpusculaire, Universit\'e de Gen\`eve, Gen\`eve; Switzerland.\\
% ...
$^{*}$ Deceased

\end{flushleft}
